\documentclass[12pt,a4paper]{book}
\usepackage{a4,amsmath,amssymb,mathbbol}
\usepackage{amsthm} 
\usepackage{amscd}
\usepackage{mathrsfs}
\usepackage{wick}
\usepackage{exscale} \delimitershortfall=-2pt
\usepackage{enumerate}
\usepackage{fancyheadings}
\usepackage{calc}

\author{Tobias Schlegelmilch\\II. Institut f\"ur Theoretische Physik\\
Universit\"at Hamburg}

\renewcommand{\chaptermark}[1]{\markboth{#1}{}}

\newcommand{\abs}[1]{\lvert #1 \rvert}
\newcommand{\norm}[1]{\left\lVert #1 \right\rVert}

\DeclareMathOperator*{\supp}{supp}
\DeclareMathOperator*{\diam}{diam}
\DeclareMathOperator{\ad}{ad}

\DeclareMathOperator{\Ima}{Im}
\DeclareMathOperator{\Rea}{Re}
\DeclareMathOperator{\Ran}{Ran}

\DeclareMathOperator*{\wlim}{w-lim}

\DeclareMathOperator{\obj}{obj}
\DeclareMathOperator{\id}{id}

\newtheorem{theorem}{Theorem}[chapter]

\newtheorem{lemma}[theorem]{Lemma}
\newtheorem{cor}[theorem]{Corollary}

\theoremstyle{definition} 
\newtheorem{definition}[theorem]{Definition}

\theoremstyle{remark} 
\newtheorem*{remark}{Remark}
\newtheorem*{example}{Example}

\newcommand{\one}{\mathbb{1}}

\newcommand{\RR}{{\mathbb{R}}}
\newcommand{\NN}{{\mathbb{N}}}

\newcommand{\CC}{{\mathbb{C}}}

\newcommand{\QQ}{{\mathbb{Q}}}

\newcommand{\cB}{{\mathcal{B}}}
\newcommand{\cD}{{\mathcal{D}}}
\newcommand{\cH}{{\mathcal{H}}}

\renewcommand{\t}{\tau}

\newcommand{\si}{\sigma}
\newcommand{\la}{\lambda}
\newcommand{\ph}{\varphi}
\newcommand{\ep}{\varepsilon}

\numberwithin{equation}{chapter}

\begin{document}

\pagenumbering{roman}
\pagestyle{plain}
\thispagestyle{empty}

\begin{center}
\ \\[5ex]
\begin{LARGE}
\textbf{Local Scattering Operators for $P(\ph)_2$ Models and the Time-dependent Schr\"odinger
Equation}
\end{LARGE}\\[15ex]
\begin{large}
          Dissertation\\
zur Erlangung des Doktorgrades\\
    des Fachbereichs Physik\\
    der Universit\"at Hamburg\\[35ex]
           vorgelegt von\\[2ex]
Tobias Schlegelmilch\\
aus Saarbr\"ucken\\[5ex]
Hamburg,\\
2005

\end{large}
\end{center}
\clearpage
\pagestyle{empty}
\newpage
\ \\
\vfill
\begin{tabular}{ll}
Gutachter der Dissertation: & Prof. Dr. K. Fredenhagen\\
                            & Prof. Dr. G. Mack\\
			    & Prof. Dr. J. Yngvason\\
			    & \\	 
Gutachter der Disputation:  & Prof. Dr. K. Fredenhagen  \\
                            & Prof. Dr. J. Louis\\
                            & \\
Datum der Disputation: & 8. November 2005 \\
                            & \\
Vorsitzender des Pr\"ufungsausschusses: & Prof. Dr. J. Bartels\\
                            & \\
Vorsitzender des Promotionsausschusses: & Prof. Dr. G. Huber\\
                            & \\
Dekan des Fachbereichs Physik: & Prof. Dr. G. Huber
\end{tabular}

\clearpage
\pagestyle{plain}
\newpage

\begin{quote}
\begin{center}
\textbf{Abstract}
\end{center}
\vspace{3ex}
We establish the existence of Bogoliubov's local scattering operators for
$P(\ph)_2$ models of constructive quantum field theory in a nonperturbative way.
To this end, we 
use the technique of evolution semigroups to prove a new result on
wellposedness of the Cauchy problem for the time-dependent Schr\"odinger
equation under very general
assumptions.
\\[10ex]
\begin{center}
\textbf{Zusammenfassung}
\end{center}
\vspace{3ex}
Wir beweisen die Existenz von lokalen Streuoperatoren im Sinn von  Bogoliubov
f\"ur $P(\ph)_2$-Modelle der konstruktiven Quantenfeldtheorie ohne Anwendung
st\"orungstheoretischer Methoden. Zu diesem Zweck beweisen wir einen neuen Satz
\"uber Existenz und Eindeutigkeit der L\"osung des Cauchyproblems der
zeitabh\"angigen Schr\"odingergleichung unter sehr allgemeinen Voraussetzungen
und verwenden hierf\"ur die Technik der Evolutionshalbgruppen.
\end{quote}

\cleardoublepage

\tableofcontents

\cleardoublepage

\pagenumbering{arabic}
\pagestyle{fancyplain}
\chapter*{Introduction}
\addcontentsline{toc}{chapter}{Introduction}
\chaptermark{Introduction}

Perturbative quantum field theory is a successful framework for the description
of high energy physics. In quantum electrodynamics, perturbative calculations
lead to amazing agreement with the experiments. For weak 
interactions the agreement is still impressive. Physical quantities are
expanded in formal power series in `physical' parameters like the coupling
constant. The terms of these expansions are in general ill-defined, but by
renormalization it is possible to obtain results which are verified by
the experiments to great accuracy.

However, on a conceptional level, it is not clear what an interacting quantum
field theory really is. The program of \emph{constructive quantum field theory}
was started to clarify this point. Beginning in the late 1960s, physicists were
looking for models which satisfy a set of mathematically clear-cut axioms
describing what a quantum field theory could be \cite{wightmanlecture}. Despite
huge efforts and a lot of interesting results for lower spacetime dimensions
and simplified interactions, the question whether a physically relevant quantum
field theory in four dimensions exists is still open. In the attempts to
construct a $SU(2)$ Yang-Mills theory, the high energy behavior can be
controlled
by advanced methods of nonperturbative renormalization \cite{konstruktiv}, but
it seems that the low energy behavior is much harder to
handle. The \emph{adiabatic limit} is out of
reach and thus also crucial quantities like Schwinger functions which rely on
global concepts.

Another irritating problem in quantum field theory is the failure to include
gravitational fields. Classical general relativity does not lead to a
renormalizable theory if `quantized' analogously to the other
fundamental forces. This might give the impression that quantum field theory
with its nonlocal concepts inherited from quantum theory is not the right kind
of theory to describe all fundamental forces. In fact, one could conjecture that
quantum
field theory is an effective theory and other approaches as for instance string
theory and its generalizations are better suited as  basic theories.

But it might well be that a formulation of quantum field theory which
incorporates locality in a fundamental way can overcome some of the
aforementioned problems. Quantum field theory on curved spacetimes turned out
to be fruitful as a testing ground.
Because of the weakness of gravitation compared to the electroweak and the
strong interaction, it is a reasonable approximation to incorporate
relativistic effects as a fixed background and to neglect back reactions.  
R.~Brunetti, K.~Fredenhagen and R.~Verch have given a definition of quantum
fields in an intrinsically local and covariant way \cite{Brunetti:2001dx}. A
rigorous formulation of perturbation theory fits into this framework. It is
independent of global concepts and allows for an extension of perturbative
renormalization to curved spacetimes \cite{FB}. A central tool are the
so-called local
scattering operators. These operators arise in the
St\"uckelberg-Bogoliubov-Epstein-Glaser renormalization theory \cite{BS,
scharf}. In the context of the generally covariant approach, they can be
considered as generalized quantum fields and connect different quantum field
theories. They allow for a construction of the algebras of local
observables for an
interacting theory from the free fields in the sense of formal power series.
Here, only local concepts come into play.
Therefore, it would be interesting to investigate also models of constructive
quantum field theory in the generally covariant context.
The strategy is to
construct the local scattering operators in a nonperturbative way. This is the
aim of the present work. The local scattering operators are defined by the time
evolution in the interaction picture for interactions localized in compact
spacetime regions and for a fixed frame of reference. Because of the
localization in
spacetime, the Hamiltonian becomes time-dependent and the time evolution is the
solution of the associated time-dependent Schr\"odinger
equation. Even for the simplest constructive models, where the localized
Hamiltonian can be defined in the Fock space of the free fields, the
existent theory on the wellposedness of the Cauchy problem of the time-dependent
Schr\"odinger equation is not sufficient for our aims \cite{wski}. Therefore we
develop a new result on wellposedness of the time-dependent Cauchy problem for
 evolution equations of the Schr\"odinger type.
We obtain
solutions which are sufficiently regular to show the existence  of
the local scattering operators for $P(\ph)_2$ models. To our knowledge, this is
 the first complete proof of existence of these objects in a quantum field
theory with nonlinear field equation. 
A different approach to a special case of our result is due to W.
Wreszinski and collaborators \cite{Wreszinski:2003ye}. By our
nonperturbative existence result for local scattering operators, we relate a
model of constructive quantum field theory to the generally covariant
formulation of quantum field theory. From the rigorous local scattering
operators one gets the algebras of local observables as $C^*$-algebras and the
interacting theory in a local way. The infrared and the ultraviolet problem are
disentangled. Hence, as an example, it is possible to obtain the local net for
a theory of free massless bosons in two dimensions, an infrared divergent theory
without vacuum state.  

The methods of constructive quantum field theory have found numerous
applications outside of quantum field theory \cite{Rivasseau:1995mm}. We
think that our new
solvability theorem for the time-dependent Schr\"odinger equation will 
also
be applicable to and useful in other branches of quantum theory.

The plan of this thesis is as follows:

In Chapter \ref{ch:lso} we give a survey of the axiomatic and constructive
approaches to
quantum field theory. We indicate the r\^ole of local scattering operators
$S(g)$ in perturbation theory. They are used to obtain the algebras of local
observables. Because this construction turns out to be independent of
perturbation theory, it opens up the possibility to construct the local net via
local scattering operators in a constructive setting, that is without using
perturbation theory.
For this reason, we propose
a nonperturbative definition of the local
scattering operators. They are defined as being
identical to the time evolution
operator in the interaction picture for a localized interaction, where the time
evolution is evaluated over an interval containing the time support of the
interaction. To this end we need a nonperturbative solution of the
time-dependent
Schr\"odinger equation.

The topic of Chapter \ref{ch:eeq} is the Cauchy problem for hyperbolic
nonautonomous
evolution equations. The time-dependent Schr\"odinger equation is a special case
of this type of equations. We investigate the properties of the autonomous and
nonautonomous Cauchy problem and define various notions of solvability. Weak
solutions exist under very general assumptions, but are not sufficiently regular
to be interpreted in terms of an evolution operator or \emph{propagator}.  We
present the theory of wellposedness established by T. Kato and some of its
implications. Due to
technical difficulties, Kato's theory is not applicable to the localized
Schr\"odinger equation arising in simple models of constructive quantum field
theory.

In Chapter \ref{sec:evolutionsemigroups} we relate the nonautonomous Cauchy
problem to an autonomous one by
a technique due to J. Howland. We give a survey and clarify the notions of
solvability which  arise naturally in this approach: mild solutions are
associated with the closure of a special operator $G_0$ generating an evolution
semigroup. A new uniqueness and
continuity
result for weak solutions follows. However, these techniques are not sufficient
for the application we have in mind. We generalize the setting by considering
the situation where the closure of $G_0$ is not necessarily a generator.
Choosing a certain extension of $G_0$ with the generation property, we arrive
at the main result in this chapter 
(Theorem~\ref{th:se}). This is new result on the wellposedness of the
time-dependent
Schr\"odinger equation, where we use approximative solutions. The main results
of
 this Chapter are obtained in collaboration with
R.~Schnaubelt and can be found in 
\cite{paper1}.

Chapter \ref{exlso} is devoted to the existence of the local scattering
operators for
$P(\ph)_2$ models. First, we demonstrate that the techniques of Chapter
\ref{ch:eeq} are not sufficient for our purpose. But with the existence theorem
for approximative solutions, we formulate sufficient conditions for the
existence of local scattering operators in Theorem~\ref{th:exlso}. In
particular, we derive an existence result for the local scattering operators
for $P(\ph)_2$ models, including cases of non-semibounded polynomials
$P(\lambda)$. This shows the disentanglement of the ultraviolet and the
infrared problem which is the main advantage of this approach. As a special
case, we get a simple construction of the algebra of local observables for the
massless scalar field in two dimensions, a theory without ground state. 
Finally, we indicate shortly what in our opinion could be interesting topics for
further investigations. 

The Appendix provides some basic facts about $P(\ph)_2$ models, that is scalar,
massive, polynomial self-interacting quantum field theories in two dimensions.
Moreover, we present an auxiliary result about the sum of maximally accretive operators which we will find useful. 

For our approach, we need a considerable amount of semigroup theory. For an
introduction to this topic we refer to Pazy's monograph \cite{pazy}. Another
recommendable reference is the book of Engel and Nagel \cite{en}. The latter
book contains also an introduction to the theory of evolution semigroups. For
the theory of vector-valued integration see \cite{zaanen} or \cite{diesteluhl}.

\section*{Notation}

By $X$ and $Y$ we denote Banach spaces, $\cB(X,Y)$ is
the space of bounded linear operators from $X$ to $Y$, and $\cB(X):=\cB(X,X)$.
In the chapters dealing mainly with quantum field theory, we work in the
setting of a
Hilbert space $\cH$.

 For intervals $I\subset \RR$, we set
$D_I=\{(t,s)\in I^2:t\ge s\}$ and $I_s=[s,\infty)\cap I$ for $s\in I$. We
consider various Banach spaces of $X$-valued functions: $L^p(I,X), 1 \leq p \leq
\infty$,
denotes the space of (equivalence classes of) strongly measurable
functions $f: I \to X$ such that $\|f\|_p^p:=\int_I \|f(t)\|^p \, dt < \infty$.
The spaces $C(I,X)$ resp.\
$C^n(I,X)$, are the sets of continuous resp.\ $n$-times continuously
differentiable
functions endowed with the appropriate sup--norms. Frequently we use the
abbreviation $E_p := L^p(I, X)$, sometimes we even omit the subscript $p$ if
the meaning is clear from the context. By
$W^{n,p}(I,X)$ we denote the Sobolev space of
 vector-valued functions whose $n$th derivative is a function in  $L^p(I,X)$.
A subscript `$0$' (e.g. $C_0(I,X)$) indicates that the
functions of the corresponding class vanish at infinity (if $I$ is unbounded)
and at finite end points of $I$ which are not contained in $I$.
A subscript `$c$' (e.g. $C_c(I,X)$) denotes a set of functions with
compact support in $I$ .
 The set $C_b(\RR, X)$
denotes the continuous functions which are bounded in norm.  By $\mathcal{S}$
we denote the Schwartz space of rapidly decreasing, smooth functions.
The translation operators are denoted by $\tau_\sigma : E_p \to E_p$, 
$$
\tau_\sigma f(t) = 
\begin{cases}
f(t-\sigma) &\text{if}\; t-\sigma \in I\\
0 &\text{if}\; t-\sigma
\not\in I.
\end{cases}
$$

The domain $D(A)$ of a closed operator $A$ on $X$ is always endowed with the
graph norm of $A$. If the meaning is clear
from the context, we omit the identity operator $\one$. For example,
we write the sum of an operator $A$ with a multiple $\lambda \one$ of the
identity operator as $A + \lambda$.
For a normal operator $A$ we denote by $\rho(A) \subset \CC$ the resolvent set
and by $\sigma(A)\subset \CC$ the spectrum. For $\lambda \in \rho(A)$, the
resolvent of $A$ is $R(\lambda,A) := (\lambda - A)^{-1}$. By $C^\infty(A)$ we
denote the intersection $\bigcap_{n \in \NN} D(A^n)$. 

The Poincar\'e group is denoted by $\mathcal{P}$ and the open forward and
backward lightcone in Minkowski space by $V_+$ resp. $V_-$.

Frequently, we designate a generic constant by the letter $c$.

\chapter{Local Scattering Operators}
\label{ch:lso}

Since the early days of quantum field theory, formal perturbation theory has
proved to be a reliable guide to high energy physics. In the last years, it was
possible to formulate perturbation theory in a way which is suitable to
an understanding in the context of axiomatic quantum field theory. The effort
leading to this considerable improvement was motivated by an investigation of
quantum field theory on curved spacetimes \cite{FB, Hollands:2001fb}. In this
context, methods relying on global symmetries are not applicable. Hence it is
crucial to emphasize locality. The appropriate framework of this approach
is algebraic quantum field theory. In particular, it is possible to define the
local net of an interacting theory without having to address the adiabatic
limit first. Thus the ultraviolet and infrared behavior of a theory can be
studied independently of each other. A crucial observation for our work is that
this
mechanism, leading to the disentanglement of high and low energy properties, is
in fact independent of perturbation theory. 

The central objects in these developments are the local scattering operators
$S(g)$ which arise in the St\"uckelberg-Bogoliubov-Epstein-Glaser formulation
of perturbation theory. In this context, they are the generating functionals of
the time-ordered products. But it is possible to characterize the local
scattering operators directly as solutions of a time-dependent Schr\"odinger
equation with an interaction localized in a compact region of spacetime.

One may take a more abstract point of view. In the formulation of
locally covariant quantum field theories as covariant functors
\cite{Brunetti:2001dx, Brunetti:2004ic}, the local scattering operators can be
interpreted as generalized quantum fields and arise as natural transformations
in the sense of category theory. 

In this chapter, we will shortly summarize the axiomatic approach to quantum
field theory. We will define the local scattering operators and describe their
relation to the algebras of local observables. 

\section{Quantum field theory}

As mentioned in the introduction, formal perturbative quantum field
theory was successfully applied to high energy physics. But there remains a
logical puzzle: Is quantum field theory the appropriate language for the
description
of nature and a mathematically consistent theory at the same time? 

To put the
discussion of this question on solid grounds, a set of axioms is a suitable
starting
point to clarify the assumptions and to test their consistency. In the approach
of G\aa{rding} and Wightman \cite{garding}, fields are distributions which take
values in the set of operators on a Hilbert
space. An alternative formulation emphasizes the algebraic structure of bounded
operators representing observables which are measurable in fixed spacetime
regions. This formulation is due to Haag and Kastler \cite{haagkastler}. The two
sets of axioms are not equivalent, there are theories which fulfill the
Haag-Kastler axioms but not the Wightman axioms. A generally covariant
formulation of algebraic quantum theory was recently proposed by R.~Brunetti,
K.~Fredenhagen and R.~Verch 
\cite{Brunetti:2001dx}.

Having an axiomatic definition of quantum field theory at hand, one might
search for examples fulfilling the axioms and leading to a
nontrivial scattering matrix. Because of huge technical and conceptional
difficulties, it was  not possible to investigate theories which are
expected to correctly describe physics with interaction. The idea of
constructive quantum field
theory is to start with simplified models which are suitable
for the
development of skills necessary to address the more difficult ones.
Despite the enormous effort which was put into this program, up to the time of
writing it did not achieve its
aim, the rigorous construction of a Yang-Mills theory with the correct gauge
group in four-dimensional spacetime. Nevertheless, it was possible to
construct interacting quantum field theories fulfilling the Haag-Kastler or the
Wightman axioms and to gain considerable physical insights, not only concerning
the existence question but also for scattering theory, particle interpretation,
phase space analysis. The methods developed in constructive quantum field
theory and nonperturbative renormalization found further applications in other
fields, for example in statistical or solid state physics or the analysis
of partial differential equations \cite{Rivasseau:1995mm}.

\subsection{Axiomatic quantum field theory}

The Wightman axioms state conditions for a quantum field theory which are close
in spirit to the traditional Hilbert space formulation of quantum physics. They
incorporate the requirements of special relativity by a unitary representation
of the Poincar\'e group. On the other hand, the Haag-Kastler axioms start with
algebras of local
observables. The Poincar\'e group acts via automorphisms of these local
algebras,
hence in this approach it is possible to discuss the implementability of a
Hilbert space representation afterwards.

\subsubsection{Wightman axioms}
\label{wightman}

The Wightman axioms set the following framework for a quantum theory of
fields:
\begin{description}
\item[Hilbert space] The pure states are rays in a Hilbert space $\cH$ with
 scalar product $(\cdot\,,\,\cdot)$ which
carries a unitary representation of the covering group $\overline{\mathcal{P}}$
of the Poincar\'e group $\mathcal{P} = \mathcal{L} \ltimes \RR^4$, where
$\mathcal{L}$ denotes the proper, orthochronous Lorentz group. There is exactly
one vacuum state, that
is a
Poincar\'e-invariant ray with $U(a,\Lambda(\alpha)) \Omega = \Omega$ where $a
\in \RR^4, \alpha \in SL(2,\CC) =
\overline{\mathcal{L}}$. The
translations $U(a,\id) = e^{iP_\mu a^\mu}$ are generated by the self-adjoint
energy-momentum operators $P^\mu$. Their spectrum is a subset of the closed
forward
lightcone $ \overline{V}_+ = \{ p \in \RR^4 : p^2 \geq 0,\ p^0 \geq 0\}$. This
is the \emph{spectrum condition}.
\item[Fields] For every Schwartz function $f \in \mathcal{S}(\RR^4)$ the field
$\ph(f)
= \int
\ph(x) f(x) d^4x$ is an unbounded operator, defined on a dense set $\mathcal{D}
\subset \cH$ common to all $\ph(f)$ and invariant under their application. We
say that the fields are operator-valued
distributions. The
domain $\mathcal{D}$ contains the vacuum $\Omega$ and is invariant under
application of $U(a,\Lambda(\alpha))$ for all $a \in \RR^4, \alpha \in
SL(2,\CC)$. In general, there are several fields (type
$i$) which may have several spinor or tensor components (index $\lambda$).
Hence the general expression for the fields as operator-valued distribution are
$$
\ph(f) = \sum_{i, \lambda} \int \ph^i_\lambda(x) f^{i,\lambda}(x)\,d^4x.
$$
The set of
fields
contains with $\ph$ also its hermitean conjugate $\ph^*$, defined as a
sesquilinear form via $(\psi_1, \ph(x)^* \psi_2) = \overline{(\psi_2, \ph(x)
\psi_1)}$, $\psi_1, \psi_2 \in \cH$.
\item[Transformation properties] Let $\alpha \in SL(2,\CC)$ and
$M^{(i)}(\alpha)$ be a
finite-dimensional representation
matrix of $\alpha$. The fields transform under $\overline{\mathcal{P}}$ as
$$
U(a,\alpha) \ph^i_\lambda(x) U(a,\alpha)^{-1} = \sum_\rho
M_\lambda^{(i)\rho}(\alpha^{-1}) \ph^i_\rho(\Lambda(\alpha) x + a)
$$
in the sense of distributions.
\item[Causality] If the supports of $f$ and $g$ are spacelike separated, then
the fields obey causal commutation relations:
$[\ph^i(f), \ph^j(g)] = 0 $ for bosonic fields or $[\ph^i(f), \ph^j(g)]_+ = 0$
with the anticommutator $[\cdot,\cdot]_+$ in the fermionic case.
\item[Completeness] Every operator on $\cH$ can be approximated by linear
combinations of products of the $\ph(f)$.
\item[Time-slice axiom] There exists a dynamical law which allows for the
computation of the fields at arbitrary times in terms of the fields in a
small time slice $\mathcal{O}_{t,\epsilon} := \{x \in \RR^4 : |x^0 - t| <
\epsilon\}$. 
\end{description}
One can formulate the axioms equally well in terms of the vacuum expectation
values $w^{(n)}(x_1, \dots, x_n) = (\Omega, \ph(x_1)\dots \ph(x_n) \Omega)$,
the so-called \emph{Wightman functions}. Given a set of tempered distributions
$\{w^{(n)}\}$, $n \in \NN$, fulfilling these axioms, one can reconstruct the
quantum fields and the Hilbert space $\cH$. The \emph{Schwinger functions}
$S^{(n)}$ are the continuation of the Wightman functions to purely imaginary
times. The spectrum condition ensures the analyticity of the $w^{(n)}$. It is
also possible to reverse the argument: Starting from the Euclidean
Schwinger functions, a Wightman quantum field theory on Minkowski space can be 
recovered if the Schwinger functions $\{S^{(n)}\}$, $n \in \NN$, satisfy the
\emph{Osterwalder-Schrader} axioms, see for example
\cite{glimmjaffe1}.

\subsubsection{Haag-Kastler axioms}
\label{haag-kastler}

To every finite, contractible open subset $\mathcal{O}$ of the Minkowski space
one assigns the set
$\mathcal{A}(\mathcal{O})$ of bounded observables which can be measured inside
of $\mathcal{O}$. The algebras of local observables $\mathcal{A}(\mathcal{O})$
are often defined in such a way that they are $C^*$-algebras. The following
axioms are imposed:
\begin{description}
\item[Isotony]
If $\mathcal{O}_1 \subset \mathcal{O}_2$ then
$\mathcal{A}(\mathcal{O}_1) \subset \mathcal{A}(\mathcal{O}_2)$.
\item[Covariance] There is a representation $\beta$ of the Poincar\'e group
$\mathcal{P}$ by automorphisms: $\{a, \Lambda\} \mapsto \beta_{\{a,
\Lambda\}}$
such that if $A \in \mathcal{A}(\mathcal{O})$ then $\beta_{\{a,\Lambda\}}(A)
\in \mathcal{A}( \Lambda \mathcal{O} + a)$.
\item[Causality] If $\mathcal{O}_1$ and $\mathcal{O}_2$ are spacelike
separated, then $\mathcal{A}(\mathcal{O}_1) \subset
(\mathcal{A}(\mathcal{O}_2))'$, that is $[A_1, A_2] = 0$ for all $A_1 \in
\mathcal{A}(\mathcal{O}_1),\ A_2 \in \mathcal{A}(\mathcal{O}_2)$.
\item[Time-slice axiom] The algebra belonging to a neighborhood of a Cauchy
surface\footnote{A \emph{Cauchy surface} is a subset of a region in spacetime,
which is intersected exactly once by every inextendible curve, which has no
spacelike 
tangent vectors.} of a region equals the algebra of the full
region (existence of a hyperbolic equation of motion).
\end{description}
The \emph{quasilocal algebra} $\mathcal{A}$ is  the inductive
limit
$\bigcup_{\mathcal{O}} \mathcal{A}(\mathcal{O})$, which can be defined if the
regions $\{\mathcal{O}\}$ form a directed set. This is the case for open,
relatively compact subsets of Minkowski space. If the algebras
$\mathcal{A}(\mathcal{O})$ are $C^*$-algebras, we define the quasilocal algebra
by closure of the inductive limit in norm. Without loss of generality we
assume that $\mathcal{A}$ contains an identity $\one$.

A \emph{state} $\rho$ is a complex-linear functional on $\mathcal{A}$ which is
positive and normalized, that is it fulfills $\rho(A^*A) \geq 0$ for all $A \in
\mathcal{A}$ and $\rho(\one) = 1$.

A state is invariant under a group $G$, represented by automorphisms $\beta$ on
$\mathcal{A}$, if it satisfies
$$
\rho(\beta_g(A)) = \rho(A).
$$
for all $g \in G$.

For a given state one can get a representation of the quasilocal algebra on a
Hilbert space. This is the \emph{GNS construction}, see \cite{haag}.
\begin{theorem}
Let $\mathcal{A}$ be a $C^*$-algebra and $\rho$ a state on $\mathcal{A}$. Then
there exist a Hilbert space $\cH_\rho$, a vector $\Omega_\rho$ and a
representation $\pi_\rho$ of $\mathcal{A}$ by bounded operators on $\cH_\rho$
such that 
$$
\rho(A) = (\Omega_\rho, \pi_\rho(A) \Omega_\rho)\quad \text{for all}\; A \in
\mathcal{A}.
$$
The vector $\Omega_\rho$ is cyclic for $\pi_\rho(\mathcal{A})$. If $\rho$ is
invariant under a group $G$ then there exists a representation $U_\rho(g)$ of
 elements $g \in G$ by unitary operators on $\cH_\rho$ such that
$$
\pi_\rho(\beta_g(A)) = U_\rho(g) \pi_\rho(A)U_\rho(g)^{-1}.
$$ 
The vector $\Omega_\rho$ is invariant under $G$: $U_\rho(g) \Omega_\rho =
\Omega_\rho,\ g
\in G$.
\end{theorem}

We see that the transition from the algebraic level to a Hilbert space
representation depends on the choice of a state. This choice is in general not
unique and one can get different inequivalent representations of the quasilocal
algebra.

Often one also wants to impose a stability condition. Then one assumes that the
local algebra is a concrete algebra of operators on a Hilbert space and the
automorphisms belonging to the Poincar\'e group are implemented by unitary
operators. The joint spectrum of the generators of the unitary representatives
of the translations  should then be a subset of the forward lightcone. This
assumption corresponds to the spectrum condition of the Wightman axioms.

\subsection{A generally covariant approach}

Algebraic quantum theory emphasizes locality. But it is not suitable to
incorporate the covariance property of general relativity. A recent approach of
R. Brunetti,
K. Fredenhagen and R. Verch \cite{Brunetti:2001dx} generalizes the setting of
algebraic quantum theory in a generally covariant way, allowing for the
definition of a quantum field theory on all spacetimes of a certain class. We
follow the presentation in \cite{Brunetti:2004ic}. A
general covariant quantum field theory is considered as a
functor between two
categories. The first one describes the local relations. Its objects
are certain topological spaces and its morphisms are structure
preserving embeddings. The second category provides the information about the
algebraic structure of observables. The standard choice for quantum physics is
the category of $C^\ast$-algebras where the morphisms are unital embeddings.
In
classical physics, one considers Poisson algebras instead of $C^\ast$-algebras.
Recently, also
perturbative quantum field theory was incorporated into this concept. Here one
deals with algebras which possess nontrivial representations as formal power
series of Hilbert space operators. The principle of algebraic quantum field
theory states that the functor $\mathcal{A}$ contains all physical
information.
Now we will put these ideas in more exact terms.
We consider the categories $\mathfrak{L}$ and $\mathfrak{O}$. 

The category $\mathfrak{L}$ is defined in the following way: The class of
objects,  $\obj(\mathfrak{L})$, consists of all $(d \geq 2)$-dimensional,
smooth,
globally hyperbolic\footnote{A spacetime is \emph{globally hyperbolic} if it
has a smooth foliation in Cauchy surfaces.}
Lorentzian\footnote{A \emph{Lorentzian} spacetime of dimension $n$ has a
Pseudo-Riemannian metric of\\ signature (1,n-1)} spacetimes $M$ which are
oriented
and
time-oriented. For two members $M_1, M_2$ of $\obj(\mathfrak{L})$ the
morphisms
$\psi \in \hom_\mathfrak{L}(M_1, M_2)$ are chosen to be isometric embeddings
$\psi: M_1 \to M_2$ which satisfy the following conditions:
\begin{enumerate}[(i)]
\item If $\gamma : [a,b] \to M_2$ is an arbitrary causal curve\footnote{A
\emph{causal curve} has no spacelike tangent vectors.} and $\gamma(a),
\gamma(b) \in \psi(M_1)$ then the whole curve lies in $\psi(M_1)$, i.e.
$\gamma(t) \in \psi(M_1)\ \forall\ t \in (a,b)$.
\item Every morphism preserves orientation and time-orientation of the
embedded
spacetime.
\end{enumerate}
Composition is defined to be the composition of maps and the unit element in
$\hom_\mathfrak{L}(M, M)$ is the identical embedding $\id_M$.

Now we define $\mathfrak{O}$: The class of objects, $\obj(\mathfrak{O})$, is
given by the unital $C^\ast$-algebras $\mathcal{A}$. The morphisms in
$\hom_\mathfrak{O}(\mathcal{A}, \mathcal{B})$ are the faithful, injective,
unit-preserving $\ast$-homomorphisms with the composition of maps. The unit
element in $\hom_\mathfrak{O}(\mathcal{A},\mathcal{A})$ for every $\mathcal{A}
\in \obj(\mathfrak{O})$ is the identical map $\id_\mathcal{A} : A \mapsto A,\ A
\in \mathcal{A}$. 

This choice of the categories $\mathfrak{L}$ and $\mathfrak{O}$ may be changed
to fit to the physical situation. In particular, for perturbation theory one
would
substitute the $C^\ast$-algebras by general topological $\ast$-algebras.

\begin{definition}
A \emph{locally covariant quantum field theory} is a covariant functor
$\mathsf{A}$ from $\mathfrak{L}$ to $\mathfrak{O}$ which has the covariance
properties (denoting $\mathsf{A}(\psi)$ by $\alpha_\psi$)
$$
\alpha_{\psi'} \circ \alpha_\psi = \alpha_{\psi' \circ \psi}, \quad
\alpha_{\id_M} = \id_{\mathsf{A}(M)}
$$
for all morphisms $\psi \in \hom_\mathfrak{L}(M_1, M_2)$, all morphisms $\psi'
\in \hom_\mathfrak{L}(M_2, M_3)$ and all $M \in \obj(\mathfrak{L})$.

Moreover, a locally covariant quantum field theory described by a covariant
functor $\mathsf{A}$ is called \emph{causal} if the following holds: Consider
two morphisms $\psi_j \in \hom_\mathfrak{L}(M_j, M),\ j=1,2,$ such that the sets
$\psi_1(M_1)$ and $\psi_2(M_2)$ are not connected by a causal curve in $M$. Then
$$
[\alpha_{\psi_1}(\mathsf{A}(M_1)), \alpha_{\psi_2}(\mathsf{A}(M_2))] = \{0\},
$$
where the element-wise commutation makes sense in $\mathsf{A}(M)$.
\end{definition}

We will see that perturbative quantum field theory fits into this context and
allows for a formulation on curved spacetimes \cite{FB}. A crucial object are
the
local scattering operators which fit into
the generally covariant context as natural transformations, as we will see in
Section~\ref{sec:lso}.

\subsection{Constructive quantum field theory}
\label{constructive}

Axiomatic quantum field theory was developed to be a rigorous foundation for the
understanding of the dynamics of elementary particles. But in the early 1960s
only the free fields were known to fulfill the axioms, thus showing their
consistency. But the main question, whether the idealizations involved in the
axioms result in a language suitable for practical purposes of elementary
particle physics, remained unanswered.
Therefore, as a first step, simplified models were examined. In the following,
we will shortly review the development of constructive quantum field theory,
see \cite{osterwalder, streater}.

The rigorous construction of examples for interacting quantum field theories is
fundamentally affected by a famous result known as \emph{Haag's Theorem}
\cite{haag55}.
Whereas in quantum mechanics every representation of the canonical commutation
relations is unitarily equivalent to the Schr\"odinger representation, this is
no longer the case in a quantum field theory dealing with a system of
infinitely many degrees of freedom. The appearance of \emph{strange
representations} can be traced back to the work of K.~O.~Friedrichs
\cite{friedrichs1}
and L.~van~Hove \cite{vanhove}.
This turned out to be a generic situation and has consequences for the proposal
to construct interacting quantum field theories starting from free fields.
\begin{theorem}
\label{haag}
Let $\ph$ be a free field on a Hilbert space $\cH$ with Hamiltonian $H_0$. Let
the space translations be implemented by unitary operators $U(\vec{x}) =
U((0,\vec{x}), \id)$. Assume that there is an operator-valued distribution
$\tilde{\ph}$ which satisfies:
\begin{enumerate}[(i)]
\item coincidence with the free field at $t=0$:
$\tilde{\ph}(x)\negthickspace\restriction_{x^0 = 0} =
\ph(x)\negthickspace\restriction_{x^0 = 0}$ and
$\partial_0 \tilde{\ph}(x)\negthickspace\restriction_{x^0 = 0} =
\partial_0\ph(x)\negthickspace\restriction_{x^0 = 0}$;
\item translation covariance: $U(\vec{y}) \tilde{\ph}(x^0, \vec{x})
U(\vec{y})^{-1} = \tilde{\ph}(x^0, \vec{x} - \vec{y})$;
\item existence of the Hamiltonian: There is a self-adjoint operator $H$ on
$\cH$ such that $\tilde{\ph}(t,\vec{x}) = e^{itH} \tilde{\ph}(0,\vec{x})
e^{-itH}$. 
\end{enumerate}
Then $H$ and $H_0$ differ only by an additive constant and $\tilde{\ph} = \ph$. 
\end{theorem}
Thus, if one wants to work in the usual Fock space and  to avoid dealing with
strange representations, it is convenient to break the translation symmetry.
This is done by placing the system under consideration in a finitely extended
box $V$ or by replacing the coupling constant by a compactly supported
smooth function $g$ on spacetime. 

But another cut-off turns out to be necessary. The models are
inspired by simple interaction Lagrangians built from the free field. To obtain
the Hamiltonian
as a well defined operator one has to introduce a high-momentum cut-off
$\kappa$ by keeping only the frequencies $\leq \kappa$ in the Fourier transform
of the free field. 

In this way the cut-off Yukawa theory $Y_4$ with the Hamiltonian
$$
H_{\kappa,V} = H_{0,B,V} + H_{0,F,V} + \lambda \int_V :\psi^+_\kappa(\vec{x})
\psi_\kappa(\vec{x}): \ph_\kappa(\vec{x})\,d^3x 
$$
was investigated by O. Lanford in \cite{lanfordphd}. Here $\psi$ is a
fermion field and $\ph$ is a boson field. By $H_{0,B,V}$ and $H_{0,F,V}$ 
we denote the free fermionic resp. bosonic Hamiltonians in a box $V$ with
periodic boundary conditions. The colons denote Wick ordering, a 
prescription for the proper multiplication of the operator-valued distributions.
An example
where
the coupling is of higher degree in $\ph_\kappa$ than the free Hamiltonian is
the cut-off $(\ph^4)_4$ model studied by A. Jaffe \cite{jaffephd}. The
Hamiltonian is given by
$$
H_\kappa = H_0 + \lambda \int g(\vec{x}) :\ph_\kappa^4(\vec{x}):\,d^3x.
$$
In both models self-adjointness and semiboundedness of the Hamiltonians have
been
established. Moreover, uniqueness of the vacuum was proved.

The next step in the construction of the quantum field theories would be the
removal of the cut-offs $V\to \RR^3$ resp. $g \to \text{const.}$
and $\kappa \to \infty$. The limiting theories should satisfy the
Wightman or Haag-Kastler axioms. But passing to this limit was
impossible without a further
significant simplification: The number of spacetime dimensions $d$ had to
 be reduced
to $d=2$ and later $d=3$. This is mainly related to the high-energy behavior of
the theories which affects the $\kappa \to \infty$ limit. One has to add
$\kappa$-depending terms to the Hamiltonian which diverge in the
limit, renormalization is necessary.

Denote by $H_{\text{ren}}$ the renormalized Hamiltonian and indicate the number
of spacetime dimensions by a subscript.
From formal perturbation theory the following behavior was predicted and
confirmed by rigorous calculation \cite{glimm68}. In the $(\ph^{2n})_2$ model on
two-dimensional spacetime,  one finds $D(H_{\text{ren}}) \subset D(H_0)$. In
this case, Wick ordering is sufficient to renormalize the Hamiltonian. Apart
from that, only a finite constant has to be added which corresponds to a finite
shift of the vacuum energy. For the mass shift model
$(\ph^2)_3$ the form domain of the renormalized Hamiltonian is still contained
in the form domain of the free Hamiltonian $D(H^{1/2}_{\text{ren}}) \subset
D(H^{1/2}_0)$, but for $H_0$ and $H_{\text{ren}}$ themselves the inclusion of
the
domains is no longer
true. For the models $(\ph^2)_4$ and $Y_2$ even the form domain of 
$H_{\text{ren}}$ is not contained in the form domain of the free Hamiltonian,
only $D(H_{\text{ren}}) \subset \cH$ remains valid. The Yukawa model $Y_2$ needs
infinite vacuum-energy and boson-mass renormalizations in the Hamiltonian. Even
more singular are $Y_3$ and $(\ph^4)_3$. These models on three-dimensional
spacetime need an infinite wave-function renormalization: The
domain of $H_{\text{ren}}$ is no longer a subset of the Hilbert space $\cH$
which is the Fock space of the free fields. 

The models we mentioned up to now are superrenormalizable, that is, the
counterterms are polynomials in the coupling constant and the degree of the
divergences gets less severe in higher orders of perturbation theory.
In this context, the \emph{Hamiltonian strategy} led to some considerable
insights. The idea is to describe an interacting theory by a construction of 
its dynamics in a Hilbert space. The easiest model where the Hamiltonian
strategy is applicable is $(\ph^4)_2$. Up to the middle of
the 1970s it was
known that the model exists without any cut-offs. It fulfills the Haag-Kastler
axioms and
most of the Wightman axioms \cite{bqft}. These results were extended to the
technically more difficult $P(\ph)_2$ models, where $P(\lambda)$ is a
semibounded polynomial of degree $\geq 4$. Moreover, some features of $Y_2$ and
$(\ph^4)_3$ were accessible via the Hamiltonian strategy \cite{glimm68a,gj}. 

Already in this work it turned out to be very useful to investigate the
Hamiltonian $H$ via its associated semigroup $(e^{-tH})_{t \geq 0}$. This can be
regarded as a Euclidean method since it follows formally from the substitution
$t \to -it$. But this was only the beginning of a powerful \emph{Euclidean
approach} \cite{symanzik68} to constructive quantum field theory. This method
is based on the fundamental correlations between boson quantum field theory and
probability theory, the analyticity properties of the Wightman functions and
the Schwinger functions as their Euclidean counterpart and, last but not least,
the connection between Euclidean quantum field theory and classical statistical
mechanics. Soon the study of Hamiltonians was abandoned in favor of the direct
examination of the Schwinger functions via a Euclidean Gell-Mann-Low formula.
The Schwinger functions are defined by functional integrals as moments of a
certain probability measure on a function space. For their rigorous
construction,
one starts again from a regularized theory with cut-offs. For the removal of the
cut-offs powerful renormalization methods were developed: \emph{correlation
inequalities} and the \emph{cluster expansion}. By these methods, up to the
beginning of the 1980s superrenormalizable models where under control. Examples
are $P(\ph)_2$, $Y_2$, the Sine-Gordon model $(\sin \epsilon \ph)_2$ and the
Hoegh-Krohn model $(e^{\alpha \ph})_2$ in two dimensions. Further examples in
three dimensions are $(\ph^4)_3$
and $Y_3$. For these examples on two- and three-dimensional spacetime the
existence of the Schwinger functions was proved,
they define a quantum field theory with nontrivial scattering operator. One can
analyze the particle spectrum and the equations of motion. Moreover, one can
investigate phase transitions and symmetry breaking, and one finds Borel
summability
of a formal power series expansion. Thus the relation to perturbation theory
is well understood. 

In four dimensions there are new challenges. The counterterms are only known
as formal power series in the coupling constant. An example for a
renormalizable model which is no more superrenormalizable is
$\lambda(\ph^4)_4$. In this situation one needs a new technique, the
\emph{renormalization group}, which goes back to ideas of Wilson and Kadanoff.
The integration over the function space is
performed by a sequence of integrals with a fixed momentum scale. One can
relate the counterterms to different momentum scales via the \emph{flow
equation}. This method works for models which are \emph{asymptotically free},
 the coupling decreases for high momenta.
However, $\lambda(\ph^4)_4$ is not asymptotically free for positive coupling
constant
$\lambda$. But for negative coupling it is, and a rigorous construction was
given
in
\cite{gawedzki85}. But it is no physical model as it seems to be impossible to
recover the quantum field theory on Minkowski space via the
Osterwalder-Schrader reconstruction.  

Renormalization group methods were successfully applied to the Gross-Neveu model
in two dimensions, a model with quartic fermion interaction of several flavors
\cite{gawedzki85a}. Moreover, it was possible to investigate gauge theories in
three and four dimensions, see \cite{Balaban:1984gp, Balaban:1985yy} and
related papers by T. Balaban. In this program a Yang-Mills theory is
investigated in a finite volume. With lattice regularization and block spin
transformations, the high-energy limit of gauge-invariant observables as
smoothed Wilson loops is tackled. But it seems that Balaban's ideas are not
directly applicable to the Schwinger functions. These are the subject of the
work of 
 J. Magnen, V. Rivasseau and R. S{\'e}n{\'e}or \cite{konstruktiv} for an $SU(2)$
Yang-Mills theory in finite Euclidean volume.
Here
the ultraviolet problem seems to be under control. But for large volumes there
are problems with the appearance of large fields,
 which at the moment seem
insurmountable. Without control of the adiabatic limit, there is 
no possibility to define the interacting theory via its Schwinger functions.

Thus a local approach is expected to lead to an improvement of the understanding
of interacting quantum field theory in the constructive context.
Despite the fact that there are a lot of technical and even conceptional
questions open, it would be interesting to develop a strategy to disentangle
the
infrared and the ultraviolet problem. This decoupling was achieved in the
context of perturbation theory, but it is possible to carry over the main idea
to constructive quantum field theory. The interacting theory
is obtained via the local net \cite{FB}, thus the theory is fixed without the
need for a vacuum state or related global concepts. 
A crucial tool are the local scattering operators
which we introduce in the next section.

\section{Local scattering operators}
\label{sec:lso}

As in constructive quantum field theory, in the
Bogoliubov-St\"uckelberg-Epstein-Glaser formulation of perturbation
theory problems with Haag's theorem are circumvented by replacing the coupling
constant by a compactly supported, smooth function $g$. The time evolution for 
the localized Hamiltonian in the interaction picture leads to the \emph{local
scattering operators} $S(g)$. They are examples for a class of generalized
quantum fields in the functorial sense and allow for a local formulation of
perturbative quantum field theory. The proposal of the present work is to
perform a similar strategy to obtain local scattering operators in a
nonperturbative way for models of constructive quantum field theory. We will
see that this is possible if we are able to find nonperturbative solutions of a
time-dependent Schr\"odinger equation with localized interaction.

\subsection{Definition of the local scattering operators}

\subsubsection{Local scattering operators and generalized quantum fields}

For the interpretation of a physical theory it is crucial to compare
measurements associated with different spacetime regions or actually with
different
spacetimes. This comparison can be done in terms of locally covariant quantum
fields. To cover this kind of general situations, the locally covariant
quantum
fields are defined as  natural transformations from the functor of quantum
field theory to another functor on the category of spacetimes $\mathfrak{L}$.
Here a standard choice is the functor $\mathsf{D}$ which associates to every
spacetime $M$ its space of compactly supported, smooth test functions
$\mathcal{D}(M)$. The morphisms are the pushforwards $\mathcal{D}(\psi) =
\psi_\ast$.
\begin{definition}
A \emph{locally covariant quantum field} $\ph$ is a natural transformation
between the functors $\mathsf{D}$ and $\mathsf{A}$. That is for any object $M
\in \mathfrak{L}$ there exists a morphism $\ph_M : \mathcal{D}(M) \to
\mathsf{A}(M)$ such that for any pair of objects $M_1, M_2$ and 
any morphism $\psi$ between them the following diagram commutes:
\begin{equation}
\begin{CD}
\mathcal{D}(M_1) @>\ph_{M_1}>> \mathsf{A}(M_1)\\
@V{\psi_\ast}VV       @VV{\alpha_\psi}V\\
\mathcal{D}(M_2) @>>\ph_{M_2}> \mathsf{A}(M_2)
\end{CD}
\end{equation}
\end{definition}
A standard example for a quantum field according to this definition is the
free
Klein-Gordon field on all globally hyperbolic spacetimes and its Wick
polynomials. 

A more general locally covariant quantum field is the local scattering
operator
of St\"uckelberg-Bogoliubov-Epstein-Glaser. In contrast to the free
Klein-Gordon field it is not linear. For $M \in \mathfrak{L}$ and $g
\in \mathcal{D}(M)$, the local scattering operators are  unitary operators
$S_M(g)$ which fulfill the conditions
\begin{gather}
S_M(0) = \one, \\
S_M(f + h + g) = S_M(f + h) S^{-1}_M(h) S_M(h + g), \label{compcaus}
\end{gather}
where in the latter \emph{causality condition} the supports of $g$ and
$f$ are separated by a Cauchy surface of $M$ and the support of $f$
lies
in the future of this Cauchy surface. There is no restriction concerning the
support of $h$.

Using the local scattering operators, it is possible to define a new quantum
field theory. This approach leads to the axiomatic perturbation theory
\cite{FB}
where the local scattering operators are defined as formal power series.
 Hence the objects of $\mathfrak{O}$ are in this context
$\ast$-algebras of operators defined as formal power series. 

\subsubsection{Local scattering operators in perturbation theory}

Let $\mathcal{A}$ be the algebra of observables of a free quantum field theory.
To be specific we could choose $\mathcal{A}$ to be the unital $\ast$-algebra
generated by the smeared fields $\ph(f), f \in \mathcal{D}(\RR^4)$, which obey
the Klein-Gordon equation $(\Box + m^2) \ph = 0$ in a distributional sense
together with the appropriate commutation relation, $[\ph(f),\ph(g)] = i
(f,\Delta \ast g)$. Here the propagator function $\Delta =
\Delta_{\text{av}} - \Delta_{\text{ret}}$ is the difference of the retarded and
advanced Green's functions of $(\Box + m^2)$. The free fields satisfy
Wightman's axioms, hence the fields have an invariant domain $\mathcal{D}$.
There are other fields $A$ which are relatively local to $\ph$, that is $[A(g),
\ph(f)] = 0$ if the support of $f$ is spacelike to the support of $g$. These
fields form the \emph{Borchers' class}. If the fields from the Borchers' class
can be evaluated at fixed times (that is, restricted to spacelike surfaces),
they serve as building blocks for local interactions. We define the interaction
Lagrangian as $\mathcal{L}_I(\vec{x}) = A(t,\vec{x})$ with $x^0 = t$. 

For a given
test function $g \in C^\infty_c(\RR^4)$, the localized Hamiltonian in the
interaction picture is 
$$
V(t;g) = - \int g(t,\vec{x}) A(t,\vec{x}) d^3x.
$$
The corresponding time evolution operator $U(t,s)$ is formally obtained by a
Dyson expansion \cite{dyson}. We evaluate it over a time interval $(\sigma,
\tau) \subset \RR$ which is chosen in such a way that $\supp g \subset
(\sigma,\tau) \times \RR^3$. As $V(\tau;g) = V(\sigma;g) = 0$, we get the
scattering
operator depending on the localization function $g$,
\begin{equation}
\label{s(g)bytproducts}
S(g) = \one + \sum_{i=1}^\infty \frac{i^n}{n!} \int T(A(x_1) \dots
A(x_n))g(x_1)\dots g(x_n) d^4x_1 \dots d^4x_n,
\end{equation}
where the operator-valued functionals $T(\dots)$ are the \emph{time-ordered
products}.

Unfortunately, the
restriction of fields from the
Borchers' class to spacelike surfaces is in general not
possible. In the above example, only the free
fields themselves together with their derivatives have this property. Hence a
direct
application of this strategy does not lead to interesting examples for
interacting
theories. 

Nevertheless,
the St\"uckelberg-Bogoliubov-Epstein-Glaser formulation of perturbation theory
is based upon the definition of the local scattering operator as in
\eqref{s(g)bytproducts}. The emphasis it put on the time-ordered products. They
are defined directly as multilinear mappings from the $n$th power of the
Borchers' class to operator-valued distributions $T(A_1(x_1) \dots A_n(x_n))$
with domain $\mathcal{D}$ such that certain conditions are fulfilled
\cite{glaser, scharf}. These conditions allow for the recursive construction of
the
time-ordered products. Attention has to be payed to the total diagonal in
$\RR^{4n}$, as in general distributions can not be multiplied at
coinciding
points. The extension to this
hypersurface is a renormalization procedure \cite{FB}. We will not go into
further detail. 

One crucial property of the time-ordered products is the causal factorization,
\begin{equation}
\label{tcausal}
T(A(x_1) \dots A(x_n)) = T(A(x_1) \dots A(x_k))T(A(x_{k+1}) \dots A(x_n)),
\end{equation}
if $(\{x_1, \dots, x_k\} + \overline{V}_+) \cap \{x_{k+1}, \dots, x_n\} =
\emptyset$. This property as well as the others have their counterpart on the
level of the local scattering operators. In particular, the causal
factorization  in the form of equation \eqref{compcaus} remains valid. 

\subsubsection{The abstract definition of local scattering operators}
Following
\cite{BS} we define local
scattering operators as a family of operators, depending on test functions,
fulfilling a set of conditions which are consistent with the requirements of
the Epstein-Glaser approach of perturbation theory as well as the generally
covariant approach. Note that
the following definition does not refer to the Dyson expansion. Moreover,
we allow spacetimes of arbitrary dimension $d \geq 2$.
\begin{definition}\label{lso}
Let $\cH$ be a Hilbert space, carrying a unitary representation
$\overline{\mathcal{P}} \to \cB(\cH), (a, \alpha) \mapsto U(a,\alpha)$ of the
universal covering group $\overline{\mathcal{P}} = \RR^d \rtimes G$ of the
Poincar\'e group, where $G$ is the universal covering group of  the proper,
orthochonous
Lorentz group $SO^+(1,d-1)$, which is the identity component of the homogeneous
symmetry group $O(1,d-1)$ of the $d$ dimensional  spacetime, $2 \leq d \in \NN$.

A family $\{S(g) : g \in \mathcal{D}(\mathbb{R}^n,\mathbb{R})\}$  of
linear operators on $\cH$ is a family of local
scattering operators if 
\begin{enumerate}[(i)]
\item $S(0) = \one$, 
\item $S(g)^\ast = S(g)^{-1}$,
\item $S(g)$ transforms covariantly under $\overline{\mathcal{P}}$: $U(a,\alpha)
S(g)
  U(a,\alpha)^{-1} = S( g_{\langle 
 a,\Lambda(\alpha)\rangle})$
with $  g_{\langle 
 a,\Lambda(\alpha)\rangle}(x) = g(\Lambda^{-1}(\alpha)(x-a))$, where
$\Lambda(\alpha) \in SO^+(1,d-1)$,
\item causal factorization holds true: 
Let
$f,g,h \in \mathcal{D}(\RR^n)$ such that $(\supp f + \overline{V}_+) \cap \supp
g = \emptyset$, then
\begin{equation}
\label{causal}
S(f+h+g) = S(f+h) S(h)^{-1}S(h+g).
\end{equation}
\end{enumerate}
\end{definition}
Notice that equation equation \eqref{causal}  is independent of the
support properties of $h$.
\begin{remark}
In this definition, we restrict ourselves to scalar theories. This is sufficient
for our application in Section~\ref{pphilso}.
If one would consider local scattering operators associated with localized
interactions constructed from free spinorial or tensorial fields, test
functions with several components $g^\lambda$ would come into play and the
covariance condition would be changed, involving a finite-dimensional
 matrix representation of $G$ in spinor space, analogously to the transformation
properties of the fields in
Section~\ref{wightman}.
\end{remark}
In the definition of the local scattering operators, we could restrict the
causality
condition associated to the causal factorization of the
$T$-products \eqref{tcausal}: $S(f + g)
  = S(f)S(g)$ if $(\supp f + \overline{V}_+) \cap \supp g = \emptyset$. But we
are aiming at a definition of the local net and therefore, the stronger relation
\eqref{compcaus} is necessary.

\subsection{Interacting fields and the adiabatic limit}
\label{algebras}

The definition of the interacting fields with the local scattering operator
goes back to Bogoliubov and Shirkov \cite{BS}. 
It has regained considerable interest for the rigorous treatment of
perturbation theory \cite{FB} and opens up the possibility to disentangle the
infrared and
the ultraviolet problem. One finds that the arguments are indeed independent
of perturbation theory, if the local scattering operators are defined without
recourse to the time-ordered products as in Definition~\ref{lso}. We
follow the presentation of \cite{FB}, see also \cite{FD}.

Let $\mathcal{A}$ be a unital $*$-algebra and assume $\mathcal{V}$ to be the
space of possible interaction Lagrangians. It is considered as an abstract,
finite-dimensional, real vector space. 

Given an assignment of test functions $f \in \mathcal{D}(\RR^n, \mathcal{V})$ to
unitary operators $S(f) \in \mathcal{A}$ which fulfill the conditions of
Definition~\ref{lso} and hence the causality condition \eqref{causal}, we can
define a new family of unitary operators which satisfies the same functional
equation by
\begin{equation}
\label{rels(f)}
S_g(f) := S(g)^{-1} S(g+f).
\end{equation}
Here the localized interaction $g \in \mathcal{D}(\RR^n, \mathcal{V})$ is fixed.
These \emph{relative scattering
operators} are local objects: as a consequence of the causality condition one
can show that
$$
[S_g(f), S_g(h)]  = 0
$$
if $(x-y)^2 <0$ for all $(x,y) \in \supp f \times \supp h$. Hence, if the 
functional derivatives of the relative local scattering operators exist, they
are local fields,
$$
A_g(x) :=\frac{\delta}{\delta h(x)} S_g(hA)|_{h=0},
$$
with respect to the interaction $g \in \mathcal{D}(\RR^n, \mathcal{V})$. In this
formula we have $h \in
\mathcal{D}(\RR^n)$ and $A \in \mathcal{V}$. For a constant interaction
extended over the whole spacetime this is
Bogoliubov's definition of the interacting field \cite{BS}.

For every finite, contractive subset $\mathcal{O}$ of the spacetime the
families $\{S_g(h) : h \in \mathcal{D}(\mathcal{O},\mathcal{V}) \}$ generate
a $*$-algebra $\mathcal{A}_g(\mathcal{O})$. This algebra is the algebra of
local observables. Notice that in perturbation theory this $*$-algebra
consists of formal power series as the local scattering operators are obtained
in this sense. If the local scattering operators are unitary operators on a
Hilbert space and as such elements of a $C^*$-algebra, also the algebra of
local observables is a $C^*$-algebra.

A crucial observation is that the algebra $ \mathcal{A}_g(\mathcal{O})$ depends
only locally on $g$.
\begin{theorem}
Let $g,g' \in \mathcal{D}(\RR^n, \mathcal{V})$ such that
$g\negthickspace\restriction_{\mathcal{O}'}=g'\negthickspace\restriction_{
\mathcal { O } ' } $ for a causally
closed
region $\mathcal{O}' \supset \mathcal{O}$. Then
there exists a unitary
operator $V \in \mathcal{A}$ such that
$$
V S_g(h) V^{-1} = S_{g'}(h)
$$
for all $h \in \mathcal{D}(\mathcal{O}, \mathcal{V})$.
\end{theorem}
For the proof see \cite{FB}. Again, the causal factorization in the form
\eqref{compcaus} enters crucially. As the structure of the algebras of local
observables is independent of the behavior of the interaction outside of a
neighborhood of an open region $\mathcal{O}$ of spacetime, the local net in
the sense of the Haag-Kastler axioms in Section~\ref{haag-kastler} is determined
if one
knows the relative scattering operators $f \mapsto S_g(f)$ for all test
functions $g \in
\mathcal{D}(\RR^n, \mathcal{V})$.

Moreover, it is possible to obtain the quasilocal algebra
$\mathcal{A}_\mathcal{L}$ for an interaction Lagrangian $\mathcal{L}$ which is
no longer localized. This purely algebraic construction corresponds to the
adiabatic limit, but in contrast to other formulations it is not necessary to
extend the support of the interaction $g$ explicitly to the whole spacetime.

The construction is based upon the following ideas (see \cite{FB}). Let
$\Theta(\mathcal{O})$ be the set of all $g \in \mathcal{D}(\RR^n)$ which
are identically equal to $1$ on a causally closed open neighborhood of
$\mathcal{O}$. This set is the base of the bundle
\begin{equation}
\label{bundle}
\bigcup_{g \in \Theta(\mathcal{O})} \{g\} \times \mathcal{A}_{g
\mathcal{L}}(\mathcal{O}).
\end{equation}
Define $\mathcal{U}(g, g')$ to be the set of all unitary operators
$V \in \mathcal{A}$ intertwining the relative scattering operators
$$
VS_{g \mathcal{L}}(h) = S_{g'\mathcal{L}}(h)V
$$
for all $h \in \mathcal{D}(\mathcal{O}, \mathcal{V})$. We define
$\mathcal{A}_\mathcal{L}(\mathcal{O})$ to be the algebra of covariantly
constant sections in the bundle \eqref{bundle}. This means that if $A \in
\mathcal{A}_\mathcal{L}(\mathcal{O})$, then $A = (A_g)_{g \in
\Theta(\mathcal{O})}$, where $A_g \in
\mathcal{A}_{g \mathcal{L}}(\mathcal{O})$ and $VA_g = A_{g'}V $
for all $V \in \mathcal{U}(g, g')$.
In particular, the algebra $\mathcal{A}_\mathcal{L}(\mathcal{O})$ contains the
elements $S_\mathcal{L}(h)$, given by the sections $(S_\mathcal{L}(h))_{g}
= S_{g\mathcal{L}}(h)$. 

To complete the construction of the net of algebras of local observables, we
have to specify the embeddings which lead to the condition of isotony in the
axioms~\ref{haag-kastler}. But the embedding $i_{21}: 
\mathcal{A}_\mathcal{L}(\mathcal{O}_1)
\hookrightarrow \mathcal{A}_\mathcal{L}(\mathcal{O}_2)$ for $\mathcal{O}_1
\subset \mathcal{O}_2$ is inherited from the inclusion
$\mathcal{A}_{g
\mathcal{L}}(\mathcal{O}_1) \subset \mathcal{A}_{g
\mathcal{L}}(\mathcal{O}_2)$ for $g \in \mathcal{D}(\mathcal{O}_2)$ by
restricting the sections from $\Theta(\mathcal{O}_1)$ to
$\Theta(\mathcal{O}_2)$. These embeddings satisfy $i_{12} \circ i_{23} =
i_{13}$ for $\mathcal{O}_3 \subset \mathcal{O}_2 \subset \mathcal{O}_1$. Hence
they define an inductive system and we define the quasilocal algebra
$\mathcal{A}_\mathcal{L}$ as the norm closure of the inductive limit of the
algebras of local observables,
$$
\mathcal{A}_\mathcal{L} := \overline{\bigcup_\mathcal{O}
\mathcal{A}_\mathcal{L}(\mathcal{O}) }.
$$

The Poincar\'e covariance of the local scattering operators implies this
property also for the relative scattering operators: Let $(a,\alpha) \in
\overline{\mathcal{P}}$, then
$$
U(a,\alpha) S_{g \mathcal{L}}(h) U(a,\alpha)^{-1} = S_{ g_{\langle
a,\Lambda(\alpha)
\rangle} \mathcal{L}}
( h_{\langle a,\Lambda(\alpha)
\rangle}),
$$
where $h_{\langle a,\Lambda(\alpha)\rangle} =  h(\Lambda^{-1}(x-a))$
and we consider again Lorentz scalars for simplicity.

We define the automorphisms which implement Poincar\'e covariance of the local
algebras, 
$$
(\beta_{\{a,\Lambda(\alpha)\}}(A))_g := U(a,\alpha) A_{g_{\langle
a,\Lambda(\alpha)
\rangle}} U(a,\alpha)^{-1},
$$
for $A \in \mathcal{A}_\mathcal{L}(\mathcal{O})$ and $g \in
\Theta(\Lambda(\alpha)\mathcal{O} + a)$ .
One has to check that $\beta_{\{a,\Lambda(\alpha)\}}(A)$ is again a covariantly
constant section as defined above. Hence $\beta_{\{a,\Lambda(\alpha)\}}$ is an
automorphism of the net of local algebras which implements the action of the
Poincar\'e group,
$$
\beta_{\{a,\Lambda(\alpha)\}}(\mathcal{A}_\mathcal{L}(\mathcal{O})) =
\mathcal{A}_\mathcal{L}(\Lambda(\alpha) \mathcal{O} + a).
$$ 

Furthermore, in perturbation theory it turns out that it is sufficient to
localize
the interaction in `small' regions.

\subsection{Local scattering operators and the time-dependent Schr\"odinger
equation}

There is a straightforward way to obtain the family of local scattering
operators
in a nonperturbative way. Instead of using a Dyson expansion to
describe the time evolution in the interaction picture formally, we investigate
the wellposedness of the Cauchy problem of the time-dependent Schr\"odinger
equation rigorously.

Assume the time evolution $U(t,s)$ of a quantum theory is generated by a
Hamiltonian of the form $H(t) = H_0 + V(t)$, hence it solves the
Schr\"odinger equation, 
\begin{equation}
i \frac{d}{dt} U(t,s) = H(t) U(t,s), \quad U(s,s) = \one.
\end{equation}
The scattering operator is defined as the strong limit
\begin{equation}
S = \lim_{t \to \infty} \lim_{s\to - \infty} e^{iH_0t} U(t,s) e^{-iH_0s}
\end{equation}
if it exists.
This formula is simplified by transformation in the \emph{Dirac} (or
\emph{interaction})
\emph{picture}:
Setting $V^D(t) = e^{iH_0t} V(t) e^{-iH_0t}$ and denoting by
$U^D(t,s)$ the solution of the Schr\"odinger equation with respect to
$V^D(t)$, one finds
\begin{equation}
\label{soperatorlimit}
S = \lim_{t \to \infty} \lim_{s\to - \infty} U^D(t,s).
\end{equation}

Similar to the approach in perturbation theory we define 
$$
V(t;g) = - \int A(0,\vec{x}) g(t,\vec{x})\,d^{d-1}x
$$
for a localized coupling $g \in C^\infty_c(\RR^d)$. The Hamiltonian in the
 interaction picture is 
$$
V^D(t;g) = e^{iH_0t} V(t;g) e^{-iH_0t} = - \int A(t,\vec{x})
g(t,\vec{x})\,d^{d-1}x.
$$ 
If the Cauchy problem of the time-dependent Schr\"odinger equation with respect
to $V^D(t;g)$ is wellposed with
propagator $U^D(t,s)$, the limit \eqref{soperatorlimit} exists trivially
because of the localization of the interaction. We define the local scattering
operator by
\begin{equation}
\label{s(g)definition}
S(g) := U(\tau, \sigma),
\end{equation}
where the time interval $(\sigma, \tau) \subset \RR$ is chosen such that $\supp
g
\subset (\sigma,\tau) \times \RR^n$. As the propagator is trivial outside of
the time support of $g$, the definition of $S(g)$ does not depend on the choice
of $\sigma$ and $\tau$ as long as the support condition is fulfilled. Moreover,
the
properties of the time evolution as discussed in the next chapter lead to the
conditions of Definition~\ref{lso}. 

The field $A(x)$ describing the interaction comes from the Borchers' class of
the free fields. Although the assumption of the restriction of $A$ to fixed
times remains problematic in four spacetime dimensions, at least for 
models on lower dimensional spacetimes this definition makes sense. As the
approach is manifestly Hamiltonian, the strategy has a similar appearance as 
the constructive quantum field theory before the `Euclidean revolution'. To
test the approach we will concentrate on models which are accessible to the
Hamiltonian strategy. But even for these models the interaction is an
unbounded operator with complicated
properties. We have to develop
advanced methods to solve the corresponding time-dependent Schr\"odinger
equation. This is our task for the next chapter. 
\newpage
\thispagestyle{empty}
\cleardoublepage

\chapter{Evolution Equations}
\label{ch:eeq}

As we have seen, the existence question for local
scattering operators outside of perturbation theory can be traced back to the
question of solvability of a time-dependent Schr\"odinger equation. 
In the present Chapter we will address the wellposedness of the Cauchy problem
for evolution equations of the type of the time-dependent Schr\"odinger
equation. There are two main sources of difficulties: the time dependence of
the Hamiltonian and the hyperbolic type of the Schr\"odinger equation.

For the time-independent case, that is for the Cauchy problem for autonomous
evolution equations in Banach spaces, there is a well-developed theory. In
fact, the wellposedness theory of the autonomous Cauchy problem is equivalent
to the theory of strongly continuous operator semigroups on Banach spaces. Every
strongly continuous semigroup is the solution of a Cauchy problem, and every
solution gives rise to a strongly continuous semigroup. The semigroup property
together with strong continuity
implies features like exponential boundedness, differentiability, closedness of
the generator. Hence wellposedness in the autonomous case can be formulated
completely using generation theorems of Hille-Yosida type.

For the nonautonomous (time-dependent) Cauchy problem, the situation is quite
different. If there exists a strongly continuous solution which depends
continuously on the initial value, it can be interpreted in terms of a
\emph{propagator}. This is a family of operators satisfying  a \emph{causal
factorization} equation, $U(t,r)U(r,s) = U(t,s)$ for $t \geq r \geq s$. But this
property is considerably weaker than the semigroup property of the solutions in
the autonomous context. It does \emph{not} imply exponential boundedness or
differentiability. In fact, not every propagator is related to a
Cauchy problem for a nonautonomous evolution equation. This makes it very
difficult to develop a general wellposedness theory in the time-dependent
situation.
Nevertheless, if the generator of the nonautonomous evolution equation is the
generator of an analytic semigroup for
every moment in time, there are quite
sophisticated existence theorems due to P. Acquistapace, B. Terreni and others,
see \cite{Survey}.
This is denoted as the \emph{parabolic} case. 

Unfortunately, the time-dependent Schr\"odinger equation is not of this type.
It is a \emph{hyperbolic} nonautonomous evolution equation, in general the
Hamiltonian
generates a unitary group which is not analytic. 

After discussing the general properties of the Cauchy problem for autonomous
and nonautonomous evolution equations, we will give a review of existence
theorems for various types of solutions in the hyperbolic case. Although weak
solutions exist under very
general assumptions, they have not enough regularity for our purposes. Moreover,
weak solutions are in general not unique. Therefore we will give better suited
concepts of solvability. The standard theorem for strong solutions is due to T.
Kato. But there the assumptions are too restrictive to discuss the existence of
local scattering operators in quantum field theory. 

The key to a well-suited existence theory will be a technique due to
J.~Howland, which relates the time-dependent Cauchy problem to a
time-independent one. But this is the subject of the next Chapter. 

The theory of evolution equations has a wide range of applications, for example
to partial differential equations with variable boundary conditions. Therefore,
we formulate the results of the present and the following chapter not only for
Hilbert spaces, but for general Banach spaces $X$ if possible.  

\section{The autonomous Cauchy problem}

For $t \geq s$ we consider the \emph{autonomous Cauchy problem}, that is
the
initial value problem
\begin{equation}
\label{acp}
\frac{d}{dt} u(t) = A u(t), \quad u(s) = x,
\end{equation} 
on a Banach space $X$, where $A$ is a closed linear operator with domain of
definition $D(A)$. 
\begin{definition}
A \emph{classical solution} of the Cauchy problem is a continuous
$X$-valued
function $t \mapsto u_x(t) \in D(A),\ t\geq s,$ which is continuously
differentiable
and satisfies \eqref{acp}.
\end{definition}
 Existence and uniqueness of solutions in general depend on
the choice of the initial value. 
\begin{definition}
The autonomous Cauchy problem is called \emph{wellposed} if
\begin{enumerate}[(i)]
\item $D(A)$ is dense in $X$ and there exists a classical solution $u_x$ of
\eqref{acp} for every $x \in D(A)$,
\item the solution is unique,
\item the solution depends continuously on the initial value: For every 
sequence $\{x_n\} \subset D(A), x_n \to 0$, we have $u_{x_n}(t) \to 0$ uniformly
on compact time intervals.
\end{enumerate}
\end{definition}
Wellposedness of the autonomous Cauchy problem \eqref{acp} can be completely
characterized by properties of the
operator $A$.
\begin{theorem} 
\label{th:acp}
Let $A$ be a densely defined linear operator with  nonempty resolvent set.
Then the Cauchy problem \eqref{acp} with $s=0$ is wellposed if and only if $A$
is the
generator of a strongly continuous
semigroup $(T(t))_{t\geq 0}$. For every $x \in D(A)$, $u_x(t) = T(t)x$ is a
classical solution.
\end{theorem}
For the proof see \cite{pazy}. 

Every strongly continuous semigroup
$(T(t))_{t \geq
0}$ is exponentially bounded. This means that there exist constants $\omega \geq
0$ and $M \geq 1$ such that $\|T(t)\|
\leq M e^{\omega t},\ t\geq 0$ \cite[Theorem~1.2.2]{pazy}. If $\omega =0$ and
$M=1$, we call $(T(t))_{t \geq 0}$ a strongly continuous semigroup of
\emph{contractions}.
The theorem of Hille and Yosida gives a criterion for an operator $A$
being the generator of a strongly continuous semigroup of
contractions.
\begin{theorem}[Hille-Yosida]
\label{th:hy}
A linear operator $A$ is the generator of a strongly continuous
semigroup $(T(t))_{t \geq 0}$, if and only if
\begin{enumerate}[(i)]
\item $A$ is closed and densely defined,
\item $\RR^+ \subset \rho(A)$ and $\|R(\lambda,A)\| \leq \frac{1}{\lambda}$ for
every $\lambda > 0$. 
\end{enumerate}
\end{theorem}
For the proof see \cite{pazy}. In practice, the following notions turn out to
be handy: For $x \in X$ denote by $F(x) \subset X^*$ the
\emph{duality set} $F(x) :=
\{ x^* \in X^* : x^*(x) = \|x\|^2 = \|x^*\|^2 \}$. A linear operator $A$
is \emph{dissipative} if for every $x \in D(A)$ there is an $x^* \in F(x)$ such
that $\Rea x^*(Ax) \leq 0$. It is \emph{accretive} if $-A$ is dissipative. A
maximally dissipative operator $A$ has no proper
dissipative extension. Maximally accretive operators are defined analogously.
The notion of dissipativity is related to the
boundedness assumption on the resolvent in the Theorem of Hille-Yosida.
\begin{theorem}
A linear operator $A$ is dissipative if and only if
$\|(\lambda - A)x\| \geq \lambda \|x\| $
for all $x \in D(A)$ and $\lambda > 0$.
\end{theorem}
Thus we arrive at the following theorem.
\begin{theorem}[Lumer-Phillips]
Let $A$ be a densely defined, linear operator. If $A$ is dissipative and there
is a $\lambda_0 > 0$ such that $\Ran(\lambda_0 - A) = X$, then $A$ generates a
strongly continuous semigroup of contractions. Conversely, if $A$ is the
generator of a strongly continuous, contractive semigroup, then $\Ran(\lambda -
A) = X$
for all $\lambda > 0$ and $A$ is dissipative.
\end{theorem}
For the proof of this and the preceding theorem, see again Pazy's monograph
\cite{pazy}.

If the initial value $x \not\in D(A)$, the Cauchy problem has in general no 
solution. However, the
orbit $u: [0,\infty)\to X,\ t \mapsto T(t)x$ of the strongly continuous
semigroup $(T(t))_{t\geq 0}$
solves \eqref{acp} in a
generalized way. 
\begin{definition}
A continuous function $u:[0,\infty) \to X$ is called a \emph{mild solution}
of \eqref{acp} if there is a sequence $\{ x_{n}\} \subset D(A)$ such that
$x_{n} \to u(0)$ and $T(t)x_{n} \to u(t)$ uniformly on bounded intervals as
$n\to \infty$. 
\end{definition}
One can show that the generalized solution is independent of the choice of the
sequence $\{x_{n}\}$ and coincides with the solution of \eqref{acp} if
$u(0)\in D(A)$. Hence, with the assumptions of Theorem~\ref{th:acp}, the Cauchy
problem has a generalized solution for every initial value $x \in X$.
All solutions are exponentially bounded.

This generalized notion of solvability is appropriate for the
application to quantum mechanics: Usually one does not restrict the possible
initial states of a quantum system to lie in the domain of the
Hamiltonian.

For the initial value problem of the corresponding nonautonomous evolution
equation there is no such easy characterization of solvability in terms of the
generator. Therefore, it
will be useful to relate the time-dependent situation to the time-independent
one, where the question of solvability can be discussed with help from theorems
like Theorem~\ref{th:hy}. In particular, we will develop also in the
nonautonomous context a
notion of solvability which is similar to the mild solutions above.

\section{The nonautonomous Cauchy problem}

For $t \geq s$ let $A(t)$ be  linear operators on a Banach space $X$
with domains of definition $D(A(t))$.  The
nonautonomous Cauchy problem is given
by
\begin{equation}
\label{ncp}
\frac{d}{dt} u(t) = A(t) u(t), \quad u(s) = x, 
\end{equation} 
where $x \in X$ is the initial value.
\begin{definition}
A classical solution of the nonautonomous Cauchy problem is a 
$C^{1}([s,\infty),X)$-function $u_{s,x}$ such that $u_{s,x}(t) \in D(A(t))$ and
\eqref{ncp} holds.
\end{definition}
An appropriate definition of wellposedness is given by R.~Nagel and G.~Nickel in
\cite{nagelnickel}.
\begin{definition}\label{wellposed}
The nonautonomous Cauchy problem \eqref{ncp} is called classically wellposed
with regularity subspaces $(Y_s)_{s \in \RR}$ if
\begin{enumerate}[(i)]
\item the subspace $Y_s = \{y \in X : \exists \;\text{a classical
solution} \;u_{s,y}\; \text{of \eqref{ncp}} \} \subset D(A(s))$ and $Y_s$ is
dense in $X$
for all $s \in \RR$,
\item  the solution $u_{s,y}$ is unique for every $y \in Y_s$,
\item the solution $u_{s,y}$ depends continuously on the initial data: Let $s_n
\to s\in \RR$, $Y_{s_n} \ni y_n \to y \in Y_s$ and for $z = y_n$ or $z=y$ set
$u_{r,z}(t) = z$ if $ t
< r$. Then $\| u_{s_n, y_n}(t) - u_{s,y}(t)\| \to 0$ uniformly for $t \in I,\ I
\subset \RR$ compact.
\end{enumerate}
\end{definition}
If moreover $\| u_{s,y}(t) \| \leq M e^{\omega (t-s)} \|y\|$, the solution is
called exponentially bounded. 

Given a wellposed nonautonomous Cauchy problem  with exponentially
bounded
solutions, these solutions give rise to \emph{strongly continuous propagators}.
Let $I \subseteq \mathbb{R}$ be an interval and $D_I = \{ (t,s)
\in I \times I : t\geq s \}$.
\begin{definition}
\label{def:propagator}
A strongly continuous, exponentially bounded {\em propagator} or {\em evolution
family} is a family of bounded
operators $\{ U(t,s)\}_{(t,s) \in D_I}$ such that
\begin{enumerate}[(i)]
\item $U(t,r) U(r,s) = U(t,s)$ for $ t\geq r\geq s$ and $
U(t,t) = \one$,
\item $D_I \ni (t,s) \mapsto U(t,s)$ is strongly continuous,
\item $\|U(t,s)\| \leq M e^{\omega(t-s)}$.
\end{enumerate}
\end{definition}
Propagators arise naturally if one considers the properties of
dynamical systems. Their properties reflect the conditions one would expect for
a causal time evolution, hence the strong continuity of $(t,s) \mapsto U(t,s)$
seems to be a reasonable assumption. However, dealing with weaker notions of
solvability and  
in the context of evolution semigroups in
Chapter~\ref{sec:evolutionsemigroups}, 
 it will turn out to be convenient to introduce also \emph{weakly measurable}
propagators. These are defined in the sense of Definition~\ref{def:propagator},
but with condition (ii) replaced by
\begin{enumerate}
\item[(ii')] $D_I \ni (t,s) \mapsto U(t,s)$ is weakly measurable.
\end{enumerate}

Propagators
generalize the concept 
of strongly continuous semigroups in the context of nonautonomous evolution
equations. Indeed, given a strongly continuous semigroup $(T(s))_{s\geq 0}$, the
definition $U(t,s) := T(t-s)$ yields a strongly continuous, exponentially
bounded
propagator.

However, the theory of the nonautonomous Cauchy problem is very different from
its autonomous counterpart. The semigroup property 
$$T(t)T(s) = T(t+s),\quad t\geq s
\geq 0,
$$
together with strong continuity already implies strong differentiability of $t
\mapsto T(t)$ on a
dense set, exponential boundedness, and it fixes the properties of the
generator $A$ (for example closedness)
together with its relation to the Cauchy problem \cite{pazy}. In contrast, the
causal condition 
$$
U(t,r) U(r,s) = U(t,s),\quad t\geq r\geq s,
$$ 
does not allow for similar statements for the propagators $U(t,s)$,
the family
of generators $A(t)$ and the nonautonomous Cauchy problem. We give some
examples which illustrate the difficulties in the time-dependent situation:
\begin{example}
Consider  a continuous, nowhere differentiable function $t \mapsto
f(t) >
0$ on $\mathbb{R}$. Then
$U(t,s) := f(t) f(s)^{-1}$ is a propagator according to 
Definition~\ref{def:propagator}, but nowhere differentiable. Hence it is not
related to a classical solution of a nonautonomous Cauchy problem
\eqref{ncp}. The propagator $U(t,s)$ has no generator $A(t)$.
\end{example}
\begin{example}
The propagator $U(t,s) := e^{t^2 - s^2}$ on $X= \CC$ satisfies (i) and
(ii),
but it is not exponentially bounded.
\end{example}
Even if the nonautonomous Cauchy problem \eqref{ncp} is wellposed, the
generators $A(t)$ may behave in a surprising way, compared to the autonomous
situation. The following examples demonstrate some pathologies:
\begin{example}
The nonautonomous Cauchy problem \eqref{ncp} is wellposed with
smooth
solutions, but the intersection
of the domains of the generators is trivial, $\bigcap_t D(A(t)) = \{0\}$, see
Section~\ref{goldstein} and \cite{goldstein}.
\end{example}
\begin{example}
The nonautonomous Cauchy problem \eqref{ncp} is wellposed, but the
regularity subspaces $Y_t$ are strictly
contained in $D(A(t))$ \cite{nickeldiss}.
\end{example}
\begin{example}
The nonautonomous Cauchy problem \eqref{ncp} is wellposed, but $A(t)$ is
not even \emph{closable}
\cite{nagelnickel}.
\end{example}
\begin{example}
The nonautonomous Cauchy problem \eqref{ncp} is 
wellposed, but this property is not stable under perturbations
of the $A(t)$ by \emph{bounded} operators or scaling $A(t) \to \alpha A(t)$,
$\alpha > 0$ \cite{nagelnickel}.
\end{example}

Hence, to deal with this situation, the notion of classical solutions turns out
to be too
restrictive. Thus we will use concepts of solvability which are strictly weaker
(if the generators $A(t)$ are unbounded).
In analogy to the mild solutions in the autonomous case, we consider the
orbits $u=U(\cdot,s)x$ for {\em
every} $x\in X$. 

\begin{definition}\label{approx-def}
Let $A(t)$, $t\in I$, be linear operators on $X$, $s < b$, $s\in I$,  $x\in X$,
and $1\le p\le\infty$.
\begin{enumerate}[(a)]
\item Let $I=(a,b]$. A function $u\in C([s,b],X)$ is an
{\em ($E_p$)-mild solution of} the nonautonomous Cauchy problem \eqref{ncp} if
\begin{enumerate}[(i)] 
\item $u(s)=x$,
\item there are
$u_n\in W^{1,p}((s,b),X)$ (respectively $C^1((s,b),X)$ if $p = \infty$),
$n\in\NN$, 
such that $u_n(t)\in D(A(t))$ for almost every
$t\in[s,b]$, $A(\cdot)u_n\in L^p([s,b],X)$ (respectively $C([s,b], X)$ if
$p=\infty$),
\item $u_n\to u$ uniformly on $[s,b]$,
\item $-\dot{u}_n+A(\cdot)u_n\to 0$
in  $L^p([s,b],X)$ as $n\to\infty$.
\end{enumerate}
\item Let $I=\RR$. A function $u\in C(\RR,X)$ is an
{\em ($E_p$)-mild solution on $\RR$} of the nonautonomous Cauchy
problem \eqref{ncp} if 
\begin{enumerate}[(i)]
\item $u(s)=x$,
\item for each $d>|s|$ there are
$u_n\in W^{1,p}((-d,d),X)$ (respectively $C^1([-d,d],X)$ if $p = \infty$),
$n\in\NN$,
 such that $u_n(t)\in D(A(t))$ for almost every
$t\in[-d,d]$, $A(\cdot)u_n\in L^p([-d,d],X)$ (respectively $C([-d,d],X)$ if
$p=\infty)$),
 \item $u_n\to u$ uniformly on $[-d,d]$, 
\item $-\dot{u}_n+A(\cdot)u_n\to 0$ in 
$L^p([-d,d],X)$
  as $n\to\infty$ for every $d > |s|$.
\end{enumerate}
\end{enumerate}
\end{definition}
In Theorem~\ref{weak} we will see that, with some special assumptions,
mild solutions are closely related to \emph{weak solutions}.
For a reflexive Banach space $X$ and positive $p,q$ with $1 = \frac{1}{p}
+\frac{1}{q}$ consider the function space $E_p =L^{p}(I,X)$ on the interval $I =
[a,b], -\infty \leq a < b \leq \infty$. We identify $L^{q}(I,X^{*})$ and
$E_p^{*}$.
By $\mathcal{T}$ we denote a space of test functions.
\begin{definition}
\label{weaksolution}
Given $p,q \in \RR$ such that $\frac{1}{p} + \frac{1}{q} = 1$, $I \subset \RR
$, and define the
test function space $\mathcal{T}$ by
\begin{multline}
\label{weaktestfunctions}
\mathcal{T} :=
\{ f \in L^q(I,X^{*}) : f(t) \in D(A^*(t))\;, f \;\text{continuously
differentiable,} \;\\ 
A^*(\cdot)f(\cdot)
\;\text{continuous,}\; 
f(b) = 0 \}.
\end{multline}
For $x \in X$ we call $u \in L^{p}(I,X)$ a \emph{weak solution} of
the nonautonomous Cauchy problem \eqref{ncp} with initial value $x$ if
\begin{equation}
\int_{I} \left[\left(\dot{f}(t), u(t)\right) + \left(A^{*}(t)f(t),
u(t)\right)\right] \, dt + \left(f(a),x\right)
= 0
\end{equation}
for all $f \in \mathcal{T}$.
\end{definition} 
 As we will see in Section~\ref{sec:weaksolution}, weak solutions of the
nonautonomous Cauchy problem exist
under very general assumptions. But, because we do not know whether weak
solutions are continuous or unique, they are not well-suited for the
definition of local scattering operators.
  
To get to a more appropriate notion of solvability, it is useful to
distinguish those solutions which can be approximated by certain sequences of
solutions of an evolution equation with bounded generator. We define the
approximation of the generator in the following way:
\begin{definition}
\label{admissible-def}
Let $A(t)$, $t\in I,\ I\subset \RR$, be operators on a Banach
space $X$.
 We call the bounded operators $A_n(t),\ n \in \NN$, an \emph{admissible
bounded 
approximation} of $A(t)$ if for $(t,s) \in D_I$ and $n \in \NN$,
\begin{enumerate}[(i)] 
\item the
function $t \mapsto A_n(t)$ is strongly continuous,  
\item $ \lim_{n \to \infty} A_n(t)y =
A(t)y$ for all $y \in D(A(t))$,
\item $\|A_n(t)y\| \leq
 c (\|A(t)y\| + \|y\|)$ for all $y \in D(A(t))$ and a constant $c > 0$,
\item $\|U_n(t,s)\| = M
e^{\omega(t-s)} $
where $\omega \in \RR$, $M
\geq 1$ and $U_n(t,s)$ is the propagator generated by $A_n(t)$.
\end{enumerate}
\end{definition}
The properties of $A_n(t)$ assure the existence of
the propagators $U_n(t,s)$ associated with $A_n(t)$,
see Theorem~\ref{th:boundedgen}.
\begin{remark}
For operators $A(t)$, $t\in I,\ I\subset \RR$, such that $Y\subset D(A(t))\
\forall t \in \RR$ with a Banach space $Y \subset X$, densely embedded in $X$,
we could define admissible bounded approximations \emph{on $Y$} by replacing
(ii) and (iii) by
\begin{enumerate}
\item[(ii')]$ \lim_{n \to \infty} A_n(t)y =
A(t)y$ for all $y \in Y$,
\item[(iii')]$\|A_n(t)y\| \leq
 c \|y\|_Y$ for all $y \in Y$ and a constant $c > 0$.
\end{enumerate}
This definition would be sufficient for the use in Theorem~\ref{extension},
but we will work with the conditions of Definition~\ref{admissible-def}, which
we find more natural.
\end{remark}
A class of operators $A(t)$ which feature admissible bounded approximations are
the so-called
Kato-stable operators, see Definition~\ref{de:katostable}. To obtain an
approximation of an unbounded operator, in general a Kato approximation is
suitable. For Hilbert spaces we obtain
admissible bounded approximations via the functional calculus.
\begin{lemma}
\label{adap}
\begin{enumerate}[(i)]
\item Let $A(t), t \in I$, be Kato-stable operators with constants $M, \omega$.
Assume that $t \mapsto R(\omega', A(t)) = (\omega' - A(t))^{-1}$ is
strongly continuous for $t \in I$ and some $\omega' > \omega$. Then $A_n(t) := n
A(t) R(n, A(t)), n > \omega$, are admissible bounded approximations of $A(t)$.
\item Let $A(t), t \in I$, be skew-adjoint operators on a Hilbert space $X$.
Assume that $t \mapsto R(1,A(t))$ is strongly continuous for $t \in I$.
Set $\ph_n(i\tau) = i\tau$ for $|\tau| \leq n$ and $\ph_n(i\tau) = \pm in$ for 
$\pm \tau \geq n$.   
Then $A_n(t) = \ph_n(A(t))$ are skew-adjoint admissible bounded approximations
of
$A(t)$. The corresponding propagators $U_n(t,s)$ are unitary. 
\end{enumerate}
\end{lemma}
\begin{proof}
(i) For arbitrary $n > \omega$ the formula
$$
R(n, A(t)) = [1 + (n-\omega')R(\omega',A(t))]^{-1} R(\omega', A(t))
$$
shows the strong continuity of $t \mapsto R(n,A(t))$, and strong continuity of
$t \mapsto A_n(t)$ follows. The remaining assertions are standard properties of
the
Kato approximation, see for example \cite[Lemma~II.3.4]{en}.\\
(ii) As above we see that $t \mapsto R(\omega ,A(t))$ is strongly continuous for
all $\omega \in \rho(A(t))$. Let $\{t_m\}$ be a sequence with $t_m \to t$ as $m
\to \infty$.  Then $R(\omega ,A(t_m)) \to R(\omega ,A(t))$ strongly, hence for
every bounded, continuous function $u$ on $i\RR$ we have $u(A(t_m)) \to u(A(t))$
strongly. With $u = \ph_n$ this is strong continuity of $t \mapsto A_n(t)$. The
remaining assertions are shown using the functional calculus for normal
operators
\cite{weidmann}.
\end{proof}

We use these generators with admissible bounded approximation to formulate the
following
notion of solvability.
\begin{definition}
\label{quasi-approx-def}
Let $A(t)$, $t\in I,\ I \subset \RR$, be linear operators on $X$ with admissible
bounded
approximation $A_n(t)$. Let $s\in I$, $s< b$,  $x\in X$,
and $1\le p\le\infty$.
\begin{enumerate}[(a)]
\item In the case $I=(a,b]$ a
weakly continuous function $u: [s,b]\to X$ is an
{\em ($E_p$)-approximative solution} of the nonautonomous Cauchy
problem \eqref{ncp} if 
\begin{enumerate}[(i)]
\item $u(s)=x$, 
\item there
are
$u_n\in C^1([s,b],X)$, $n\in\NN$,  such that 
\begin{enumerate}[(1)]
\item$u_n(s) = x$, 
\item $u_n(t)\to u(t)$ weakly as $n\to\infty$ all for $t \in [s,b]$, 
\item $-\dot{u}_n+A_n(\cdot)u_n\to
0$
in  $L^p([s,b],X)$ as $n\to\infty$.
\end{enumerate}
\end{enumerate}
\item In the case $I=\RR$ a
weakly continuous function $u: \RR \to X$ is an
{\em ($E_p$)-approximative solution on $\RR$} of the nonautonomous Cauchy
problem \eqref{ncp} if 
\begin{enumerate}[(i)]
\item $u(s)=x$,
\item
for each $d>|s|$ there are
$u_n\in C^1(\RR,X)$, $n\in\NN$,   such that 
\begin{enumerate}[(1)]
\item $u_n(s) = x$, 
\item $u_n(t)\to u(t)$ weakly as $n \to \infty$ for  $ t \in \RR$, 
\item $-\dot{u}_n+A_n(\cdot)u_n\to 0$ in  $L^p([-d,d],X)$
  as $n\to\infty$ for every $d > |s|$.
\end{enumerate}
\end{enumerate}
\end{enumerate}
\end{definition}

We will see in Section~\ref{existencequasiapprox} that the unique approximative
solution of the time-dependent Schr\"odinger equations can be obtained with
very general requirements. At the same time, this notion of solvability is
considerably more regular than weak solutions and suitable for the definition
of local scattering operators from the time evolution in the interaction
picture, see Chapter~\ref{exlso}.

\subsection{Goldstein's example}
\label{goldstein}

In the course of this work, we will review some theorems about the wellposedness
of the Cauchy problem for nonautonomous evolution equations. As we will see in
the following section, weak solutions exist under very general assumptions.

However, if one aims at  strong solutions one has to impose quite restrictive
assumptions on the
generators $A(t)$. In particular, one could ask whether it is necessary that the
domains
of the generators have a dense intersection. In other words: If
the nonautonomous Cauchy problem \eqref{ncp} has a strong solution for all
initial values from a dense
set, and if the generators $A(t)$ are maximally dissipative for all times $t$,
is
the intersection $\bigcap_{t \geq 0} D(A(t))$ necessarily dense in
$X$? Goldstein's example \cite{goldstein} shows that
this is indeed not the case.

Goldstein shows that there are self-adjoint operators $H(t) \geq 0$ for $t \geq
0$ on a Hilbert space
$X$ such that the following conditions are fulfilled: 
\begin{enumerate}[(i)]
\item The resolvent of $H(t)$ depends smoothly on $t$, that is for all $\lambda
\in \CC \setminus [0, \infty)$ and $x \in X$ the mapping $t \mapsto R(\lambda,
H(t))^{-1}x$ is an element from $C^\infty([0,\infty), X)$.
\item There is a dense set $Y \subset X$ such that the nonautonomous
Cauchy problem \eqref{ncp} with $A(t) = -iH(t)$ has a unique
solution $u \in C^\infty([0,\infty), X)$ with $u(0) = y$ for all $y \in Y$. The
solution depends continuously on the initial value.
\item $ \bigcap_{t \geq 0} D(H(t)) = \{0\}$.
\end{enumerate}

Let $S \geq 0$ be a self-adjoint unbounded operator with dense domain $D(S)
\subset X$. According to a theorem of J.~von~Neumann,
there is a unitary
operator $U_1$ such that the nonnegative self-adjoint operator $T = U_1 S U_1^*$
satisfies $D(T) \cap D(S) = \{0\}$. Now we use the spectral theorem to represent
$U_1$ as $U_1 = \int_0^{2 \pi} e^{i \lambda } \;dE_\lambda$ and define the
bounded, nonnegative and self-adjoint operator $L$ by $L := \int_0^{2 \pi}
\lambda \;dE_\lambda $. Choose nonnegative, nontrivial smooth functions $\ph,
\psi$ and
$\eta$ with $\supp \ph \subset [0,1)$, $\supp \eta \subset (1,2)$, $\supp
\psi \subset (2,3)$ and $\int_0^2 \eta(t)\;dt = 1$. Define
\begin{equation}
\label{goldsteinH}
H(t) := \ph(t) S + \eta(t) L + \psi(t) T.
\end{equation}
Then we have following result.
\begin{theorem}
\label{th:goldstein}
Let $Y := C^\infty(S)$. For
every $y \in Y$ the nonautonomous Cauchy problem \eqref{ncp} has  a
unique solution $t \mapsto u(t)$ which fulfills the
conditions (i), (ii) and
(iii). It is given by
\begin{equation}
u(t) :=
\begin{cases}
e^{i \int_0^t \ph(s)ds\; S}y &\text{for}\; 0\leq t \leq 1\\
e^{i \int_0^t \eta(s)ds\; L}u(1) &\text{for} \; 1 \leq t \leq 2\\
e^{i \int_0^t \psi(s)ds\; T}u(2) &\text{for} \; 2 \leq t \leq \infty.
\end{cases}
\end{equation}
\end{theorem}
For the proof see \cite{goldstein}.

This simple example describes a situation which we expect to find also in the
context of quantum field theory. As we have seen in
Section~\ref{constructive},
even if the localized interaction
Hamiltonian can be defined in the Fock space of the free fields, it will have a
domain of definition which has trivial intersection with the domain of the free
Hamiltonian if a nontrivial renormalization is necessary. Moreover, it may
happen
that the interaction term itself has
a wildly varying domain for different moments in time. Goldstein's example
gives a hint in the direction that the time evolution and hence the local
scattering operators may
exist nevertheless. We will return to this example in
Section~\ref{goldsteindressing}.

\subsection{Existence of weak solutions} 
\label{sec:weaksolution}

Discussing the time-dependent Schr\"odinger equation and the interaction picture
in quantum theory, M.~Reed and B.~Simon state in \cite{RSII} that weak solutions
always exist if the Hamiltonian on a Hilbert space $X$ can be written as $H(t) =
H_0 + V(t)$ with a self-adjoint, possibly unbounded operator $H_0$ and
$V(\cdot):
\RR \to \mathcal{B}(X)$ strongly continuous.  

However, weak solutions of the nonautonomous Cauchy problem \eqref{ncp} exist
under much more general
conditions. We cite a result due to H.~Sohr \cite{Sohr75} together with its
proof which demonstrates the application of abstract methods known from 
partial
differential equations to the Cauchy problem \eqref{ncp}. We will return to
the topic of weak solvability in Theorem~\ref{weak}.
\begin{theorem} \label{satz1}
Let $X$ be a reflexive Banach space and let $ A(t)
:D(A(t)) \to X$ be a generator of a strongly continuous semigroup of contractions for every $t
\in I=[a,b]$. For every given initial value $x \in X$ there exists a weak
solution $u \in E_p$ of the nonautonomous Cauchy problem \eqref{ncp} in
the sense of
Definition~\ref{weaksolution} over the test function space $\mathcal{T}$ of
equation \eqref{weaktestfunctions}.  
\end{theorem}
\begin{proof} 
By the theorem of Hille and Yosida (Theorem~\ref{th:hy}), the resolvent $
\left(1 - s A(t)\right)^{-1} $ is
bounded and bijective 
from $\mathcal{H}$ to $D(A(t))$ with norm less or equal to $1$ for every $s
\geq 0$.  With an equidistant partition of the interval $I$ we can
define an approximate solution of the nonautonomous Cauchy problem \eqref{ncp}.
Let $\delta_n := \frac{1}{n} (b - a)$,
$t^{(n)}_{\nu} := a + \frac{\nu}{n} (b - a)$ and 
$$
C_{n,\nu} := \left(
1 - \delta_n A(t^{(n)}_{\nu})\right)^{-1}\!,\; C_{n,0} :=
\one \; \text{for} \; n 
\geq 1,\; 0 \leq \nu \leq n.
$$
With $W_{n,\nu} := C_{n,\nu} \cdot \ldots \cdot
C_{n,0} $ define $u_n \in L^{p}(I,X)$ by
$u_{n}(t) :=W_{n, \nu -1}x$ for $t \in [ t^{(n)}_{\nu - 1}, t^{(n)}_{\nu}
], 1\leq \nu \leq n $. 

Due to the boundedness of $u_n(t)$ by $\|x\|$, we have
$$
 \|u_n\|_{p} \leq \norm{u_0}(b - a)^{1/p}. 
$$
With this sequence of approximations $u_n$ we can
establish the existence of the weak solution $u$.

Let $\{ \mathbb{C} \}$ be the space of bounded, complex-valued
sequences $\{c_n\}_{n \in \NN}$. Define a semi-norm on $\{ \mathbb{C} \}$ by
$|\|
\{c\} \||:=  \limsup \abs{ c_n} $. By dividing out the
zero-space $N = \{ \{c\} \in \{\mathbb{C}\} :	 |\|\{c\}\||
= 0 \}$  one gets a Banach space $\left[ \mathbb{C} \right] =
\{\mathbb{C}\} / N$. 

The constant sequences form a  closed subspace of $\{ \mathbb{C} \}$,
which is the image of an  isometric embedding $i$: 
$$
\mathbb{C} \ni c \overset{i}{\hookrightarrow} \left[c\right] \in
\left[ \mathbb{C} \right]    
$$

By the theorem of Hahn-Banach, the bounded linear form $\tilde{F} :
i(\mathbb{C}) \to
\mathbb{C},\ \tilde{F}( \left[ c \right] ) = c$ is extendable to $F:
\left[\mathbb{C} \right]  \to
\mathbb{C},\ F\negthickspace\restriction_{i(\mathbb{C})} = \tilde{F},\ \|F\| =
\|\tilde{F}\|$. Note that
this extension is not unique.

For every $f \in L^{q}(I,X^{*})$ the sequence $ \{ ( f , u_n )
\}_{n \in
  \mathbb{N} }$ is an element of $\{\CC\}$, because the absolute values of its
elements are bounded
by $\|f\|_{q} \|x\|
(b -a )^{1/p}$. Hence $|F([\{(f,u_n)\}_{n \in \NN}])| \leq \|F\| \|f\|_q \|x\|
(b-a)^{1/p}$, and $ f \mapsto F( [ \{( f, u_n
  )\}_{n \in \NN} ] ) $ defines a bounded antilinear form on
$E_p^* = L^q(I,X^*)$. Therefore, by the reflexiveness of the spaces involved
and application of Riesz' Lemma, there exists a $u \in E_p$ such that $F(
[\{(f, u_n )\}_{n\in\NN}]) = ( f , u )$ for all $f \in L^q(I,X^*)$. 

Now one shows that the function $u$ solves the nonautonomous Cauchy problem \eqref{ncp} in the weak
sense.  
Let $f \in \mathcal{T}$. Define $\Delta_n f$ and $T_n f \in L^q(I,X^*)$ by
$$
\Delta_n f(t) = \delta_n^{-1} \left( f( t^{(n)}_{\nu}) - f(
  t^{(n)}_{\nu - 1}) \right) \quad \text{and} \quad
T_n f(t) = f( t^{(n)}_{\nu})
$$       
for $ t \in [ f( t^{(n)}_{\nu - 1}), f( t^{(n)}_{\nu})
] $. By the assumptions  on $f$, one has $\lim_{n \to \infty} T_n f =
f$, $\lim_{n \to \infty} T_n A(\cdot) f = A(\cdot) f$  and $ \lim_{n \to \infty}
\Delta_n f =
\dot{f} $ in the strong sense.

Because the norm on the space of sequences depends on a limes superior, 
one can use the approximate expressions instead of the original ones
in the argument involving the linear form $F$:
$$
(\dot{f} ,u )  = F( [ \{(\dot{f} ,u_n )\}_{n \in \NN}] 
) = F([\{(\Delta_n f ,u_n)\}_{n \in \NN}]).
$$ 
Moreover, we calculate
$$
( \Delta_n f, u_n ) = - (  f(a), x )  -
\sum_{\nu=1}^n \delta_n \left(A^*(t^{(n)}_{\nu}) f(t^{(n)}_{\nu}), W_{n,\nu}
x \right).  
$$ 
By assumption, $t \mapsto A^*(t)f(t)$ is continuous and one gets for the equivalence classes,
\begin{multline*}
\left[ \left\{ 
\sum_{\nu=1}^n \left( A^*(t^{(n)}_{\nu}) f(t^{(n)}_{\nu}), W_{n,\nu} x
\right)\delta_n\right\}_{n \in \NN}\right]
=\\
\left[ \left\{\sum_{\nu=1}^n \left( A^*(t^{(n)}_{\nu}) f(t^{(n)}_{\nu}),
W_{n,\nu-1}
x\right)\delta_n\right\}_{n \in \NN}\right],
\end{multline*} 
because the difference of both sequences converges to zero. 
With this equality we find
\begin{multline*}
\left[ \left\{\sum_{\nu=1}^n\left( A^*(t^{(n)}_{\nu}) f(t^{(n)}_{\nu}), u_n
\right)
\delta_n \right\}_{n \in \NN}\right] 
= \left[\left\{ \left( T_n A^*(\cdot)f , u_n\right) \right\}_{n \in \NN} \right]
=\\ \left[ \left\{ \left( A^*(\cdot)f , u_n\right)\right\}_{n \in \NN}
\right],
\end{multline*} 
and one arrives at the assertion:
\begin{align*}
( \dot{f},u ) &= F([\{( \Delta_n f,u_n) \}_{n \in \NN}
])\\
&=  -F([ \{( f(a), x)\}_{n \in \NN} 
]) -  F( [ \{ \sum ( A^*(
t^{(n)}_{\nu} ) 
f(t^{(n)}_{\nu}), u_n) \delta_n\}_{n \in \NN} ])\\
&= -( f(a) ,x) - F( [ \{ ( A^*(\cdot)f, u_n)\}_{n \in \NN} ])\\
&= -( f(a) ,x) -( A^*(\cdot)f ,u) 
\end{align*}
$$
\Longleftrightarrow
\int_a^b  (\dot{f}(t), u(t)) \;dt+  \int_a^b  (A^*(t)
f(t), u(t))\;dt+ (f(a), x) = 0.
$$ 
\end{proof}

Evidently, the crucial point is to ensure that the test function space
$\mathcal{T}$ is sufficiently large, at least not trivial. This gives the
important conditions on the generators $A(t)$. 

The theorem ensures the existence of weak
solutions for a wide class of
nonautonomous evolution equations. However, we have no information on
uniqueness and regularity properties of the solution. In particular, it is not
clear if the solution can be interpreted in terms of a propagator. Hence, in the
present form this approach to nonautonomous evolution equations is not
appropriate for our purpose.

A more recent result on weak solvability is for instance
\cite{bbl}. Although there are some statements about uniqueness and
regularity, the context of this work is different from ours. For example, the
domains of the generators are assumed to be closed subspaces. It does not seem to be 
possible to use similar methods to the setting we have in mind. A good starting
point for a systematic discussion about results on weak methods mainly in the
parabolic case is Carroll's
book \cite{carroll}.

In the following, we return to stronger notions of solvability which come with
uniqueness and regularity statements for the solutions.

\subsection{Bounded generators}

If the generators $A(t)$ are bounded operators, it is easy to solve the nonautonomous Cauchy problem
\eqref{ncp}. The method is closely related to the Dyson-Phillips expansion
\cite{dyson, dysonphilips}.
\begin{theorem}
\label{th:boundedgen}
Let $X$ be a Banach space and let $A(t)$ be  bounded operators for $t \in I$,
where $I\subset \RR$ is a compact interval with $s \in I$. Assume that $t
\mapsto A(t)$ is a strongly continuous function. Then
for every $x \in X$ there is a solution $t \mapsto u(t)$ of the nonautonomous Cauchy problem \eqref{ncp}.
\end{theorem}
\begin{proof}
We cite the proof of \cite[Theorem~5.1]{pazy}. Note that in this theorem
continuity of $t \mapsto A(t)$ in the operator norm is assumed. However, this
assumption is not necessary for the argument. 

Set $I=[0,T]$ and $\alpha = \max_{t \in I} \|A(t)\|$. Define a mapping $S:
C(I,X) \to C(I,X)$ by
\begin{equation}
(Su)(t) = x + \int_s^t A(\tau) u(\tau) \;d\tau.
\end{equation}
Clearly,
$$
\|Su(t) - Sv(t))\| \leq \alpha (t-s) \|u - v\|_\infty,
$$
and, by induction,
$$
\|S^nu(t) - S^nv(t)\| \leq \frac{1}{n!} \alpha^n(t-s)^n \|u-v\|_\infty.
$$
Hence, for $n$ large enough, $b = \frac{1}{n!} \alpha^n(T-s)^n < 1$ and
$$
\|S^nu - S^nv\|_\infty \leq b \|u-v\|_\infty,
$$
so a generalization of Banach's contraction principle for the vector-valued
context implies the existence of a unique fixed point $u \in C(I,X)$ which obeys
\begin{equation}\label{fp}
 u(t) = x + \int_s^t A(\tau) u(\tau) \;d\tau.
\end{equation}
The continuity assumptions on $A(\cdot)$ and $u(\cdot)$ together with the
estimate $\|A(t) u(t) - A(s)u(s)\| \leq \|(A(t) - A(s))u(t)\| +
\|A(s)(u(t)-u(s))\|$ result in the differentiability of the right-hand side of
\eqref{fp}. Therefore $u$ is the unique solution of the
nonautonomous Cauchy problem \eqref{ncp}.
\end{proof}

\subsection{The Theorem of Kato}

The first article of T.~Kato concerning the nonautonomous Cauchy problem
goes back  to 1953 \cite{kato1953}. In 1970, Kato published a more rigorous and
detailed version of his ideas \cite{kato1970}. Kato's work is still the
reference method for the treatment of the wellposedness problem for
nonautonomous evolution equations of hyperbolic type, hence we cite the main
statements in detail. Although there are  lots of extensions and
improvements of the original result, the main assumptions remain essentially
unchanged, see \cite{Survey}.

The Theorem of Kato allows nonconstant domains of definition $D(A(t))$, but
the
intersection of these domains for different times has to contain a joint dense
subspace with special features leading to the invariance of the subspace with
respect to the propagator.

For the proofs of all statements in this section which we do not give
explicitly we refer to Pazy's monograph \cite[Chapter~5]{pazy}. Another
reference is Tanabe's book \cite{tanabe}. In the following, we will define
\emph{admissible subspaces} and \emph{Kato-stable} families of operators before
we come to the existence theorem in Section~\ref{kex}.

\subsubsection{Admissible subspaces}
Let $A$ be a linear operator on a Banach space $X$ and let $Y$ be a subspace of
$X$. 
\begin{definition}
$Y$ is an \emph{invariant subspace} of $A$ if $A$ maps $D(A) \cap Y \to Y$. 
\end{definition}
\begin{definition}
The \emph{part of $A$ in $Y$} is the linear operator $\tilde{A}$ given by
$\tilde{A} y = A y$
on the domain $D(\tilde{A}) = \left\{ y \in D(A) \cap Y: Ay \in
Y\right\}$.
\end{definition}
The restriction $A\negthickspace\restriction_Y $ is an extension of $\tilde{A}$.
We have
$A\negthickspace\restriction_Y = \tilde{A}$ if $Y$ is an invariant subspace.

In the following we assume that $Y$ is a Banach space with norm
$\norm{\cdot}_Y$ and $Y$ is continuously embedded in $X$, that 
is the norm $\norm{\cdot}_Y$ is stronger than the norm of $X$:
There is a constant $c$ such that $\|y\| \leq c\|y\|_Y$ for $y \in Y$.

\begin{definition}
Let $(T(t))_{t\geq 0}$ be a strongly continuous semigroup with generator $A$.
The subspace $Y$ is \emph{$A$-admissible} if it is an invariant subspace of
$T(t),\ t\geq 0$, and the restriction of $(T(t))_{t\geq 0}$ to $Y$ is again a
semigroup in $Y$, strongly continuous with respect to $\|\cdot \|_Y$.  
\end{definition}

Criteria for admissible subspaces are given in the following theorems.
\begin{theorem}
Let $(T(t))_{t\geq 0}$ be a strongly continuous semigroup with generator $A$ and
$(\omega, \infty) \subset \rho(A)$, $\omega>0$.
The subspace $Y$ is $A$-admissible if and only if $Y$ is an invariant subspace
of $R(\lambda,A),\ \lambda > \omega$, and $\tilde{A}$, the part of $A$ in $Y$,
generates a strongly continuous semigroup in $Y$. Moreover, if  $Y$ is
$A$-admissible, then $\tilde{A}$ is the infinitesimal generator of
$(T(t))_{t\geq 0}\negthickspace\restriction_Y$. 
\end{theorem}

\begin{theorem}
Let $Y$ be dense in $X$ and let $S: Y \to X$ be an isomorphism. The subspace
$Y$ is $A$-admissible if and only if $A_1 = S A S^{-1}$ is the generator of a
strongly continuous semigroup $(T_1(t))_{t \geq 0}$. The semigroup is given
by $T_1(t) =  S T(t) S^{-1}$.
\end{theorem}

For Hilbert spaces there is a related result due to Okazawa \cite{okazawa1993}.
\begin{theorem}
Let $A$ be a
linear operator in a Hilbert space $X$ and $c \geq b \in \RR$. Assume, that  the closure of
$A + b$ is maximally accretive.  Let $S$ be a self-adjoint, strictly
positive
operator with $D(S) \subset D(A)$ such that $D(S)$ is a core for the closure of
$A+ b$ and 
\begin{equation*}
\Rea(Au, Su) \geq -c (u,Su)
\end{equation*}
for all $u \in D(S)$.
Then $D(S^{1/2})$ is $A$-admissible.
\end{theorem}
For the proof see \cite{okazawa1993}. Under the assumptions of the last theorem one can think of $S^{1/2} A S^{-1/2} + c$ as being maximally  accretive.

\subsubsection{Stable families of generators}
\label{katostable}
Now let $\{ A(t)\}_{t \in [0,T]}$ be a family of generators of
strongly continuous semigroups on a Banach space $X$. 
\begin{definition}
\label{de:katostable}
The family $\{ A(t)\}_{t \in [0,T]}$ of generators of
strongly continuous
semigroups is said to be \emph{stable} or \emph{Kato-stable} if for any finite
family $\{t_j\}$
with $ 0\leq t_1 \leq \dots \leq t_k\leq T,\ k \in \mathbb{N}$, there are
positive constants $M,\ \omega$ such that $(\omega, \infty) \subset \rho(A(t))$
for all $t \in [0,T]$ and 
\begin{equation}
\norm{ \prod_{j=1}^k R(\lambda,A(t_j))} \leq M(\lambda-\omega)^{-k}, \quad
\lambda>\omega.
\end{equation}
\end{definition}
An example for Kato-stable operators is easily obtained: Every family  
$\{A(t)\}$ of maximally dissipative operators fulfills the
Definition~\ref{de:katostable} because of the Hille-Yosida theorem.
An alternative formulation of Kato-stability due to H.~Neidhardt is given in the
following Lemma.
\begin{lemma}\label{lem:katostable}
The family $\{ A(t)\}_{t \in [0,T]}$ is Kato-stable with
constants $M, \omega$
if and only if there are norms $\| \cdot \|_t$ on $X$ and a
constant $M \geq 1$ such that $\|x\| \leq \|x\|_t \leq \|x\|_s \leq M \|x\|$
and $\|R(\lambda, A(t))x\|_t \leq (\lambda - \omega)^{-1} \|x\|_t$ for $(t,s)
\in D_I, \lambda > \omega$ and $x \in X$.
\end{lemma}
\begin{proof}
Apply a rescaling $A(t) \to A(t) - \omega$ and observe that Kato-stability
can be formulated in terms of the semigroups instead of resolvents
(\cite[Theorem~5.2.2]{pazy}). Then the assertion follows from 
\cite[Proposition~1.3]{nickel}.
\end{proof}

Concerning stability in subspaces we have the following theorem:
\begin{theorem}
Let $Y \subset X$ be a Banach space, continuously embedded  in 
$X$. Assume there is a family of isomorphisms $Q(t): Y \to X$, $t \in [0,T]$,
such
that $Q(t) \in \cB(Y,X)$ and $Q^{-1}(t) \in \cB(X,Y)$ are uniformly bounded
by a
constant $c$ and the map $t \mapsto Q(t)$ is of bounded variation in the norm
of $\cB(Y,X)$. Let $\{A(t)\}_{t \in [0,T]}$ be a stable family of generators and
set $A_1(t) = Q(t)A(t)Q^{-1}(t)$. If $\{A_1(t)\}_{t \in [0,T]}$ is a stable
family of generators in $X$, then $Y$ is $A(t)$-admissible for $t \in [0,T]$,
and
$\{\tilde{A}(t)\}_{t \in [0,T]}$ is a stable family of generators in $Y$.
\end{theorem}
The next theorem gives a perturbation result.
\begin{theorem}
Let $\{A(t)\}_{t \in [0,T]}$ be a stable family of generators of strongly
continuous semigroups with stability constants $M, \omega$, and let
$\{B(t)\}_{t \in [0,T]}$ be a
family of bounded linear operators such that $ \|B(t)\| \leq c$ uniformly in
$t$. Then $\{A(t) + B(t)\}_{t \in [0,T]}$ is a stable family of generators of
strongly continuous semigroups with stability constants $M, \omega + cM$.
\end{theorem}

\subsubsection{The existence theorem}
\label{kex}
Now we can state Kato's existence theorem.
\begin{theorem}
\label{kato1}
Let $X$ and $Y$ be Banach spaces, $Y \subset X$ dense and continuously
embedded. For every $t \in [0,T]$
let $A(t)$ be the generator of a strongly continuous semigroup. Assume that:
\begin{enumerate}[(i)]
\item $\{ A(t)\}_{t \in [0,T]}$ is a stable family
with constants $M,
\omega$.
\item $Y$ is $A(t)$-admissible for every $t \in [0,T]$. Moreover, 
$\{ \tilde{A}(t)\}_{t \in [0,T]}$, the family given by the parts of $A(t)$ in
$Y$, is
a stable family in $Y$ with constants $\tilde{M}, \tilde{\omega}$.
\item $Y \subset D(A(t))$ for every $t \in [0,T]$. $A(t)$ as an operator from $Y
\longrightarrow X$ is bounded. The function $t \mapsto A(t)$ is
norm-continuous with respect to $\cB(Y,X)$.
\end{enumerate}
Then there exists a unique propagator $U(t,s),
(t,s) \in D_{[0,T]}$, such that
\begin{enumerate}[(a)]
\item $\|U(t,s)\| \leq M e^{\omega (t-s)}$,
\item  the right derivative $\frac{\partial}{\partial t}^+
U(t,s)y |_{t=s} = A(s) y$ for every $y \in Y, s \in [0,T]$,
\item $\frac{\partial}{\partial s}U(t,s) y = -U(t,s)A(s)y$ for
every $y
\in Y, 0 \leq s\leq t \leq T$.
\end{enumerate}
The derivatives are taken in the strong sense in $X$.
\end{theorem}

This theorem enables us to define a unique propagator associated with the family
of generators $\{ A(t)\}_{t \in [0,T]}$, but it is not
strong enough to
establish the
classical wellposedness of the nonautonomous Cauchy problem \eqref{ncp}. This is
the point of the next theorem,
where the solutions even take values in $Y$ if the initial value stems from $Y$.
\begin{theorem}\label{kato2}
Given the assumptions of Theorem~\ref{kato1}, where $(ii)$ is replaced by
\begin{enumerate}
\item[(ii')] There is a family of isomorphisms $\{Q(t)\}$ where $Q(t): Y \to
X$
such that for every $y \in Y, t\mapsto Q(t)y$ is continuously differentiable in
$X$
and $Q(t)A(t)Q^{-1}(t) =  A(t) + B(t)$ for a family $\{B(t)\}$ of bounded
operators which is continuous in $t$.
\end{enumerate}
Then the conclusion of Theorem~\ref{kato1} holds true. Moreover, 
\begin{enumerate}
\item[(d)]$U(t,s) Y \subset Y$,
\item[(e)]$(t,s) \mapsto U(t,s)y$ is continuous in $Y$ for $y\in Y$.
\end{enumerate}
This implies that for $y \in Y$, $u(\cdot) = U(\cdot,s)y \in C^1((s,T], X)$
solves
the nonautonomous Cauchy problem \eqref{ncp} with $u(s) = y$.
\end{theorem}

Kato proves these results by approximation of the generators $A(t)$ and solving
the corresponding Cauchy problem \eqref{ncp}. The strategy is similar to the
proof
of Theorem~\ref{satz1}, but, instead of an abstract existence argument based on
Riesz' Theorem, a limit calculation with Duhamel's integral formula is used.
It is possible to reproduce this
proof using evolution semigroups and an approach similar to Theorem~\ref{char}.
This is done in G.~Nickel's thesis \cite{nickeldiss}.

With regard to the application in quantum field theory, we remark that Kato's
Theorem is not very well-suited.
In the simple $P(\ph)_2$ models there are indeed common dense cores $Y$ for the
Hamiltonians $H(t)$, but invariance of these subspaces is in general
problematic,
due to the lacking smoothing properties of the time evolution. Even if one could
establish invariance, Kato-stability in $Y$ seems unlikely.

\subsection{Time-independent domain}

A simple reformulation of Theorem~\ref{kato2} is possible if the generators
$A(t)$ have a common time-independent domain. This theorem can also be found in
the books of Yosida \cite{yosida} and Reed and Simon \cite{RSII} with
independent proofs. For Hilbert spaces it is possible to apply the result using
a
scale of spaces and thus to extend it to the case where the domain of a
\emph{quadratic form} is independent of time. This idea is due to Kisy\'nski
\cite{kisynski}. It allows for an application to a special situation of the
$(\ph^4)_2$ model, but it is not sufficient for the proof of the existence of
local scattering
operators.
 
\begin{theorem}\label{timeind}
Let $X$ be a Banach space and  $I \subset \mathbb{R}$ an open interval. Let
$A(t)$ be maximally dissipative and assume that $0 \in \rho(A(t))$ for every
$t \in I$. Moreover, assume that
\begin{enumerate}[(i)]
\item the operators $A(t),\ t \in I$, have a common domain of definition,
$D(A(t)) = D$;
\item for each $x \in X$, $(t,s) \mapsto (t-s)^{-1} \left( A(t)A(s)^{-1} -
1 \right)
x$ 
is uniformly strongly continuous for $t \neq s$ in any fixed subinterval of $I$;
\item $\lim_{s \to t} (t-s)^{-1} \left( A(t)A(s)^{-1} - 1 \right) x$
exists uniformly for $t$ in every fixed subinterval of $I$. Moreover, the limit
is bounded and continuous in $t$.
\end{enumerate}
Then there exists a strongly continuous, bounded propagator $U(t,s)$ such that
$t \mapsto U(t,s)x$ solves the nonautonomous Cauchy problem \eqref{ncp} for all initial values $x \in D$.
\end{theorem}
\begin{proof}
The operators $A(t)$ are maximally dissipative, hence Kato-stable. By
assumption they are isomorphisms from $D \to X$. The conditions (ii) and (iii)
imply
the continuous differentiability of $t \mapsto A(t)$ as a mapping with values in
$\cB(D,X)$. So we can set $Q(t) = A(t)$ and apply Theorem~\ref{kato2} with
$B(t) = 0$. 
\end{proof}
In the case of a Hilbert space $X$ the assumptions are fulfilled if for
instance $A(t) = -iH(t) = -i(H_0 + V(t))$, $H_0$ is self-adjoint, $V(\cdot): \RR
\to
\mathcal{B}(X)$ and $V(t)$ maps $D(H_0)$ into $D(H_0)$, $[H_0, V(t)]$ is a
bounded operator and $t \mapsto \|[H_0, V(t)]\|$ is locally bounded, see
\cite{RSII}.

\subsubsection{Application: The Theorem of Kisy\'nski}
\label{sec:kis}
As an application of the existence theorem for time-independent
domains, we
present a theorem due to Kisy\'nski \cite{kisynski} which gives conditions of
solvability for the nonautonomous Cauchy problem \eqref{ncp} in a Hilbert space
if the form domain of $H(t)$ is
constant. Note that the domain of definition of $H(t)$ as an operator in
$\cH$ may well depend on time. The price we have to pay for the 
improvement over Theorem~\ref{timeind} is that the solution of the evolution
equation takes its value in a much larger space than the original Hilbert
space. We present a simplified version of Kisy\'nski's Theorem, which
nevertheless shows the underlying idea.

\begin{theorem}
\label{kiszynskisimple}
Let $H$ be a positive self-adjoint operator in a Hilbert space $\cH$ and
$\cH_{+1} \subset \cH \subset
\cH_{-1}$ the scale of spaces with respect to $H$. For $t\in [0,T],\ T>0$, let
$H(t)$
be a symmetric bilinear form on $\cH_{+1} \times \cH_{+1}$ such that 
\begin{equation}\label{equiv}
c^{-1} (H + 1) \leq H(t)+ 1 \leq c (H + 1)
\end{equation}
 for a constant $c>0$. Suppose further that
$H(t) \in \cB(\cH_{+1}, \cH_{-1})$ is continuously differentiable in norm and
$\pm \dot{H}(t) \leq c(H+1)$. Then there is a unitary propagator $U(t,s)$ on
$\cH$ such that for $x \in \cH_{+1},\ t\mapsto  U(t,s) x \in \cH_{+1}$ is
continuous  in $t$ and such that for  $ x \in \cH_{+1}$ one has $\frac{d}{dt}
U(t,s) x = -i
H(t) U(t,s) x$.
\end{theorem}
\begin{proof}
The main idea is contained in \cite[Chapter~7]{kisynski}. The space $\cH_{+1}$
is a Hilbert space; it equals $D(H^{1/2})$ together with the scalar product
$(\cdot,\cdot)_{+1} = (\cdot,(H+1)\cdot)$ which induces a norm
$\|\cdot\|_{+1}$. By equation \eqref{equiv}, the bilinear form $(\cdot,\cdot)_t
= (\cdot, (H(t) + 1) \cdot)$ induces a norm $\|\cdot\|_t$ on $\cH_{+1}$
which is equivalent to $\| \cdot \|_{+1}$. Hence the bilinear form $H(t)$ is
closed as it gives an equivalent scalar product on $\cH_{+1}$. The associated
operator $H(t) : \cH_{+1} \to \cH_{-1}$ has the domain of definition $D(H(t)) =
\{x \in \cH_{+1}: y \mapsto (y,x)_{t}\; \text{is continuous in}\;\cH_{-1} 
\}$. Again by  \eqref{equiv} we conclude that $ D(H(t)) =
\cH_{+1}$, because by definition $\cH_{-1}$ consists of the conjugate linear
forms on
$\cH_{+1}$. By the form representation theorem, $H(t)$ is a
self-adjoint, positive operator which maps $\cH_{+1} \to \cH_{-1}$. 

Now set $X= \cH_{-1}, D=\cH_{+1}, A(t) = -i H(t)$, then the assumptions on the
differentiability of $H(t)$ as a bounded operator from $\cH_{+1}$ to $\cH_{-1}$
imply that the assumptions of Theorem~\ref{timeind} are fulfilled and the
assertions follow.
\end{proof}

In \cite{dimock}, J.~Dimock investigates the $(\ph^4)_2$ model with a coupling
of the form $g = g_0 + g_1(t,x)$, where $g_0$ is a positive coupling
constant and $g_1 \in C^\infty_c(\RR^2)$ is small and localized: It is assumed
that  $(\diam(\supp g_1) +1)\|g_1\|_\infty$ is sufficiently
small. With this
assumption it is possible to use Kisy\'nski's Theorem in the aforementioned
formulation to prove the existence of the time evolution. Dimock uses this fact
to establish the time evolution for the $(\ph^4)_2$ model on a
two-dimensional, curved spacetime. We return to this topic in Section~\ref{cst}.

The crucial point about the splitting of the coupling into a constant and a
small, localized term is that the scale of spaces $\cH_{+1} \subset \cH \subset
\cH_{-1}$ is built with respect to the Hamiltonian $H = H_0 + g_0 \int :\ph^4
(x):\,dx$ which describes a fixed $\ph^4$ background. The background
Hamiltonian is understood either to be contained in a sufficiently large box or
it may result from a Euclidean construction, hence the underlying Hilbert space
is not the Fock space of the free fields. In any case, the approach is not
well-suited to tackle the existence question of local scattering operators. It
is
not possible to choose $g_0 = 0$ and to use Kisy\'nski's Theorem with the space
triple corresponding to the free Hamiltonian. The problem is that the
bilinear form $H(t) = H_0 + \int g_1(t,x) :\ph^4(x):\, dx$ is not closed on
$\cH_{+1} = D(H_0)$. Furthermore, we did not succeed in performing
the limit $g_0 \to 0$ in a way which would allow to patch together the time
evolution inside of $\supp g_0$ and the free time evolution outside. A similar
problem occurs in
Section~\ref{ph4background}, where we are in fact able to make the interacting
background arbitrarily small, but it remains impossible to connect to the case
of a vanishing background in a continuous way.

We overcome these difficulties in Section~\ref{existencequasiapprox} and
Section~\ref{pphilso} by applying a new wellposedness result. It is this
result which enables us to show the existence of the time evolution for
$P(\ph)_2$ for curved spacetimes without the restrictions on $g_1$ as well as
to prove the existence of the local scattering operators for $P(\ph)_2$ models.

\chapter{Evolution Semigroups}
\label{sec:evolutionsemigroups}

Kato's approach to the Cauchy problem for hyperbolic nonautonomous evolution
equations is inspired by the theory of ordinary differential equations. The
strategy is to  approximate the evolution equation by a sequence of equations
which are easy to solve. Then the convergence of the solution operators is
investigated.

There is  a different approach to the present problem.
Similar to the abstract methods for the study of the existence of weak solutions
which are developed by Lions (see e.g.~\cite{carroll}),
the evolution equation \eqref{ncp} is considered as a functional equation
involving an operator sum.

In \cite{dapratogrisvard}, DaPrato and Grisvard investigate  the
sum $-\frac{d}{dt} + A(\cdot)$ as an operator on the
function
spaces
$E_p = L^p([0,T],X)$ or $C([0,T],X)$
with a suitable domain of definition. They
derive conditions for the
operator sum being densely defined and closable and apply their theory to
parabolic and hyperbolic evolution equations. These ideas were improved in a
paper by
DaPrato and Iannelli \cite{dapratoiannelli}. Here the inhomogeneous Cauchy
problem, $\dot{u}(t) -A(t) u(t) = f(t),\ u(0)=x$, is mapped on the equation
$G u := \{\dot{u} - A(\cdot) u, u(0)\} = \{f,x\}$ and conditions
implying invertibility of $G$ are
studied. The results extend Kato's theory in some respects, but the main
restrictions which limit the usefulness of Kato's results in the context of
quantum field theory do persist.

In this chapter we consider the so-called \emph{evolution semigroups}. By
functional methods similar to the aforementioned, it is possible to relate the
nonautonomous Cauchy problem to an autonomous one. 
This idea goes
back to J.~Howland \cite{howland74} for the study of scattering theory. It
was generalized by D.~Evans
\cite{evans} and others, see \cite{Survey}. The attractive feature of
evolution semigroups is that the powerful theory of strongly continuous operator
semigroups applies. From this one is enabled to derive results for the
time-dependent situation. In recent years evolution semigroups have attracted
new
interest: They turned out to be useful for the study of spectral and asymptotic
properties of propagators. The reader is again referred to \cite{Survey} for a
survey. 

However, most of the results deal with the parabolic situation, and the question
of existence of solutions has not been the subject of the main efforts. 
This is the starting point for the present work.

In the following, we define evolution semigroups and describe their properties.
We show that the existence theory for solutions of the nonautonomous
Cauchy problem \eqref{ncp} can be related to
properties of generators of evolution semigroups. In Section~\ref{wellp} we see
that the generator property of the closure of the operator $G_0 = -\frac{d}{dt}
+ A(\cdot)$ implies
the existence of unique mild solutions of the nonautonomous Cauchy
problem \eqref{ncp}, and we
investigate their relation to weak solutions. In the Hilbert space case,
Section~\ref{sohrproof} contains sufficient requirements for this
theory of solvability. In Section~\ref{existencequasiapprox} we generalize these
considerations to the situation where
the closure of $G_0$ is not necessarily a generator. Under quite general assumptions, $G_0$
has an extension which is an evolution generator and which corresponds to a certain limit of bounded operators. For
Hilbert spaces, the evolution groups associated to this extension lead to
unique approximative solutions of the nonautonomous Cauchy
problem \eqref{ncp}. This is a new existence and uniqueness result for
approximative solutions.
It is this approach
which we will find applicable to the quantum field theoretical problem we have
in mind. 

The main results of this chapter are obtained in collaboration with
R.~Schnaubelt
and can be found in \cite{paper1}.

\section{Definition and properties}

For the definition of evolution semigroups as well as for their basic
properties we follow the presentation in \cite{schnaubeltdiss, paper1}.

Given an arbitrary interval $I \subset \RR$ and a Banach space $X$, we define
the function spaces
$E_p=L^p(I,X),\ 1\leq p<\infty$, with the usual $L^p$-norm as well as $E_\infty=
C(I,X)$ with the $\sup$-norm. For details on this kind of spaces of
Banach-space-valued functions see \cite{diesteluhl, zaanen}. 

Now let  $\{ U(t,s)\}_{(t,s) \in D_I}$ be a strongly continuous,
exponentially bounded
propagator
according to Definition~\ref{def:propagator}.
The \emph{evolution semigroup} $(T(\sigma))_{\sigma \geq
0}$ on $E_{p}$
is the strongly continuous semigroup formally defined by 
\begin{equation}\label{eq:evsgr}
(T(\sigma)f)(t)=\begin{cases}
           U(t,t-\sigma)f(t-\sigma)&\;\text{if}\;  t,t-\sigma \in I,\\
           0 &\;\text{if}\; t\in I, t-\sigma \notin I.
         \end{cases}
\end{equation}
We denote its generator by $G$.

The semigroup property of $(T(\sigma))_{\sigma \geq
0}$ is obvious. In the following we show strong continuity. To this end we
consider
$p < \infty$ and $p =
\infty$ separately.
It is sufficient to consider left
half-open intervals:
Assume that $I$ is left closed and denote its left endpoint by $a$. Set $I' =
I\setminus \{a\}$ if $I$ is bounded from below and $I'=\RR$ if $I= \RR$. Given
an exponentially bounded propagator on $D_I$, its
restriction on $D_{I'}$ induces the same evolution semigroup. But in
general a propagator on $D_{I'}$ has no continuous extension to
$D_{\overline{I'}} = D_{I}$, as seen in the example $X=\CC,
I=(0,\infty)$ and $U(t,s) = p(t)/p(s)$ for $p(t) = 2 + \sin(1/t)$ \cite{rrsv}.
We denote by $C_{c, I'}(I,X)$ the set of
continuous functions compactly supported in $I'$. 

\begin{theorem}
\label{th:evsgr}
Equation \eqref{eq:evsgr} defines a strongly continuous semigroup on $E_{p}$,
the \emph{evolution semigroup} $(T(\sigma))_{\sigma \geq 0} = (e^{\sigma
G})_{\sigma \geq 0}$. 
\end{theorem}
\begin{proof}
First we consider the case $p = \infty$. Let $f \in C_{c, I'}(I,X)$. Boundedness
of the propagator $U(t,s)$ in norm (by $M e^{\omega(t-s)}$) implies $\|T(\sigma)
f\|_{\sigma} \leq M e^{\omega \sigma}\|f\|_{\infty}$, so $(T(\sigma))_{\sigma
\geq 0}$ is a family of bounded operators. Furthermore, we see that $\|T(\sigma)
f -
f \| \to 0$ for $\sigma \to 0$ because of the uniform continuity of the function
$(t,s) \mapsto U(t,s)x$ on compact sets. We conclude that $(T(\sigma))_{\sigma
\geq
0}$ is a strongly continuous semigroup. This extends to $C_{0, I'}(I,X)$. 

Second, let $1 \leq p < \infty$ and $f \in E_{p}$ . We observe that $C_{c,
I'}(I,X)$ is dense in $E_{p}$. Denote by $\tau_{\sigma}$ the semigroup of right
translations on $E_{p}$: $(\tau_{\sigma} f)(t) = \chi_I(t-\sigma)f(t-\sigma)$. 
The function $T(\sigma)f$ is measurable and $\| (T(\sigma)f)(t)\|_{X} \leq M
e^{\omega \sigma}\|(\tau_{\sigma}f)(t)\|_{X}$. So $T(\sigma)f \in E_{p}$ and
$\|T(\sigma)\| \leq M e^{\omega \sigma}$. Furthermore we find that $T(\sigma)$
converges to $\one$ on $C_{c, I'}(I,X)$ for $\sigma \to 0$. Hence, $T(\sigma)
\to \one$ as an operator on $E_{p}$ and strong continuity follows. 
\end{proof}

Note that the translations $\tau_\sigma$ constitute the evolution semigroup on
$E_p$
which is associated with the trivial propagator $U(t,s) = \one$.

We remark that the spaces $E_{p} = L^p(I,X)$ are not the most general
function spaces which are  admissible for the definition of evolution
semigroups. For a given Banach function space $F$ over a measure space
$(\Omega, \Sigma, \mu)$, we define the space $F(X)$ as the set of strongly
measurable functions $f: \Omega \to X$ with the property that $\|f(\cdot)\|_{X}
\in F$ with
the Banach space norm $\|f\|_{F(X)} = \|\|f(\cdot)\|_{X}\|_{F}$. The
conditions we have to impose on $F$ are translation invariance, strong
continuity of the semigroup of translations $\tau_{\sigma}$ and order continuity
of
the norm on $E_p$. For details see \cite{rrs}. Examples for admissible spaces
$F$
are $L^{p}(\RR) \cap L^{q}(\RR), 1\leq p,q < \infty$ or the Lorentz spaces
$L_{p,q}(\RR), 1 < p < \infty, 1 \leq q < \infty$.

We state several simple properties of evolution semigroups
which follow easily from the definition \eqref{eq:evsgr}. The operator norm of
the evolution semigroup can be obtained via the propagators
\begin{equation}
\|T(t)\|_{\cB(E)} = \sup_{s,s-t\in I} \|U(s,s-t)\|_{\cB(X)} \label{norm-eq}
\end{equation}
for $t\ge0$. Hence, by the exponential boundedness of the
propagators, we conclude $\la\in\rho(G)$ if $\Rea\la>\omega$
for the exponent $\omega\in\RR$ in \eqref{def:propagator}. If $I$ is a bounded
 interval, the choice $\omega = 0$ is possible and it follows
$\rho(G)=\CC$. This fact will be of importance later.

The Laplace transform of $T(\cdot)$ is the resolvent of $G$. Thus, by a change
of variables, we get the formula
\begin{equation}
\label{resolv}
(R(\la,G)f)(t)= \int_a^t e^{-\la(t-s)}U(t,s)f(s)\,ds,
   \quad t\in I,\, \Rea\la>w,\, f\in E.
\end{equation}

We recall that a strongly continuous semigroup $S(\cdot)$ with generator $B$
can be embedded
in a strongly continuous group if some operator $S(t)$, $t>0$, is invertible,
see e.g.\
\cite[\S 1.6]{pazy}.
Moreover, $(S(t)^{-1})_{t\ge0}$ is then generated by $-B$. An evolution
semigroup on a function space on a bounded interval $I$ is never invertible
because of the boundary condition. For the case of an unbounded time interval,
we
get the following relation between the propagator and the evolution semigroup.
\begin{lemma}\label{invertible}
Let $U(\cdot,\cdot)$ be a propagator with time interval $I=\RR$ and let
$T(\cdot)$  be the associated evolution semigroup
on $E_p$, $1\le p\le \infty$. Then the following assertions hold.
\begin{enumerate}[(i)]
\item $T(t)$ is an isometry for some $t>0$ if and only if $U(s,s-t)$ is an
isometry for $s\in\RR$
         and any $t>0$.
\item $T(t)$ is invertible for some (and hence for all) $t>0$
      if and only if $U(\cdot,\cdot)$ is invertible.  Then
$(T(t)^{-1}f)(s)=U(s,s+t)f(s+t)$,
        $s\in\RR$.
\item Let $p=2$ and $X$ be a Hilbert space. Then $T(t)$ is unitary if and only
if
       $U(s,s-t)$ is unitary for $s\in\RR$.
\end{enumerate}
\end{lemma}
\begin{proof}
We observe that assertion (iii) follows from (i) and (ii).
In (i) and (ii) the implications `$\Leftarrow$' are easy to check.  The converse
implications
are shown for $p\in[1,\infty)$; the case $p=\infty$ can be established in a
similar way.

Assume that $T(\t)$ is an isometry for some $\t>0$. Take $x\in X$, $s\in\RR$,
and
$\ep>0$. Set
$f=\chi_{[s-\t,s-\t+\ep]}x$ with the characteristic function $\chi_{J}$ of
an interval $J \subset \RR$. Then we obtain
\begin{equation}\label{isometry}
\|x\|^p=\frac{1}{\ep}\,\|f\|^p_p =
\frac{1}{\ep}\int_\RR\|(T(\t)f)(\si)\|^p\,d\si
= \frac{1}{\ep}\int^{s+\ep}_s \|U(\si,\si-\t)x\|^p\,d\si.
\end{equation}
In the limit $\ep\to0$, we arrive at $\|x\|=\|U(s,s-\t)x\|$. Thus (i) holds.

Assume that $T(\t)$ is invertible for some $\t>0$. There are constants
$M'\ge0$ and $w'\in\RR$
such that  $\|f\|_p\le M'e^{w'\t}\|T(\t)f\|_p$ for $f\in E_p$.  As in
\eqref{isometry},
one verifies that $\|x\|\le  M'e^{w'\t}\|U(s,s-\t)x\|$ for $s\in\RR$ and $x\in
X$.
Observe that $T(\t)D(G)=D(G)$. As stated below in Proposition~\ref{char0},
$D(G)$
is dense
in $C_0(\RR,X)$. Hence, $T(\t)D(G)$ is dense  in $C_0(\RR,X)$ so that
 $U(s,s-\t)$ has dense range for $s\in\RR$. Therefore we have established
assertion (ii).
\end{proof}

To address the question of wellposedness of evolution equations with evolution
semigroups as our main tool, we need an abstract characterization of these
special semigroups without referring to a propagator. The idea for this abstract
characterization goes back to the original paper of Howland \cite{howland74}. 
Let $X$ be a Hilbert space and  $\ph \in C^1(\RR)$ be a continuously
differentiable function. Assume that there is a  skew-adjoint operator $G$ on
$L^2(\RR, X)$ such that
$\ph D(G) \subset D(G)$ and
$G (\ph f) - \ph G f = - \dot{\ph} $ for all $f \in D(G)$. Moreover, let $Q$ be
the multiplication operator which is induced by the variable $t$, that is
$(Qf)(t) = t f(t)$ for $f \in D(Q)$. Clearly, $Q$ is self-adjoint. Howland
observes that $G$ has the same commutation relation with $Q$ as $
-\frac{d}{dt}$. Hence $iG$ and $Q$ constitute a representation of the
canonical commutation relations. In the Weyl form this means
$$
e^{\sigma G} e^{-i \sigma' Q} = e^{i \sigma \sigma'} e^{-i \sigma' Q}e^{\sigma
G}
$$
for $\sigma, \sigma' \in \RR$, see e.g. \cite{emch}. By uniqueness of the
Schr\"odinger representation of the CCR, there is a unitary multiplication
operator $U$ on $L^2(\RR, X)$ such that $UQU^* = Q$ and
$e^{\sigma G} = U\tau_\sigma U^* $. Setting $U(t,s) := U(t) U(s)^*$, we find 
$$
(e^{\sigma G}f)(t) = U(t,t-\sigma) f(t-\sigma),
$$
thus, according to
Theorem~\ref{th:evsgr}, $(e^{\sigma G})_{\sigma \geq 0}$ is an evolution
semigroup.
Further
generalizations of this idea can be found in \cite{evans, neidhard1, rrs}.

Crucial for this characterization is the notion of a multiplication operator.
For $p =\infty$, a bounded operator $M \in \mathcal{B}(E_{p})$ is a
multiplication operator if and only if $M\ph f = \ph M f$ for all $\ph \in
C_b(\RR)$ and $f \in E_p$,
see \cite{evans}. For $p <\infty$ and without
assuming separability of $X$, we need special subspaces \cite{neidhard1, rrs}.
\begin{definition}
Let $\mathcal{M} \in \mathcal{B}(E_{p}), 1 \leq p < \infty$. A subspace $F$ of
$E_{p}$ is called $\mathcal{M}$\emph{-determining} if
\begin{enumerate}[(i)]
\item $F$ and $\mathcal{M}F$ consist of continuous functions,
\item $F^{2} = \{f(s): f \in F\}$ is dense in $X$,
\item $F$ is dense in $E_{p}$.
\end{enumerate}
\end{definition}
Now we can characterize multiplication operators in a similar way as in the
situation of spaces of continuous functions.
\begin{theorem}
Let $I \subset \RR$ be an interval. Consider  $\mathcal{M} \in
\mathcal{B}(E_{p}), 1 \leq p < \infty$. Then $\mathcal{M} = M(\cdot) \in
C_{b}(I, \mathcal{B}(X))$ is an operator-valued function, continuous in the
strong topology if and only if
there is an $\mathcal{M}$-determining subspace $F \subset E_{p}$ such that
$\mathcal{M}(\ph f) = \ph \mathcal{M}(f)$ for all $f \in F$ and $\ph \in
L^{\infty}(I)$. 
\end{theorem}
For the proof see \cite{rrs}.
Now, Howland's idea for the characterization of evolution semigroups should be
compatible with the Definition~\ref{def:propagator} of propagators. Therefore,
one has to include a condition which leads to strong continuity of $(t,s)
\mapsto U(t,s)$. A suitable requirement for this purpose is that $G$
generates a strongly continuous semigroup not only on $E_p$, but also on
$E_\infty$.
\begin{theorem}   \label{char0}
Let $G$ generate a $C_0$--semigroup $T(\cdot)$ on $E_p$ for some $1\le p\le
\infty$.
Then the following assertions are equivalent.
\begin{enumerate}[(i)]
\item  $T(\cdot)$ is an evolution semigroup given by an evolution family
$U(\cdot,\cdot)$.
\item  $T(\sigma)\,(\ph f)= (\tau_\sigma \ph)\,T(\sigma)f$ for $f\in E$, $\ph\in
C_b(I)$, $t\ge0$.
      $D(G)$ is  densely and continuously embedded in $C_0(I,X)$.
\item For all $f$ contained in a core of $G$ and $\ph \in C^1_c(I)$
    we have   $\ph f\in D(G)$ and  $G (\ph f)=\ph\, Gf-\dot{\ph}f$.
    $D(G)$ is densely and continuously embedded in $C_0(I,X)$.
\end{enumerate}
\end{theorem}
For the proof see \cite{rrsv}. This theorem relates the characterisation of
evolution semigroups to properties of multiplication operators: The evolution
semigroup behaves like a semigroup of multiplication operators, up to a
translation.
In the last theorem, $D(G)$ is endowed with the graph norm of $G$.
Notice that the second condition in (ii) and (iii) is trivially satisfied
if $p=\infty$. However, if $p<\infty$, (ii) does not imply (i). Consider for
example a function $p$ such that $p$ and $1/p$ are discontinuous elements of $L
^\infty(\RR)$, but $p(s) \to p(t)$ as $s\nearrow t$ for a.e. $t \in \RR$.
Then set $X=\CC$, $E=L^1(\RR)$ and use $U(t,s) = p(t)/p(s)$ to
define a strongly continuous semigroup which satisfies (ii)
without the second condition, but which does not lead to an evolution
semigroup associated with a strongly continuous propagator \cite{rrsv}. However,
dropping the inclusion $D(G)\subset C_0(I,X)$ in (ii) and (iii)
still allows for the characterization of an evolution semigroup associated with
a
strongly \emph{measurable} propagator, see \cite{howland74, evans}. 
In the context of the next section this gives no additional
freedom, but in Section~\ref{existencequasiapprox} the use of
evolution semigroups which do not fulfill the second condition in (ii) and (iii)
will be crucial.

For the application of our results to models of quantum field theory, the
following result will be useful:
\begin{theorem}
\label{similarity}
Let $R = R(\cdot) \in C_b(I, \mathcal{B}(x))$ be a bounded, invertible
multiplication operator with bounded inverse. If $(T(\sigma))_{\sigma \geq 0}$
is an evolution semigroup with generator $(G, D(G))$, then the similar semigroup
$(\tilde{T}(\sigma))_{\sigma \geq 0}$, defined by $\tilde{T}(\sigma) = R T
(\sigma) R^{-1}$, is an evolution semigroup with generator
$(\tilde{G},\ D(\tilde{G}))$ given by $D(\tilde{G}) = R D(G), \tilde{G} = R G
R^{-1}$. 
\end{theorem}
\begin{proof}
Clearly, $(\tilde{T}(\sigma))_{\sigma \geq 0}$ is an evolution semigroup. 
Observe that $R^{-1}$ is also a multiplication operator, hence it commutes with
scalar multiplications. For $\ph\in
C_b(I)$ we have $\tilde{T}(\sigma)\,(\ph f)= R T(\sigma) R^{-1} (\ph f) =
(\tau_\sigma \ph)\,\tilde{T}(\sigma)f$ for $f\in E$. 
The continuous embedding of $D(\tilde{G})$  in $C_0(I,X)$ follows by boundedness
of $R$ and continuity of $t \mapsto R(t)$.
\end{proof}
In the course of this chapter, we will need a technical result concerning cores of the generator $G$ of an evolution semigroup.
\begin{lemma}\label{lem:fc}
Let $T(\cdot)$ be an evolution semigroup with generator $G$ on $E_p$, 
$1\le p\le \infty$, with $I=\RR$.
Let $D$  be a core for $G$ such that $\ph f\in D$ for $f\in D$ and
$\ph\in C_c^1(\RR)$.  Then $D_c=D\cap C_c(\RR,X)$ is also a core for $G$.
If $f\in D(G)\cap C_c(\RR,X)$, then we can take approximating functions $f_n\in
D_c$
whose supports are contained in a bounded interval $J \supset \supp f$.
\end{lemma}
\begin{proof}
Choose $\ph_n\in C^1_c(\RR)$ with $0\le \ph_n\le 1$, $\ph_n=1$ on $[-n,n]$, and
$\|\dot{\ph}_n\|_\infty\le 2$ for $n\in\NN$.
Let $f\in D$ and set $f_n=\ph_n f\in D_c$.
Then $f_n\to f$ and $Gf_n= \ph_n Gf -\dot{\ph}_n f
\to Gf$ in $E_p$ as $n\to\infty$. Thereby we have used that $Gf$ is the limit of $(T(\t)f-f)/\t$ as $\t\to0$ for $f\in D(G)$ by definition, and that we obtain
$\ph u\in D(G)$ and
$G(\ph u)=-\dot{\ph} u+\ph Gu$ if  $u\in D(G)$ and $\varphi\in C^1_c(I)$. So we
have
shown
the first assertion. To verify the second one, take $f\in D(G)\cap C_c(\RR,X)$.
Then there are $g_n\in \cD$, $n\in\NN$, converging in $D(G)$ to $f$.
Let $\ph\in C^1_c(\RR)$ be equal
to $1$ on the support of $f$. Then the supports of the
functions $f_n=\ph g_n\in \cD_c$, $n\in\NN$, are contained in $J:=\supp \ph$.
Moreover, $f_n\to f$ and $Gf_n\to Gf$ as $n\to\infty$, analogously to above.
\end{proof}

\section{Wellposedness and mild solutions}
\label{wellp}

For $I=\RR$ and $p = \infty$, G.~Nickel has investigated
the relation between wellposedness of the nonautonomous Cauchy
problem \eqref{ncp} and properties of
evolution
semigroups in \cite{nickeldiss, nickel97}. Let $C^1 = \{f \in C(\RR,X) :
f,\ d/dt f \in C_0(\RR,X)\}$.
\begin{theorem}
\label{nickelth}
Let $X$ be a Banach space and let $\{A(t)\}_{t \in \RR}$ be a family of
linear
operators. The nonautonomous Cauchy problem \eqref{ncp} is wellposed if and
only if there exists a unique evolution semigroup $T(\cdot)$ on $E_\infty(\RR)
= C_0(\RR,X)$  with generator
$(G, D(G))$ and an invariant core $D \subset C^1 \cap D(G)$ such that
$Gf= - \frac{d}{dt} f + A(\cdot)f$
for $f \in D$.
\end{theorem}
For the proof see \cite{nickeldiss}.

Assuming the hypotheses of the Theorem of Kato (Theorem~\ref{kato1}), Nickel
uses the preceding theorem to prove wellposedness in the hyperbolic case,
thereby giving a simplified proof of Kato's theorem using evolution
semigroups. In doing so, he utilizes an approximating evolution semigroup. A
similar strategy leads to Theorem~\ref{char}. 

The last Theorem~\ref{nickelth} is remarkable as it allows us to characterize
wellposedness of the nonautonomous Cauchy problem \eqref{ncp} by properties of
the generator $G$ of an evolution
semigroup. It establishes a close analogy to the time-independent situation.
Unfortunately, for an application to concrete examples we again need 
assumptions of Kato type.

For the general case $p < \infty$ and $I \subset
\RR$ an analogous equivalence result is not known. But there is a theorem
giving the implication in one direction: If the nonautonomous Cauchy
problem \eqref{ncp} is wellposed, the
generator of the evolution semigroup is the closure of $-\frac{d}{dt} +
A(\cdot)$.
\begin{theorem}
Assume the nonautonomous Cauchy problem \eqref{ncp} is wellposed in the sense of
Definition~\ref{wellposed} with
regularity subspaces $Y_t$ and bounded propagator $U(t,s)$. Moreover, let $(G,
D(G))$ be the generator of the evolution semigroup $T(\sigma)$ on
$E_p(I) = L^p(I,\RR)$ associated with
$U(t,s)$. Then
\begin{equation}
\label{F}
F_p=
\{f\in W^{1,p}_0(I,X): f(t)\in D(A(t)) \,
\text{for a.e.\ } t
\in I,\;
      A(\cdot)f(\cdot)\in L^p(I,X)\}           
\end{equation} 
 is a core for $G$ and $G f= -\frac{d}{dt}f + A(\cdot)f $ for all $f \in F_p$.
\end{theorem}
For the proof see \cite{Survey, schnaubeltdiss}. 

Our strategy proceeds in the opposite way. We define an operator $G_0$ and investigate under which assumptions it has extensions that are
the
generators of evolution semigroups. With regard to Howland's characterization
of evolution semigroups in Hilbert spaces, we could paraphrase our strategy in
the
following way: We start with an incomplete Weyl pair $\{iG_0, Q\}$, and the
construction of the evolution semigroup amounts to complete $\{iG_0, Q\}$ to a
Weyl pair $\{iG, Q\}$ with $G_0 \subset G$, see for example
\cite{joergensenmuhly}. This point of view is advocated in Neidhardt's article
\cite{neidhardtextensions}, where he uses von Neumann's theory of defect indices
to classify the self-adjoint extensions of $iG_0$. This interesting analysis
unfortunately does not lead to conditions which meet the concreteness
requirements needed to obtain wellposedness results for nonautonomous evolution
equations.

Given the nonautonomous Cauchy problem \eqref{ncp}, we endow the multiplication
operator $f\mapsto A(\cdot)f(\cdot)$ on $E_p$
with the maximal domain
\begin{equation}
D(A(\cdot))=\{f\in E_p:  f(t)\in D(A(t)) \text{\ for a.e.\ } t \in I,
   \; A(\cdot)f(\cdot)\in E_p\}.  
\end{equation}

We define the sum
$$
G_0=-\tfrac{d}{dt} + A(\cdot), \qquad D(G_0)=F_p,
$$
 in $E_p$ on the maximal domain from equation \eqref{F}.

As remarked above, wellposedness of the nonautonomous Cauchy
problem $\eqref{ncp}$ would imply that the
closure of $G_{0}$ generates an evolution semigroup (analogous to
\cite[Prop.~4.1]{Survey}). 

We want to prove the opposite implication, saying that $G_0$ possesses a closure
$G$ which generates a semigroup
$T(\cdot)$. 
The following theorem is due to R.~Schnaubelt. It is
formulated in a rather
general way in
order to make
clear under which circumstances a Kato-stable family of generators $A(t)$ on $X$
`generates'
an evolution semigroup on $E_p$. The idea of this theorem can already be found
in \cite{Survey}. 
\begin{theorem} \label{char}
Let $A(t)$, $t\in I$, be Kato-stable generators on $X$ with
constants $(M,w)$
such that $t\mapsto R(w',A(t))$ is strongly continuous for some $w'>w$. Let
$1\le p\le \infty$.
Assume that the space $(w'-G_0)F_p$ is dense in $E_p$ and that $F_p$ is dense in
$E_p$ and $E_\infty$.
Then $G_0$ with domain $F_p$ possesses a closure $G$ in $E_p$ which generates
an evolution semigroup $T(\cdot)$, given by an evolution family $U(\cdot,\cdot)$
on $X$.
\end{theorem}
\begin{proof}
The proof is closely related to Nickel's treatment of
the hyperbolic evolution equations in the $p=\infty$--situation
\cite{nickeldiss}. 
Consider the Yosida approximation $A_n(t)=nA(t)R(n,A(t))=n^2R(n,A(t))-n$ for
$n>w$ and $t\in I$. Notice that $t\mapsto A_n(t)$ is strongly continuous 
 and that $A_n(\cdot)$ is the Yosida approximation
of the generator $A(\cdot)$ on $E_p$, see
for instance Theorem~III.4.8 and Paragraph III.4.13 in \cite{en}.
Because $A_n(\cdot)$ is a bounded perturbation of $-\frac{d}{dt}$, it is then clear that
$G_n=-\frac{d}{dt}+A_n(\cdot)$ with domain $W^{1,p}_0(I,X)$ generates an
evolution
semigroup
$T_n(\cdot)$
on $E_p$ which is given by the evolution family $U_n(\cdot,\cdot)$ generated by
$A_n(\cdot)$.
For $u\in F_p$ we have $G_nu\to Gu$ in $E_p$.
 Due to the Kato stability of $A(\cdot)$,
there are norms $\|\cdot\|_t$ on $X$ satisfying the conditions of
Lemma~\ref{lem:katostable}. In particular,
$\|R(\la,A(t))\|_t\le (\la-w)^{-1}$ for $\la>w$.
This fact yields
$$
\|e^{\tau A_n(t)}\|_t = e^{-n\tau} \|\exp(\tau n^2R(n,A(t))) \|_t\\
     \le  e^{-n\tau} \exp(\tau n^2 (n-w)^{-1}) \le e^{w_1\tau}
$$
 for $ w_1:=(w+w')/2$,  all $n\ge n_0$, and some $n_0\ge w$. Hence the operators
$A_n(t)$, $t\in I$, satisfy Lemma~\ref{lem:katostable} with the same norms and 
the exponent $ w_1$.
Thus they are Kato-stable with uniform constants $M$ and $w_1$. Kato's
existence theorem (Theorem~\ref{kato1}) then shows that
\begin{equation}\label{un-est}
\|U_n(t,s)\|\le Me^{w_1(t-s)}, \;\; (t,s)\in D_I\,;\, \text{hence \ }
  \|T_n(r)\|\le Me^{w_1r},\;\; r\ge0.
\end{equation}
The Trotter--Kato theorem now implies that the closure $G$ of $G_0$ exists and
generates a semigroup $T(\cdot)$ on $E_p$, see \cite[Thm.~III.4.9]{en} or
\cite[Thm.~3.4.5]{pazy}.
Observe that the first condition of Theorem~\ref{char0}(iii) holds on the core
$F_p$ of $G$
on $E_p$. To check the second condition, it suffices to consider
$u$ contained in the core $F_p\subset C_0(I,X)$
and to show that $\|u\|_\infty=\|R(w',G)f\|_\infty\le c \,\|f\|_p$ where
$f:= R(w',G_0)u$.
Recall that $D(G)$ is dense in $C_0(I,X)$ by assumption. We use the
approximation
$G_n$ once more. Due to \eqref{resolv} and \eqref{un-est}, we have
$\|R(w',G_n)g\|_\infty \le c\,\|g\|_p$
for $g\in E_p$ and a constant $c>0$. This estimate implies that
$$
R(w',G_n)f-R(w',G)f = R(w',G_n)(G_n-G_0)u
            =R(w',G_n)(A_n(\cdot)-A(\cdot))u,
$$
and
$$
\|R(w',G_n)f-R(w',G)f\|_\infty  \le c\, \|(A_n(\cdot)-A(\cdot))u\|_p
\;\longrightarrow \; 0
$$
as $n\to\infty$.  As a result, $\|R(w',G)f\|_\infty \le c\,\|f\|_p$.
Theorem~\ref{char0} thus shows that $T(\cdot)$ is an evolution semigroup.
\end{proof}

If one wants to apply the above result, the Kato stability can possibly be checked
using dissipativity of $A(t)$, hence it is not problematic in the Hilbert space
situation with a self-adjoint Hamiltonian.
The density of $F_p$ in $L^p(I,X)$ and $C_0(I,X)$ can be established in two
situations. First, if the resolvents $R(\la,A(t))$, $\la>w$,
 are strongly continuously differentiable in $t$.
Second,
if there is a dense subset $Y$ of $X$ contained in all $D(A(t))$ and $A(\cdot)y$
is continuous
for $y\in Y$ (then $C_c^1(I,Y)\subset F_p$).  Later on we will work in the
latter setting.
In this case,  $R(\la,A(\cdot))$ is strongly continuous if
in addition $Y$ is a core for all $A(t)$.
The most difficult problem is the verification of the range condition:
\begin{equation}
\label{rangecond}
(w'-G_0)F_p\;\text{is dense in}\;L^p(I,X).
\end{equation}
Before giving sufficient conditions leading to the range condition to be
fulfilled in Theorem~\ref{th:sohr77}, in the next result we establish 
differentiability properties on spaces $Y$ as above.
\begin{theorem}\label{diff}
Suppose that the assumptions of Theorem~\ref{char} hold. Let $Y\subset D(A(t))$
for all $t\in I$
and let $A(\cdot)y$ be continuous in $X$ for $y\in Y$. Then the derivatives
\begin{align}
\frac{\partial}{\partial s} \, U(t,s)y&= -U(t,s)A(s)y, \label{diff-s}\\
 \frac{\partial^+}{\partial t} \, U(t,s)y|_{t=s}&= A(t)y, \label{diff-0}
\end{align}
exist for $(t,s)\in D_I$ and $y\in Y$.
(In \eqref{diff-s} one has to take the one--sided derivative
if $t=s$.) If $U(\cdot,\cdot)$ is invertible and $I=\RR $, then one may take
$t,s\in \RR$
in \eqref{diff-s} and two-sided derivatives at $t=s$.
\end{theorem}
\begin{proof}
Take  $t,s,s'\in I$ with $t\ge s$, $t\ge s'$, $s\neq s'$, $y\in Y$,
 and $\ph \in C^1_c(I)$ which is equal to 1 on an interval containing  $s$ and
$s'$.
Set $f=\ph y$. Then $f\in F_p$ and $Gf= -\dot{\ph} y+\ph A(\cdot)y$. Thus
standard
semigroup theory yields
\begin{align*}
U(t,s)y-U(t,s')y &= (T(t-s)f)(t) - (T(t-s')f)(t) = \int_{t-s'}^{t-s}
(T(\tau)Gf)(t) \,d\tau\\
      &=  \int_{t-s'}^{t-s} U(t,t-\tau)(-\dot{\ph}(t-\tau) y+\ph(t-\tau)
A(t-\tau)y)\,d\tau\\
      &= -\int_{s'}^s U(t,r)A(r)y\,dr.
\end{align*}
Multiplying by $(s-s')^{-1}$ and letting
$s-s'\to0$ we deduce  \eqref{diff-s}.
Using this result,  we conclude
$$U(t,s)y-y = -\int_s^t \frac{\partial}{ \partial \tau} U(t,\tau) y\, d\tau
    = \int_s^t U(t,\tau)A(\tau)y\,d\tau,$$
which implies \eqref{diff-0}. The final assertions are verified in the same way.
\end{proof}

If we knew that $U(t,s)Y\subset Y$, then wellposedness of the
nonautonomous Cauchy problem \eqref{ncp} would
follow from
the above proposition and the equality
$U(t+h,s)y-U(t,s)y=(U(t+h,t)-1)U(t,s)y.$
Unfortunately, the invariance of $Y$ is hard to verify.  Again one has to impose
the
restrictive
conditions necessary in Kato's theory. It seems that this
problem is not tackled more easily in the framework of evolution semigroups,
see~\cite{nickeldiss}. 

It turns out to be fruitful to attenuate the notion of solvability and to ask
for
\emph{mild solutions} in the sense of Definition~\ref{approx-def}. The
next theorem due to R.~Schnaubelt shows that mild solutions arise
naturally in the context of Theorem~\ref{char}. They coincide with
the orbits $u=U(\cdot,s)x$ for {\em every}
$x\in X$. 
In the following we will use cut-off functions of the following form: For
$s\in I=(a,b]$
and  $\ep\in(0,b-s]$ we take a function $\ph_\ep\in C^1_c((a,b])$ such that
$0\le \ph_\ep\le 1$, $\ph_\ep=0$ on $(a,s]$  and $\ph_\ep=1$ on $[s+\ep,b]$.
For $I=\RR$ and $d>|s|$, we take $\psi_d\in C^1_c(\RR)$ such that $0\le
\psi_d\le 1$
and $\psi_d=1$ on $[-d,d]$. Set $I_s = I \cap [s, \infty)$. 
\begin{theorem}\label{approx-thm}
Suppose that the assumptions of Theorem~\ref{char} hold. Let $s\in I$, $x\in X$,
and define $\ph_\ep$ and $\psi_d$ as above.
\begin{enumerate}[(i)]
\item Let $I=(a,b]$. Then $u=U(\cdot,s)x\in C([s,b],X)$
is the unique $(E_p)$-mild solution of \eqref{ncp},
where one may take $u_n\in F_p$ in Definition~\ref{approx-def}.
Moreover, $u$ is the only function in $C(I_s,X)$ such that $u(s)=x$,
$\ph_\ep u\in D(G)$, and $G(\ph_\ep u)=0$ on  $[s+\ep,b]$ for all
$\ep\in(0,b-s]$. 
\item Let $I=\RR$ and $U(\cdot,\cdot)$ be invertible.
Then  $u=U(\cdot,s)x\in C(\RR,X)$ is the  unique $(E_p)$-mild solution
on $\RR$ of \eqref{ncp}.
Moreover, $u$ is the only function in $C(\RR,X)$ such that $u(s)=x$,
$\psi_d u\in D(G)$, and $G(\psi_d u)=0$ on  $[-d,d]$ for  all $d>|s|$.
\end{enumerate}
\end{theorem}
\begin{proof}
Set
$$ \tilde{U}(t,s)=\begin{cases} U(t,s)& (t,s)\in D_I,
\\ 0 & t<s, \,t,s\in I.
\end{cases} $$
Since $Gf$  is the limit of $(T(t)f-f)/t$ as $t\to0$ for $f\in D(G)$, we obtain
$\ph u\in D(G)$ and
\begin{align}
G(\ph u)&=-\dot{\ph} u+\ph Gu, \; \text{if \ } u\in D(G) \text{ \ and \
}\varphi\in C^1_c(I),
    \label{g-eq1}\\
G(\ph u)&=-\dot{\ph} u, \; \text{ if \ }\varphi\in C^1_c(I) \text{ with }
           \varphi(t)=0, \; a<t\le s, \text{ \ and \ }
                 u=\tilde{U}(\cdot,s)x, \label{g-eq2}
\end{align}
where $x\in X$ and $s\in I$ in the second line. In the invertible
case,
\eqref{g-eq2} also holds for $v(t)=\ph(t) U(t,s)x$, $t\in I$,  with $\ph\in
C_c^1(I)$.
Now consider case (i). Let $u=U(\cdot,s)x$ and set
$v_n=\tilde{\ph}_n\,\tilde{U}(\cdot,s-\frac{1}{n})x$
for a function $\tilde{\ph}_n\in C^1(I)$ with $0\le \tilde{\ph}_n\le 1$,
$\tilde{\ph}_n=1$
on $[s,b]$ and  $\tilde{\ph}_n=0$ on $(a,s-\frac{1}{n}]$. Then
$$\sup_{s\le t\le b}\|u(t)-v_n(t)\|\le c\,\|x-U(s,s-\tfrac{1}{n})x\|
\;\longrightarrow \; 0$$
as $n\to\infty$. Moreover, $v_n\in D(G)$ and $Gv_n(t)=0$ for $t\ge s$, due to
\eqref{g-eq2}.
There are $w_n\in F_p$ such that $\|v_n-w_n\|_p+\|Gv_n-Gw_n\|_p\le 1/n$.
Since $D(G)$ is continuously embedded in $C_0(I,X)$, we obtain
$$\|v_n-w_n\|_{L^\infty([s,b],X)} \le \tfrac{c}{n}\quad \text{and}\quad
\|Gw_n\|_{L^p([s,b],X)} =\|-\dot{w}_n+A(\cdot)w_n\|_{L^p([s,b],X)}\le
\tfrac{1}{n}\,.$$
Hence $u$ is a mild solution of the nonautonomous
Cauchy problem \eqref{ncp}.

Let $v$ be  another mild solution with
approximating functions $v_n$ as in Definition~\ref{approx-def}. Take
$s<t-r<t\le b$ and a function $\ph\in C^1_c(I)$ which is equal to 1 on $[t-r,t]$
and equal to 0 on $(a,s]$. Then $\ph v_n\in F_p$ and
$$G(\ph v_n)= -\dot{\ph} v_n+\ph (-\dot{v}_n+A(\cdot)v_n).$$
 This identity implies that
\begin{align}
U(t,t-r)v_n(t-r)- v_n(t)&= (T(r)\ph v_n)(t) -(\ph v_n)(t)
        =\int_0^r [T(\tau)G(\ph v_n)](t)\,d\tau \notag\\
        &= \int_{t-r}^t
U(t,\si)(-\dot{v}_n(\si)+A(\si)v_n(\si))\,d\si.\label{unique}
\end{align}
Since $v_n\to v$ uniformly and the integrand converges to 0 in $L^p$ as
$n\to\infty$,
we arrive at
$v(t)=U(t,t-r)v(t-r)$ for all $t>t-r>s$. We thus obtain $u=v$ taking the limit
$r\to t-s$.

The function $u=U(\cdot,s)x$ satisfies $G(\ph_\ep u)=0$ on  $[s+\ep,b]$,
due  to \eqref{g-eq2}. Conversely, let $v\in C(I_s,X)$ be given with $v(s)=x$ 
and
$G(\ph_\ep v)=0$ on  $[s+\ep,b]$.
We can approximate $\ph_\ep v$ in the graph norm of $G$ by $v_n\in F_p$. As in
\eqref{unique},
this fact implies that $v(t)=U(t,s+\ep)v(s+\ep)$ so that again $u=v$.

The assertions in the invertible case (ii) can be shown in a similar way. Here
one starts
with $v_n(t)=\psi_n(t)U(t,s)x$ for  $n\ge |s|$ and $t\in\RR$.
\end{proof}

Under some additional assumptions we can also prove that $u=U(\cdot,s)x$
is the
{\it unique, strongly continuous} weak solution of the nonautonomous Cauchy
problem \eqref{ncp}. 
As we have seen in Section~\ref{weaksolution}, weak solutions exist in a rather
general setting.  We point out that uniqueness and continuity are the
crucial results which are not guaranteed in the general situation. We
now assume that $X$ is reflexive,
that
$1<p<\infty$,
and that the operators $A(t)$  and their adjoints $A(t)^*$, $t\in I$,
 satisfy the assumptions of Theorem~\ref{char}. Let $q=p/(p-1)$.
Replacing $-d/dt$ and the right shift by
$+d/dt$ and the left shift, one can repeat the above proofs for
$G_0'=d/dt+A(\cdot)^*$
defined on
\begin{multline*}
F'_q=\{f\in W^{1,q}(I,X): f(t)\in D(A(t))\text{ for a.e. }t\in I,\;\\
      A(\cdot)f(\cdot)\in L^q(I,X), f(b)=0\}.
\end{multline*}
If $I = \RR$ the condition $f(b) = 0$ has to be dropped.
In particular, $G_0'$ has a closure $G'$ in $L^q(I,X^*)=E_p^*$ which generates
a strongly continuous semigroup. Since $G$ is the closure of
$G_0$, it is straightforward  to check that
$G_0'\subset G^*$. Consequently,  $G'\subset G^*$, and thus $G^*=G'$.
\begin{theorem} \label{weak}
Under the above assumptions, let $s\in I$ and $x\in X$.
\begin{enumerate}[(i)]
\item Let $I=(a,b]$. Then $u=U(\cdot,s)x$ is the only function in $C(I_s,X)$
with
$u(s)=x$
such that
\begin{equation}\label{weak-eq}
\int_s^b \left( u(\tau), \dot{v}(\tau)+A(\tau)^*v(\tau) \right) \,d\tau
 = \left( x, v(s)\right) \qquad  \text{for all \ } v\in F_q'.
\end{equation}
\item Let $I=\RR$ and $U(\cdot,\cdot)$ be invertible.
 Then $u=U(\cdot,s)x$ is the only function in $C(\RR,X)$ with $u(s)=x$ such that
$$ \int_s^b \left( u(\tau), \dot{v}(\tau)+A(\tau)^*v(\tau) \right) \,d\tau=
0\qquad
  \text{for all \ } v\in F_q'\cap C_c(\RR,X^*).$$
\end{enumerate}
\end{theorem}
\begin{proof}
(i) Let $v\in F_q'$, $u=U(\cdot,s)x$, and $u_n$ be the approximating functions
from Definition~\ref{approx-def}. Integrating by parts we then obtain
\begin{align*}
\int_s^b &\left( u(\tau), \dot{v}(\tau)+ A(\tau)^*v(\tau) \right) \,d\tau
= \lim_{n\to\infty} \int_s^b \left( u_n(\tau), \dot{v}(\tau)+A(\tau)^*v(\tau)
\right) \,d\tau \\
&= \lim_{n\to\infty} \int_s^b \left( -\dot{u}_n(\tau)+A(\tau)u_n(\tau), v(\tau)
\right)\,d\tau
         +\left( u_n(s), v(s)\right)\\
&=  \left( x, v(s)\right) .
\end{align*}
Conversely, assume that $u\in C(I_s,X)$ with $u(s)=x$
satisfies \eqref{weak-eq} for all $v\in F_q'$.
Take $\ph_\ep\in C^1(I)$ as above. Using \eqref{weak-eq} for $\ph_\ep v\in
F_q'$, we deduce
$$
 \int_a^b \left( \ph_\ep(\tau)u(\tau), \dot{v}(\tau)+A(\tau)^*v(\tau) \right)
\,d\tau
= -\int_a^b  \left( \dot{\ph}_\ep(\tau)u(\tau), v(\tau) \right) \,d\tau,
 $$
because $\varphi_\ep(t)=0$ for $a<t\le s$. Since $F_q'$ is a core for $G^*$,
this equality
yields
$$
\left( \ph_\ep u, G^*v\right)_{E_p}=- \left( \dot{\ph}_\ep u,
v\right)_{E_p}\qquad
\text{ for all \ } v\in D(G^*).$$
This implies $\ph_\ep u\in D(G)$ and $G(\ph_\ep u)=-\dot{\ph}_\ep u$.
Theorem~\ref{approx-thm}
then shows that $u=U(\cdot,s)x$.

(ii) This assertion can be established in the same way, now using the functions
$\psi_d$.
One only has to verify that  $F_q'\cap C_c(\RR,X^*)$ is a core of $G^*$,
proceeding
 as in the proof of Lemma~\ref{lem:fc}.
\end{proof}

\subsection{Perturbations of evolution semigroups}

The simplest examples for evolution semigroups are obtained if the generator
$A(t) = A$ does not depend on time. Provided that the corresponding autonomous
Cauchy problem \eqref{acp} is wellposed, the generator $G$ of the associated
evolution semigroup is the closure of $-\frac{d}{dt} + A$. One might ask wether
it is
possible to obtain new evolution semigroups as perturbations of these or other
evolution semigroups belonging to solvable nonautonomous evolution equations. 

In \cite{schnaubeltdiss, rrs, rrsv} bounded perturbations and
Myadera perturbations (see \cite{en}) are investigated. The latter formulation
of perturbation theory applies in the parabolic situation.

For Hilbert spaces Howland gives a perturbation result suitable for the
application to the time-dependent Schr\"odinger equation in \cite{howland74}.
In the following we discuss the Theorem resulting from his approach.

Recall that, if $S$ is a closed, densely defined operator with $\sigma(S)
\subset \RR$, then a closed, densely defined operator $A$ is $S$-smooth if and
only if $D(S) \subset D(A)$ and 
$$
\sup_{\epsilon > 0} \int_{- \infty}^\infty \|A(S r \pm i\epsilon)^{-1}f\|^2
\,dr < \infty
$$
for all $f \in E$. For properties of such operators see for instance
\cite{rsIII}.
\begin{theorem}
Let $X$ be a Hilbert space and $E= L^2(\RR, X)$. Let $G$ be the skew-adjoint
generator of an evolution group on $E$.
Assume that $A$ and $B$ are closed,
densely defined operators on $E$ which are $iG$-smooth.
Moreover, assume that
$\ph A \subset A\ph$ and $\ph B^* \subset B^* \ph$ for every bounded scalar
function $t \mapsto \ph(t)$ and that $(Af,Bg) = (Bf, Ag)$ for all $f,g \in D(A)
\cap D(B)$. Let $R_\lambda := -R(G,\lambda)$ for $\lambda \in \CC$ with
$\Ima \lambda \neq 0$ and let there be a constant $c$ such that 
$$
\|A R_\lambda B^* f\| \leq c \| f\|
$$ 
for all $f \in D(B^*)$. Then the unique bounded extension $Q(\lambda)$ of $A
R_\lambda B^* $ has strong non-tangential boundary values $Q(r \pm i0)$
for almost every $r \in \RR$. Assume that $Q(r \pm i0)$ is quasi-nilpotent
almost everywhere and that there is a finite increasing function $t \mapsto
\rho(t)$
such that
$$
\|(I + \kappa Q(\lambda))^{-1}\| \leq \rho(|\kappa|)
$$
for real $\kappa$ and uniformly in $\lambda$. Under these assumptions,
the bounded operator
$$
\tilde{R}(\lambda,\kappa):= R_\lambda - \kappa (B R_{\overline{\lambda}})^* (I
+ \kappa Q(\lambda))^{-1} A R_\lambda
$$
is the resolvent of a self-adjoint extension  $i\tilde{G}(\kappa)$ of $ iG +
\kappa B^\ast A$ which is the generator of a unitary evolution group. 
\end{theorem}
Howland applies this theorem to the time-dependent Schr\"odinger equation
\eqref{ncp} with $A(t) = -i(-\nabla + \kappa q(x,t))$ on $X = L^2(\RR^n)$, where
$q(t,x)$ is a potential such that for some $\infty > p > n/2$ and $\epsilon >
0$, 
$$
v_p(t) := (\int_{\RR^n} |q(t,x)|^p \,d^nx)^{1/p}
$$
is an element of $L^{r+\epsilon}(\RR) \cap L^{r-\epsilon}(\RR)$,  $r :=
\frac{2p}{2p - n}$.
However, this perturbation result is clearly not applicable in a quantum field
theoretical context. The $iG$-smoothness of the perturbation implies relative
$iG$-boundedness with arbitrarily small bound \cite{rsIII}[Theorem~XIII.22].
Even for the $(\ph^4)_2$ model,
the interaction Hamiltonian is not bounded relative to $H_0$, and on the level
of the evolution semigroups the situation does not improve. This can be seen
in the following way: Assume $H_0 >0$ is self-adjoint, $V(t)$ is symmetric,
$D(V(t))\supset D(H_0)$ and it is relatively $H_0$-bounded with bound smaller
than $1$, that is $\|V(t)x\| \leq a(t)\|H_0 x\| + b(t)\|x\|$, where $0 < a(t)
<1$ uniformly in $t$,  $0 \leq b(t)$. Then 
$H(t) = H_0 + V(t)$ is self-adjoint on $D(H(t)) = D(H_0)$.  Let $I= (a,b]$ be
a bounded interval and $E = L^2(I,X)$. With additional smoothness conditions for
$t\mapsto a(t)$ and $t \mapsto b(t)$, one might conjecture that on $E$, the
dissipative multiplication operator $iV(\cdot)$ is relatively $G$-bounded with
bound smaller than $1$, where $G$ is the closure of $G_0 := - \frac{d}{dt}
-iH_0$. Thus $G - iV(\cdot)$ with domain $D(G)$ would be the generator of an
evolution semigroup \cite[Corollary 3.3.3]{pazy}. Now the autonomous Cauchy
problem with respect to $A = -iH_0$ is wellposed, hence the generator $G$ of the
evolution semigroup associated with the propagator $U(t,s) = e^{-i(t-s)H_0}$ is
indeed the closure of $G_0$. Hence we would like to deduce boundedness of the
operator $iV(\cdot) R(\lambda, G)$ from the boundedness of $V(\cdot)(H_0 +
c)^{-1}$. This would be possible if $iH_0 R(\lambda, G)$ is a bounded operator on 
$E$. But, in general, this is wrong:
In the special case under consideration, formula \eqref{resolv} yields
$$
(R(\lambda,G)f)(t)= \int_a^t e^{-\la(t-s)}e^{-i(t-s)H_0}f(s)\,ds,
$$
and we see that $R(\lambda,G)$ in general does not map arbitrary $f \in E$ into
the domain
of $H_0$ considered as a multiplication operator on $E$. The reason is the lack
of an appropriate smoothing property of the Schr\"odinger group $(e^{-itH_0})_{t
\in \RR}$. This expresses the fact that $G_0$ is not closed on $F_{p=2}$. These
difficulties are typical for the hyperbolic context. In contrast, in the
parabolic case one can find the situation where $G_0$ is already closed
on $F_2$, hence boundedness of $iH_0 R(\lambda, G)$ follows. The Schr\"odinger
semigroup $(e^{-tH_0})_{t \geq 0}$ is smoothing \cite{simonsgr}.

However, also in the hyperbolic context the domain of a Hamiltonian obtained
from a small, relatively $ H_0$-bounded perturbation equals $D(H_0)$ and this
is independent of time. Hence, with suitable smoothness
assumptions, we can apply Theorem~\ref{timeind} to show wellposedness of the
time-dependent Schr\"odinger equation and we obtain the evolution semigroup
from the associated propagator. Thus we find perturbation techniques not
to be promising for the topic of our work. In the next Section, we return to a
more
direct approach.

\subsection{The Theorem of Sohr}
\label{sohrproof}

In \cite{sohr77} H.~Sohr develops a link between parabolic and hyperbolic
evolution equations for Hilbert spaces $X$. In the parabolic context it is
possible to define $G_0$ as an operator sum on the intersection of the domains
of the time derivative and the generators. The sum is closed and accretive on
this domain. Then it is possible to recover the hyperbolic evolution equation
by a perturbation argument and a strategy similar to `Konrady's trick'
\cite{RSII}.
We present a simplified version of Sohr's theorem, but for this setting we
extend the result showing unitarity for the propagators.
\begin{theorem}
\label{th:sohr77}
Let $H(t)$,  $t\in\RR$, be self-adjoint operators on a Hilbert space $X$.
 Set  $A(t)=-iH(t)$. We assume that for every $r>0$ there
 are positive constants $\beta=\beta(r)$ and $k=k(r)$ such that $H(t)$
is bounded from below by $1-\beta$, $t\mapsto (\beta+H(t))^{-1}$
is weakly continuously differentiable,  and
\begin{equation}\label{onesided}
\frac{1}{2}\frac{d}{dt} (x, (\beta+H(t))^{-1}x) + k (x, (\beta+H(t))^{-1}x)
\geq 0
\end{equation}
for  all $x \in X$ and  $|t|\le r$. Then the conditions of Theorem~\ref{char}
and Theorem~\ref{weak} with $p=2$ are fulfilled. In particular,
there exists a unitary evolution family
$U(t,s)$, $t,s\in\RR$, such that $U(\cdot,s)x$ is the unique
mild, thus unique, continuous weak  solution of the nonautonomous Cauchy problem
\eqref{ncp} on $I=\RR$.
\end{theorem}
\begin{proof}
Since $A(t)$ is skew-adjoint, the family $\{A(t)\}_{t\in\RR}$ is Kato-stable.
Let $r>0$, $t\in I:=(-r,r]$, and $\la>0$. We first observe that
$t\mapsto (\beta(r)+H(t))^{-1}$ is Lipschitz on $[-r,r]$ since
\begin{multline*}
\left|\left(\left[(\beta(r)+H(t))^{-1}x-(\beta(r)+H(s))^{-1}x\right],
y\right)\right|\\
= \Big|\int_s^t \frac{\partial}{\partial \tau
}((\beta(r)+H(\tau))^{-1}x,y)\,d\tau\Big|
\le c\,|t-s|\,\|x\|\,\|y\|
\end{multline*}
for $-r\le t,s\le r$ and $x,y\in X$, where $c=c(r)$ does not
depend
on $x$ and $y$
by the principle of uniform boundedness. We further compute
\begin{align*}
R(\la,A(t))&= i (i\la-\beta(r)+ \beta(r)+H(t))^{-1}\\
  &= i[1+ (i\la-\beta(r))( \beta(r)+H(t))^{-1}]^{-1}  (\beta(r)+H(t))^{-1}
  \end{align*}
using \cite[Thm.IV.1.13]{en}. After multiplication with $\beta(r)+H(t)$,
the above equation  implies that
the operators $[\cdots]^{-1}$ are uniformly bounded for $t\in I$.
Hence $t\mapsto R(\la,A(t))$ is Lipschitz on $[-r,r]$.
 We take $p=2$.
For $f\in C_c^1(I,X)$ and $n\in \NN$, we define $f_n=nR(n,A(\cdot))f$.
Then $f_n$ is Lipschitz and thus $f_n\in F(I)=W^{1,2}_0(I,X)\cap D(A(\cdot))$.
Moreover, $f_n\to f$ in $E(I)$ and $C_0(I,X)$ as $n\to\infty$, see the proof of
Theorem~\ref{char}. Hence $F(I)$ is dense in $E(I)$ and in $C_0(I,X)$.

The derivative $\frac{d}{dt}$ is a maximally accretive operator on
$W^{1,2}_0(I,X)$. Its adjoint is $\frac{d}{dt}^*$ on $D(\frac{d}{dt}^*)
= \{f \in
W^{1,2}(I,X) : f(r) = 0\}$. Using the assumptions, we calculate
\begin{multline}
\Rea \left( (\frac{d}{dt} + k)^{\ast} f, (\beta + H(\cdot))^{-1}f \right)  = \\
\Rea \int_{-r}^{r}\left[ \left( -\frac{d}{dt} f(t), (\beta +
H(t))^{-1}f(t)\right) + k
\left( f(t), (\beta + H(t))^{-1} f(t)\right) \right] \; dt  \geq \\
\int_{-r}^{r}\left[ \frac{1}{2} \left( f(t), \frac{d}{dt}(\beta + H(t))^{-1}
f(t)\right) +
k
\left( f(t), (\beta + H(t))^{-1}  f(t)\right) \right] \;dt \geq 0
\end{multline} 
for $f \in D((\frac{d}{dt} +
k)^{\ast})$.
By Lemma~\ref{sohr} we conclude maximal accretivity of $\frac{d}{dt} + H(\cdot)
+ \beta +
k$ on $F = W^{1,2}_0(I,X) \cap D(H(\cdot))$.
Using the transformation $\tilde{f}(t) = e^{kt} f(t)$, we see that the constant
$k$
can be arbitrarily adjusted and so also $\frac{d}{dt} + H(\cdot) +\beta$ on $F$
is maximally
accretive.

Positivity and self-adjointness of $\beta + H(t)$ imply $\Rea
\left(\frac{d}{dt} f, (\beta + H(\cdot)) f \right) \geq \frac{1}{2} \left(
(\beta + H(\cdot))f, \frac{d}{dt}(\beta + H(\cdot))^{-1}
(\beta + H(\cdot))f\right)$. Assumption \eqref{onesided} now implies
$$
\Rea \left( (\frac{d}{dt} + i H(\cdot) + k) f, (\beta + H(\cdot))f\right) \geq
0,
$$
and we conclude
$$
\left\|\left( \frac{d}{dt} + H(\cdot) +i H(\cdot)+\beta+ k \right)\right\|^{2}  \geq
\|(\beta + H(\cdot))f\|^{2}
$$
for all $f \in F$. Consequently, $-(\beta + H(\cdot))$ is bounded relative to
$\frac{d}{dt} + H(\cdot) +i H(\cdot)+\beta+ k$ with bound $1$. By a semigroup
version of W\"ust's Theorem (\cite[Corollary 3.3.5]{pazy}), we conclude that
$  -\left(\frac{d}{dt} + i H(\cdot) + k\right)$ defined on $F$ has a closure which is
maximally dissipative and generates a strongly continuous semigroup, hence it is
surjective. This ensures the range condition for $G_0=-\frac{d}{dt}+A(\cdot)$
defined on $F(I)$, and
Theorem~\ref{char} shows that its closure $G_I$ 
exists and generates an evolution semigroup $T_I(\cdot)$ on $L^2(I,X)$
corresponding to
a propagator $U_I(t,s)$, $(t,s)\in D_I$. Since $G_0$
is dissipative, $T_I(\t)$ and $U_I(t,s)$ are contractive.

Using the uniqueness result from  Theorem~\ref{approx-thm}, we can define
the propagator $U(t,s)=U_I(t,s)$ for $(t,s)\in D_\RR$,
where $(t,s)\in D_I$.
The corresponding evolution semigroup $T(\cdot)$ with generator $G$
satisfies $T(\t)f(s)=T_I(\t)f(s)$ for $0\le \t\le 1$ and $s\in I=[-r,r]$,
if $f$ has compact support in $[-r+1,r]$
for some $r>1$. This shows that (the restriction of)
 $f\in D(G)\cap C_c(\RR,X)$ belongs to $D(G_I)$ for some $I$
 whose interior contains the support of $f$,
 and $Gf=G_If$ on $I$. Thus there are $f_n\in F(I)$ such that
 $G_0f_n\to Gf$ in $L^2(I,X)$ and $f_n\to f$ in $C_0(I,X)$ as $n\to\infty.$
 Take $\ph\in C^1_c((-r,r))$ with $\ph=1$ on the support
 of $f$, and extend the functions $ \ph f_n$ by $0$ on the complement of $\supp
f$.
 Then $\ph f_n\in F_c(\RR)=F(\RR)\cap C_c(\RR,X)$. Since
  $G_0(\ph f_n) = -\ph' f_n +\ph G_0f_n$  on $I$, the functions
  $G_0(\ph f_n)$  converge to $Gf$  in $L^2(\RR,X)$.
On the other hand, $D(G)\cap C_c(\RR,X)$ is a core of $D(G)$ by
Lemma~\ref{lem:fc}.
As a result, $F_c(\RR)$ is a core for $G$ and $Gf=G_0f$
for $f\in F_c(\RR)$. Moreover, $G_0$ is skew-symmetric on
$F_c(\RR)$, so that $G$ is skew-adjoint.
Therefore the assertions follow from
Lemma~\ref{invertible}, Theorem~\ref{approx-thm}, and Theorem~\ref{weak}.
\end{proof}

\subsubsection{Application: Quantum mechanics}

Define the \emph{Rollnik class} by
\begin{multline*}
R=\{V:\RR^3\to\CC : V\text{ is measurable, \ }\\ \|V\|_R^2:= \iint
|V(x)V(y)|\,|x-y|^{-2} \,dx\,dy < \infty\}.
\end{multline*}
It is
a Banach space with the norm $\|V\|_R$, and it contains physically reasonable
potentials which are not covered by the Kato class. The main significance of
the Rollnik condition is that it assures the Kato-smallness of $|V|^{1/2}$
with respect to $H_0^{1/2}$ and the Hilbert-Schmidt property for operators as
$(H_0 + \beta)^{-1/2}|V|(H_0 + \beta)^{-1/2}$ or $|V|^{1/2}(H_0 +
\beta)^{-1}|V|^{1/2}$ for suitable constants $\beta$. We refer to \cite{Si}
for further
properties of $R$. In quantum mechanics one would allow
potentials $V \in R + L^{\infty}(\RR^3)$. For simplicity we restrict ourselves
to the Rollnik class, but the example can easily  be modified to include an
$L^{\infty}$-part.
\begin{theorem}\label{schroedinger}
Let $V:\RR\times\RR^3\to \RR$ be a measurable function such that
$t \mapsto V(t,x)$ is continuously differentiable
for a.e.\ $x\in\RR^3$ with  partial
 derivative $\frac{\partial}{\partial t} V(t,x) = \dot{V}(t,x)$.
Suppose that  $|V(t,x)|, |\dot{V}(t,x)| \le W(x)$ for $t\in (a,b)$ and a
function $W\in R$
possibly depending on the interval $(a,b)$.
Then the assertions of Theorem~\ref{th:sohr77} hold.
\end{theorem}
\begin{proof}
Let $H_0=-\Delta$ with $D(H_0)=W^{2,2}(\RR^3)$.
Since $|V(t)| \leq W$,  $|V(t)|^{1/2}$ is a small
perturbation
of  $H_0^{1/2}$ locally uniformly in $t$, that is,  we find a constant
$b_1>0$, uniformly on compact intervals $[-r,r]$, such that
$$
\|\,|V(t)|^{1/2} x\|^2 \leq \tfrac{1}{4} \|H_0^{1/2}\|^2 + b_1 \|x\|^2
$$
for all $x \in D(H_0^{1/2})$ and $|t|\le r$, see the proof of
\cite[Thm.~I.21]{Si}. So we conclude from
the KLMN theorem \cite[Thm.~X.17]{RSII} that there is a self-adjoint operator
$H(t)$ such
that $D(|H(t)|^{1/2}) = D(H_0^{1/2})$ and $H(t) = H_0 + V(t)$ as a quadratic
form on $D(H_0^{1/2})$. Moreover, $H(t) \geq -b_1$ is semibounded from below.
Since also
 $\dot{V}(t) \in R$, we can repeat this argument to define a self-adjoint
operator
$\tilde{H}(t)$
with $D(\tilde{H}(t)^{1/2}) = D(H_0^{1/2})$ and $\tilde{H}(t) = H_0 + V(t) -
\dot{V}(t) + b_1$ as a quadratic form on $D(H_0^{1/2})$. There is a $b_2=b_2(r)
> 0$
such that $\tilde{H}(t) \geq -b_2$  for $|t|\le r$. Choosing a
sufficiently large  $c \geq 1$, we set $\beta = \beta(r) = b_1 + b_2 + c$. Then
we find
$H(t) \geq 1- \beta$ and $H(t) + \beta \geq \dot{V}(t) + c$ for $|t| \leq r$.
Moreover, Tiktopoulos' formula \cite[Thm.~II.12(a)]{Si} shows that
\begin{equation}\label{tik}
(H(t) + \beta)^{-1} = (H_0 + \beta)^{-1/2}(1 + B(t))^{-1}(H_0 +
\beta)^{-1/2},
\end{equation}
where $B(t) = (H_0 + \beta)^{-1/2}V(t)(H_0 + \beta)^{-1/2}$ is a bounded
operator with $0 \leq \|B(t)\|\le q<1$ for $|t|\le r$ and  sufficiently large
$\beta=\beta(r)$. We observe that then 
$(1+B(t))^{-1}$ is uniformly bounded for $|t|\le r$.
Our assumptions and Lebesgue's Theorem  imply the differentiability of 
$t\mapsto (f,B(t)g)$ for $f,g\in L^2(\RR^3)$. 
Therefore, $t\mapsto B(t)$  is Lipschitz continuous on $[-r,r]$ by the principle of uniform
boundedness. It follows from
$$
(1+B(t))^{-1} -(1+B(s))^{-1} = (1+B(t))^{-1} (B(s)-B(t))
(1+B(s))^{-1}
$$
that also $(1+B(\cdot))^{-1}$  is Lipschitz-continuous. Furthermore, the
equation
\begin{multline*}
(t-s)^{-1} ( [(1+B(t))^{-1} - (1+B(s))^{-1}]f , g )\\
= ( [B(s)-B(t)] (1+B(s))^{-1} f, (t-s)^{-1}\,[(1+B(t))^{-1} -
(1+B(s))^{-1}] g)
\\
+ ( (t-s)^{-1}\,[B(t) -B(s)] (1+B(s))^{-1}f, (1+B(s))^{-1} g)
\end{multline*}
implies the weak differentiability of $t \mapsto (1+B(t))^{-1}$, as the first
term on the right-hand side vanishes and the second
one converges in the
limit $t \to s$ .
By \eqref{tik} we conclude that $(H(t) +\beta)^{-1}$ is weakly continuously
differentiable. 
Finally, we calculate
\begin{align*}
-\frac{d}{dt}(x,(H(t) +\beta)^{-1}x ) &\leq ((H(t)
+\beta)^{-1}x,(\dot{V}(t) + c)(H(t) +\beta)^{-1}x)\\
&\leq  (x,(H(t) +\beta)^{-1}x),
\end{align*}
so that the assumptions of Theorem~\ref{th:sohr77} with $k=\frac{1}{2}$ are
fulfilled.
\end{proof}

\subsubsection{Application: $(\ph^4)_2$ with spatial localization}
\label{sec:phi4}

We apply Theorem~\ref{th:sohr77} to the $(\ph^4)_2$ model.
For the notation and some basic
facts see Appendix~\ref{qftbasics}. In the context of this
chapter we consider the case $P(\lambda) = \lambda^{4}$, that is the
interaction Hamiltonian is formally given as
$$
V(t;g) = \int g(t,x) :\ph(x)^4:\,dx
$$
 Under some additional assumptions on the localization function $g$ in
the interaction, we can show the existence of the time evolution with
Theorem~\ref{th:sohr77}.

Set $Y = D(H_0)
\cap D(N^2)$. We allow localization functions which are not compactly
supported in $t$ and choose them from the set $\tilde{\mathcal{S}}(\RR^2)$,
where 
$$
\tilde{\mathcal{S}}_I = \{ g \in \mathcal{S} : x \mapsto g(t,x) \in
C^{\infty}_{c,I}(\RR)\},
$$
and $C^{\infty}_{c,I}(\RR)$ are the smooth functions with
support in a fixed compact interval $I$.
\begin{theorem}
\label{th:phi^4}
Let $g \in \tilde{\mathcal{S}}_I$ be given in such a way that for every $r >0$
there exists a $k>0$ such that 
\begin{equation}
\label{eq:gcond}
k g(t,x) - \frac{1}{2} \dot{g}(t,x) \geq 0 
\end{equation}
for all $x \in \RR, |t| \leq r$.
Then the time-dependent Schr\"odinger equation
 with $H(t) = H_{0} + V(t;g)$ has a
unique mild solution with corresponding evolution family
$U(t,s)$.
\end{theorem}
We remark that compactly supported localization functions do not fulfill
\eqref{eq:gcond} in
general. An example for an admissible localization $g_\alpha \in
\tilde{\mathcal{S}}_I$ is obtained by starting with $g \in
C^\infty_c(\RR^{2})$, $\supp g \subset I \times I$, and convoluting it with a
smooth, fast decreasing
approximation of the
$\delta$ distribution in time, for example $g_{\alpha}(t,x) = (\delta_{\alpha}
\ast_{t} g)
(t,x)$  where  $\delta_{\alpha}(t) =
\frac{1}{\alpha \sqrt{\pi}} e^{- \frac{t^2}{\alpha^{2}}}$ for $\alpha
>0$.  

For the proof of the Theorem, we need two Lemmas.
\begin{lemma}
\label{lem:boundedderivative}
Let $g,h \in \tilde{\mathcal{S}}_{\RR}$. For $\beta>0$ sufficiently large,
$$(\beta +
H(t))^{-1} V(t;h) (\beta + H(t) )^{-1}$$ extends  to a 
bounded operator for every $t$.
\end{lemma}
\begin{proof}
Fix $t \in \RR$ and set $H := H(t) = H_0 + V(t;g), V(g) = V(t;g)$ and $V(h) =
V(t;h)$.
$H$ is self-adjoint on $D(H_0) \cap D(V(g))$, so in particular it is closed.
Notice that this fact is not yet proved for 
interactions with higher powers of $\ph$ than $4$. In this case only
essential self-adjointness on the  aforementioned domain is proved in the
literature.

Choose $\beta > E>0$ with a lower bound $E$ of $H$, which exists according to
Lemma~\ref{semib}.
From closedness of $H$ it follows that there is a constant $c_1$ such
that
\begin{equation}
\label{eq:ph4closed}
H_0^2 + V(g)^2 \leq c_1(\beta + H^2 )
\end{equation}
as a quadratic form on $Y \times Y$. So $( \beta + H_0 )^2 \leq c_1((\beta + H
)^2 +
\beta)$. Moreover, by
Theorem~\ref{th:n}, there is another
constant $c_2$ such
that
$V(h) \leq c_2 (H_0 + \beta)^2$. It follows that
$$
V(h) \leq 2 c_1 c_2 (\beta + H )^2.
$$
Because $Y$ is a core for $H$, the set $Z = ( \beta + H )Y$ is dense. This
implies
that 
the operator
$(\beta +H )^{-1} V(h) (\beta + H )^{-1}$ is bounded on a dense set and thus
extends to a bounded operator.
\end{proof}
\begin{lemma}
\label{lem:phi4/1}
Let $I \subset \RR$ be a compact interval. The mapping $t \mapsto H(t)y$ on $I$
is differentiable for all $y \in Y$. The
derivative is
given by
the Wick polynomial $\dot{H}(\cdot) = V(\cdot;\dot{g})$ with $Y
\subset
D(\dot{H}(\cdot))$. Moreover, $(\beta + H(t))^{-1}$ is weakly
differentiable, and 
$$
\frac{d}{dt} \left( x, (\beta + H(t))^{-1} x \right) = - \left( x, (\beta + H(t)
)^{-1} \dot{H}(t) (\beta + H(t))^{-1} x \right)
$$
for all $x \in X$ and $\beta>0$ sufficiently large. The function $t
\mapsto \frac{d}{dt} \left( x, (\beta + H(t))^{-1} x \right)$ is 
bounded.
\end{lemma}
\begin{proof}
We show that the first assertion is implied by the smoothness of $g$ in the
standard way,
using Theorem~\ref{th:n}:
Let $w(t,k_{1},\dots,k_{4})$ be the scalar kernel as defined in equation
\eqref{eq:w}. We estimate 
$\| V(t;g)y \| \leq c \| w(t,\cdot)\|_{L^{2}(\RR^{4})}
\|(N+1)^{2}x\|$ for $y \in Y$, and thus $H(t)\negthickspace\restriction_{Y}$
is uniformly
bounded as an operator from $Y \to X$.

Choose $\beta$ to be larger than the locally uniform lower bound $E$
of $H(t)$, according to Lemma~\ref{semib}.
The difference $H(t+h)-H(t) $ is defined on $Y$, as well as
$V(t;\dot{g})$. The latter operator has the same
structure as $V(t;g)$, but $g$
is replaced by its time derivative.
For $y \in Y$ and $s >0$ we find
\begin{align*}
\| &\left(\frac{1}{s}(H(t+s) - H(t)) - V(t;\dot{g})\right)y 
\| \\
 &\leq \| \left( \frac{1}{s} (V(t+s;g) - V(t;g)) -
 V(t;\dot{g})\right)(N+ 1)^{-2}\| \| (N+ 1)^{2}y \| \\
&\leq c \| \frac{1}{s} ( w(t+s,\cdot) -
w(t, \cdot) - \frac{\partial}{\partial t}
w(t, \cdot) \|_{L^{2}(\mathbb{R}^{4})}
\|(N+1)^{2} y  \|
\end{align*}
Now we turn to the second assertion. Using Lemma~\ref{lem:boundedderivative}
with $h = \dot{g}$, one sees that it is
sufficient to show weak differentiability of $(\beta + H(t))^{-1}$ on a
dense
set.

Let $y \in Y$. By the differentiability of $H(t)$ on $Y$ and the continuity of
its resolvent in $y$, one has
$$
\lim_{s \to 0} \frac{-1}{s} (\beta H(t+s) )^{-1}\left( H(t+s) -
H(t)\right)y = -
(\beta + H(t))^{-1} \dot{H}(t)y.
$$ 
For all $x \in Z = (\beta + H(t) )Y$ it follows
\begin{multline*}
\lim_{s \to 0} \frac{-1}{s} \left( x, (\beta + H(t+s) )^{-1}(H(t+s) -
H(t))(\beta + H(t))^{-1} x \right)\\
- \left(x, (\beta + H(t))^{-1}\dot{H}(t)(\beta + H(t))^{-1} x \right) = 0,
\end{multline*}
and this implies differentiability on $X$. Local boundedness of 
$$
t \mapsto
\frac{d}{dt} \left(x,(\beta + H(t) )^{-1}x \right)
$$
 follows from its continuity,
which is clear for $ x \in Z$ and extends to $x \in X$ by boundedness.
\end{proof}
\begin{proof}[Proof of Theorem~\ref{th:phi^4}]
Fix $r>0$ and $g\in \tilde{\mathcal{S}}_I$ which satisfies \eqref{eq:gcond}. By
Lemma~\ref{semib} we can find an $E = E(r)$ locally uniformly on intervals $
[-r,r]$, such that $H(t) \geq -E$
for $|t| \leq r$. Thus we set $\beta = 1 + E + c$ with a constant $c >0$ which
 will be fixed in a moment. The weak continuous differentiability of $t
\mapsto (\beta + H(t))^{-1}$ is fulfilled according to Lemma~\ref{lem:phi4/1}.
 Because of \eqref{eq:gcond} and by eventually increasing $k$, it is
possible to plug $ g -
\frac{1}{2k} \dot{g}$ instead of $g$ into the interaction term $V(t;g)$.
Lemma~\ref{semib} assures the existence of a $c>0$ such that 
\begin{equation}
H_{0} + \int :\ph^{4}(x): \left( g(t,x) - \frac{1}{2k} \dot{g}(t,x)\right) \;dx
= H(t) + V(t;- \frac{1}{2k} \dot{g}) \geq c
\end{equation} 
as a quadratic form on a core of $H(t)$ and $t \in [-r,r]$. Representing this
core as the image of a dense set under $(\beta + H(t))^{-1}$ and using the
formula for the weak derivative of $t \mapsto (\beta + H(t))^{-1}$, we can
reformulate this equation as
$$
k (\beta + H(t))^{-1} + \frac{1}{2} \frac{d}{dt}(\beta + H(t))^{-1} \geq 0 
$$
on a dense set and this extends to all of $X$ by boundedness. Hence the
assumptions of
Theorem~\ref{th:sohr77} are satisfied.
\end{proof}
Unfortunately, we are not able to localize the interaction also in time. We
will see in Section~\ref{ph4background} that we can already get scattering
operators in this situation, but this is not sufficient for a proof of the
existence of local scattering operators. The problem is due to the condition
\eqref{eq:gcond}, which was necessary to ensure the range condition
\eqref{rangecond} in the context of Theorem~\ref{th:sohr77}. This motivates us
to go further on to avoid the range condition.

\section{Wellposedness and approximative solutions}
\label{existencequasiapprox}

Up to this point, we have considered a notion of solvability which leads to
unique,
continuous
solutions of the nonautonomous Cauchy problem. If quantum field theory comes
into play, one would not expect to have complete control over the time 
evolution for every moment, due to particle creation and annihilation.
Only scattering situations, where the time evolution asymptotically  equals the
free time evolution, are considered as being meaningful.   Hence we can 
abandon some regularity properties of the solutions and  we
attenuate
the conditions on a solution of
the nonautonomous Cauchy problem \eqref{ncp} further. In the language of
Section~\ref{wellp} this means that $G_0$ is allowed to have several extensions
which are generators of
evolution semigroups. We choose the one which can be approximated in a certain
way. For Kato-stable generators on a separable Banach space we obtain operator
families $U(t,s)$ fulfilling the properties (i) and (iii) of
Definition~\ref{def:propagator}. In the special case of the time-dependent
Schr\"odinger equation in a separable 
Hilbert space we obtain also weak continuity and strong continuity up to a null
set, hence approximative
solutions in the sense of
Definition~\ref{quasi-approx-def}.

\begin{theorem}
\label{extension}
Let $I= (a,b], -\infty < a<b <\infty$, be a bounded interval.
Assume that there is a family of Kato-stable generators with constants
$(M,\omega)$
on a separable, reflexive Banach space $X$ such that 
\begin{enumerate}[(i)]
\item $ t \mapsto R(\lambda, A(t))$ and $t \mapsto R(\lambda,A(t)^{*})$ are
strongly
continuous for some $\lambda \geq 0$; 
\item there are Banach spaces
$Y$ and $Y'$ which are densely embedded in $X$ resp. $X^{*}$ such that $Y
\hookrightarrow D(A(t))$ and $Y' \hookrightarrow D(A(t)^{*})$, where the
embeddings are uniformly bounded in $t$;
\item the mappings $t \mapsto A(t)y$ and
$t \mapsto A(t)^{*}y^{*}$ are strongly continuous on $I$ for all $y \in Y$ and
$y^*
\in
Y'$.
\end{enumerate}
Then there exists an operator family $U(t,s), (t,s) \in D_{I}$ such that
$$
\|U(t,s)\| \leq M e^{\omega(t-s)},\quad U(s,s)=\one. 
$$
Moreover,
\begin{equation}
\label{cfac}
U(t,r) U(r,s) = U(t,s)
\end{equation}
for $(t,s) \in D_I$ and $r \in [t,s] \setminus N$ for a set $N = N(I)$ of
measure $0$.
The mapping $D_{I}
\ni (t,s) \mapsto U(t,s)$ is weakly continuous and strongly measurable, $s
\mapsto
U(t,s)$ is strongly continuous uniformly in $t$ and $U(t,s) \to \one$ as $(t,s)
\to (r,r)$ in $I$. 

Furthermore, we find   
\begin{align}
\frac{\partial}{\partial s} \, U(t,s)y&= -U(t,s)A(s)y, \label{qapproxdiff1}\\
 \frac{\partial^+}{\partial t} \, U(t,s)y|_{t=s}&= A(t)y\label{qapproxdiff2}
\end{align}
for $(t,s) \in D_{I}$ and $y \in Y$. The operators $U(t,s)$ are weak limits of
the propagators $U_{n_{l}}(t,s)$ which are generated by an admissible bounded
approximation $A_{n_l}(t)$ of $A(t)$.    

The operator $G_{0} = -\frac{d}{dt} + A(\cdot)$ defined on $\tilde{F} =
W_{0}^{1}(I,X) \cap L^{2}(I, Y)$ has an extension $G$ which generates a strongly
continuous semigroup $(T(\sigma))_{\sigma \geq 0}$ defined by
\begin{equation}
(T(\sigma)f)(t)=\begin{cases}
           U(t,t-\sigma)f(t-\sigma)&  t,t-\sigma \in I,\\
           0 & t\in I, t-\sigma \notin I,
         \end{cases}
\end{equation}
in analogy to an evolution semigroup.
\end{theorem}
\begin{proof}
Choose an admissible bounded approximation $A_{n}(t)$ of $A(t)$ such that, at
the
same time, $ -A_{n}(t)^{*}$ is an admissible bounded approximation of
$-A(t)^{*}$
on $X^{*}$, for example as in Lemma~\ref{adap}. We organize the proof in
several steps. Starting from weak limit points, we use evolution semigroup
techniques to establish \eqref{cfac} and the other assertions.
\subparagraph{Step 1.}
For $i,j,k \in \NN$ we start with a numbering of the pairs of rational times
$(t_{i}, s_{j}) \in D_{I} \cap \mathbb{Q}$ and a dense sequence $\{x_{k}\}$ in
$X$. 
Let $q \in \NN$ be a numbering of $\NN^{3}$. We fix $(i,j,k)$ and consider the
sequence $\{z_{n}(q)\}_{n \in \NN} = \{U_{n}(t_{i},s_{j}) x_{k}\}_{n \in \NN}$.
This sequence is bounded, hence we can choose a weakly convergent subsequence
$\{z_{\nu_{m}(q)}(q)\}_{m \in \NN}$. Now we define the diagonal sequence
$\{n_{l}\}_{l\in\NN} = \{\nu_{l}(l)\}_{l\in \NN}$ and conclude that
$U_{n_{l}}(t_{i}, s_{j}) x_{k}$ converges weakly for $l \to \infty$ and every
$(i,j,k)$. We denote the limit by $U(t_{i},s_{j}) x_{k}$. By boundedness, we
extend $U(t_{i}, s_{j})$ to all $x \in X$ and conclude $\wlim_{l \to
\infty}U_{n_{l}}(t_{i}, s_{j})x = U(t_{i},s_{j})x$ as well as $\|U(t_{i},
s_{j})\| \leq M e^{\omega(t_{i}-s_{j})}$ for all $x \in X, i,j \in \NN$.
\subparagraph{Step 2.}
For the propagators associated with the admissible bounded approximation
$A_{n}(t)$ we have
\begin{equation}\label{rationalestimate}
U_{n_{l}}(t,s) y - U_{n_{l}}(t,r) y = - \int_{r}^{s}
U_{n_{l}}(t,\tau)A_{n_{l}}(\tau)y \; d\tau.
\end{equation}
It follows that
\begin{equation}
\|U_{n_{l}}(t,s) y - U_{n_{l}}(t,r) y\| \leq c |s - r| \|y\|_{Y}
\end{equation}
with a constant $c$ which can be chosen uniformly in $t$ and independent of
$l,r,s$ and $y$.  To obtain an estimate involving the weak limit of
$U_{n_l}(t,s)$, we take $t_i, s_j, s_k \in \mathbb{Q}$, $s_j \leq t_i$ and
$s_k\leq t_i$ and find 
$$
\|U(t_i, s_j)y - U(t_i, s_k)y\| \leq \limsup_{l \to \infty} \|U_{n_l}(t_i, s_j)y
- U_{n_l}(t_i, s_k)y\| \leq c |s_j - s_k| \|y\|_Y.
$$
For fixed $t_i$ this estimate enables us to extend $s \mapsto U(t_i, s)y$ from
$s \in (a,t_i] \cap \mathbb{Q}$ to a continuous function on $(a,t_i]$, which is
bounded by $M e^{\omega(t_i - s)} \|y\|$. Hence we can extend $U(t_i,s)$ by
linearity to an operator on $X$, bounded by $M
e^{\omega(t_i-s)}$. Next we show the strong continuity of $s
\mapsto U(t_i,s)y$ on $(a,t_i]$, uniformly in $t_i$. To this end, we estimate
\begin{multline*}
(y, U_{n_l}(t_i,s)x - U(t_i,s)x) \leq (y,U_{n_l}(t_i, s_j) x - U(t_i,
s_j) x) \\
 + (\|U_{n_l}(t_i,s) x - U_{n_l}(t_i,s_j) x
\| + \|U(t_i,s) x - U(t_i, s_j)x \| ) \|y\|
\end{multline*}
for rational $s_j$ such that $|s - s_j| < \epsilon$. Thus we see that
$U_{n_l}(t_i,s)
x \to U(t_i,s) x$ weakly in the limit $l \to \infty$. Since, by assumption,
$A_n(\tau)y \to A(\tau)y$ for $y \in Y$ and $n \to \infty$, we find
$U_{n_l}(t_i, \tau) A_{n_l}(\tau)y \to U(t_i, \tau)A(\tau) y$ weakly for $l \to
\infty$ and $\tau \in (a,t_i]$. Hence \eqref{rationalestimate} leads to
\begin{equation}
U(t_i,s) y - U(t_i,r) y = - \int_{r}^{s}
U(t_i,\tau)A(\tau)y \; d\tau
\end{equation}
for $y \in Y$ and $s,r \in (a,t_i]$. This implies
\begin{equation}
\|U(t_i,s) y - U(t_i,r) y\| \leq c |s - r| \|y\|_{Y}.
\end{equation}

\subparagraph{Step 3.}
Up to this point, we have considered rational $t$. To define $U(t,s)$ for $t
\notin
\mathbb{Q}$, we use a solution of the backward Cauchy problem associated with
the
admissible bounded approximation of $A(t)^*$ on $X^*$:
$$
\dot{v}(s) = - A_n(s)^* v(s), \quad s \in (a,t], \; v(t) = x^*.
$$
The solution of this Cauchy problem is given by $v(s) = U_n(t,s)^* x^*  =:
V_n(t,s) x^*$ with $(t,s) \in D_I$. The equality
$$
U_{n_l}(t,s)^* y^* - U_{n_l}(r,s)^* y^*  = \int_r^t U_{n_l}(\tau,s)^*
A_{n_l}(\tau)^* y^* \; d\tau
$$   
for $y^* \in Y'$
together with the observation that
$U_{n_l}(t_i, s)^* \to
U(t_i,s)^*$ weakly allows us to apply the argument of Step 2 with $t$ and
$s$ interchanged. Using the assumptions on $A(t)^*$, we extend the propagator
$U(t,s)^*$ to an exponentially bounded propagator $V(t,s)$ on $X^*$ defined on
$D_I$. 
Furthermore, the mapping $t \mapsto V(t,s)$ is strongly continuous and
$V_{n_l}(t,s) \to V(t,s)$ weakly for $l \to \infty$ and $(t,s) \in D_I$. 

This enables us to define $U(t,s) := V(t,s)^*$ for $(t,s) \in D_I$ with $t \in
\RR \setminus \mathbb{Q}$. So we have $U(t,s) = V(t,s)^*$, $\|U(t,s)\|
\leq M e^{\omega(t-s)}$, $t \mapsto U(t,s)$ is weakly continuous, $U_{n_l}(t,s)
\to U(t,s)$ weakly as $l \to \infty$ on $D_I$. For coinciding times $U(t,t) =
\one$ for $t \in I$. Our last norm estimate extends to real $t$,
\begin{equation}
\|U(t,s) y - U(t,r) y\| \leq c |s - r| \|y\|_{Y},
\end{equation}
for $(t,s), (t,r) \in D_I$ and $y \in Y$. The constant $c$ remains independent
of $t$. Using the density of $Y$ in $X$ and again an $\epsilon/3$ argument, we
show that $s \mapsto U(t,s)$ is strongly continuous uniformly in $t$. From our
previous considerations for rational $t_i$ we deduce
 \begin{equation}\label{eq:integral}
U(t,s) y - U(t,r) y = - \int_{r}^{s}
U(t,\tau)A(\tau)y \; d\tau
\end{equation}
for $y \in Y$ and $a < s,r \leq t \leq b$. This formula enables us to conclude
that $U(t,s) \to \one$ strongly for $(t,s) \to (r,r)$ and
\eqref{qapproxdiff1},
\eqref{qapproxdiff2} hold. The latter fact follows analogous to the proof of
Theorem~\ref{diff}.

\subparagraph{Step 4.}
As we have seen in Step 1 to Step 3, it is possible to deduce some properties
of $U(t,s)$ from its weak approximants $U_{n_l}(t,s)$. But one crucial property
of the propagators does not survive the weak limit: The causal factorization 
$U_{n_l}(t,s) = U_{n_l}(t,r)U_{n_l}(r,s),\ t\geq r \geq s$, does not imply a
similar property for $U(t,s)$, as the
multiplication of operators is not continuous in the weak topology. 

However, we are able to prove the causal factorization using an extension
of $G_0$ which generates a strongly continuous semigroup
$(T(\sigma))_{\sigma \geq 0}$. This semigroup has a similar relation to
$U(t,s)$ as an evolution semigroup has to a strongly continuous propagator: The
difference to the definition of an evolution semigroup is
that $U(t,s)$ lacks not only the causal factorization, but also strong
continuity in $t$.

Let $\lambda \in \CC$ and $f \in E = L^2(I,X)$. For $t \in I$ we define the
operator $R_\lambda$ by
\begin{equation}
(R_\lambda f)(t) = \int_a^t e^{-\lambda(t-s)} U(t,s) f(s)\; ds.
\end{equation}
Fubini's theorem assures strong measurability of $R_\lambda f$, so $R_\lambda:
E\to E$ and it is bounded.

By assumption, the operators $A(t)$ are Kato-stable. By
Lemma~\ref{lem:katostable}, there is a monotone family of norms $\|\cdot \|_t
$ such that
\begin{equation} \label{t-norm}
\| x \| \leq \| x \|_t \leq \| x \|_s \leq M \|x\| \; \text{and} \; 
  \|R(\la,A(t))x\|_t\leq  (\la-w)^{-1}\|x\|_t
\end{equation}
for $(t,s) \in D_I, \lambda > \omega$ and $x \in X$. 

We now introduce an equivalent norm on $E$ which is constructed in such a way
that one has dissipativity of $A(\cdot)$ and $-\frac{d}{dt}$ and hence of
$G_0$ with respect to this norm.
Let $f$ be a simple function, then $\alpha(t)
= \|f(t)\|_t$ is measurable as a sum of decreasing functions with disjoint
supports. By approximation, $\alpha$ is measurable for each $f \in E$ and we
are able to introduce the norm
$$
|\| f\|| = \left(\int_I \|f(t)\|^2_t \;dt \right) 
$$
which fulfills $\|f\| \leq |\| f \|| \leq M \|f\|$. As in the proof of Kato's
Theorem in \cite[Theorem~5.3.1]{pazy}, we have
$$
U_n(t,s)x = \lim_{k\to\infty} e^{d_k A_n(t_k)}  e^{d_k A_n(t_{k-1})}
\cdot \ldots \cdot  e^{d_k A_n(t_0)}x,
$$
where $d_k=(t-s)/k$ and $t_j= s+ j(t-s)/k$ for $j=0,1, \cdots,k$.
With \eqref{t-norm} we estimate
$$
\|e^{d_l A_n(t_l)}\|_{t_l} \leq e^{-n d_l} \|\exp(d_l n^2
R(n,A_n(t_l)))\|_{t_l} \leq e^{d_l n \omega (n-\omega)^{-1}}.
$$
This implies that
$\|U_n(t,s)x\|_t \le e^{w_n(t-s)}\|x\|_s$ with $w_n = n \omega(n-\omega)^{-1}
\to w$ as $n\to\infty$,
and thus $\|U(t,s)x\|_t \le e^{w(t-s)}\|x\|_s$ for $(t,s)\in D_I$. As a result
we obtain
\begin{equation}\label{rlambda-est}
|\|R_\la f \|| \le (\la-w)^{-1}\,|\|f\||, \qquad \la>w, \;f\in E,
\end{equation}
by use of Young's inequality.

\subparagraph{Step 5.}
On $E$ and $E^*=L^2(I,X^*)$ we define
\begin{equation}
\label{eq:g0}
G_0 = -\frac{d}{dt} +A(\cdot) \quad\text{and}\quad G_0'=\frac{d}{dt} +A(\cdot)^*
\end{equation}
with $D(G_0) = \tilde{F}=W^1_0(I,X)\cap L^2(I,Y)$
and $D(G_0')=\tilde{F}':=\{f\in W^1((a,b),X^*)\cap L^2(I,Y') : f(b)=0\}$,
respectively.
Since
 $\tilde{F}'$ is dense in $E^*$ by our assumptions, we can further  set
\begin{equation}
\label{eq:gtilde}
G_1= (G_0')^\ast.
\end{equation}
Clearly, $G_0\subset G_1$. By Step 3 of this proof,  equation 
\eqref{qapproxdiff1}
holds
which implies 
\begin{equation}
\label{leftinv}
R_\lambda(\lambda - G_0) u = u \qquad \text{for}\; u \in \tilde{F}.
\end{equation}
Define $G_n = -\frac{d}{dt} + A_n(\cdot)$ on $D(G_n) = W^1_0(I,X)$. Its
resolvent $R(\lambda, G_n),\ \lambda \in \CC$, can be described as the Laplace
transform of the corresponding evolution semigroup, leading to a formula
analogous to \eqref{resolv}. Using the Theorem of Dominated Convergence we
verify that $R(\lambda, G_{n_l}) \to R_\lambda$ weakly as $l \to \infty$. 
Moreover, $A_n(t)^*y^*\to A(t)^*y^*$ for $y^*\in D(A(t)^*)$
and $t\in I$ as  $n\to\infty$. For $v \in \tilde{F}'$ and $f \in E$ we thus
obtain
\begin{align*}
(G_{n_l}R(\lambda, G_{n_l})f, v) &= (R(\lambda, G_{n_l})f, G_{n_l}^*v)
     \;\longrightarrow\; (R_\lambda f, G_0'v) \qquad \text{as \ } l\to \infty,\\
|(R_\lambda f, G_0'v)| &\le c\,\|f\|_E\, \|v\|_{E^*}.
\end{align*}
These facts imply that $\Ran R_\lambda \subset D(G_1)$ and
\begin{equation}
\label{rightinv}
(\lambda - G_1) R_\lambda f = f \qquad \text{for}\; f \in E.
\end{equation}
From these equations we can easily check that the kernel of $R_\lambda$ is
trivial, hence $R_\la$ is injective. We define $D_\lambda = \Ran R_\lambda$ and
 the operators
 $B_\lambda = R^{-1}_\lambda$ with domain $D(B_\lambda) = D_\lambda$.
The operators $B_\lambda$ are closed as the inverses of bounded operators. 
Then we see that $(\lambda -G_0) \subset B_\lambda
\subset (\lambda - G_1)$ because of \eqref{leftinv} and \eqref{rightinv}.

\subparagraph{Step 6.}
Our strategy  is to define $G$ by inverting $B_\lambda$ for $\lambda =0$. 
This is possible if $D_\lambda$ does not depend on $\lambda$. Aiming at a proof
of this assertion, we show that multiplication with $\alpha \in C^1([a,b])$ maps
$D_\lambda$ into itself.

The {\em extrapolation space} $E_{\lambda, -1}$ is the closure of $E$
with respect to the norm
$\|f\|_{\lambda,-1} = \|R_\lambda f\|$. The operator $B_\lambda:D_\la\to E$
can be
extended to an isometric operator $B_{\lambda,-1}: E \to E_{\lambda,-1}$ which
is
isomorphic to $B_\lambda$. The crucial fact about extrapolation spaces which
will be used in the following is that $u \in D_\lambda$ if and only if
$B_{\lambda,-1}u \in E$.  Details can be found in \cite[\S II.5]{en}.

The density of $\tilde{F}$ in $E$ implies the density of $B_\lambda \tilde{F}$
in
$E_{\lambda,-1}$.
For $u\in \tilde{F}$, $f = B_\lambda u$, and $\alpha \in C^1([a,b])$, we find
the following equation
$$
B_\lambda (\alpha u) = (\la-G_0)(\alpha u) = \dot{\alpha} u + \alpha B_\lambda
u,
$$
which enables us to estimate
$$ 
\|\alpha f\|_{\la,-1} = \|\alpha u - R_\la (\dot{\alpha} u)\| \le c\, \|u\| =
  c \,\|R_\la f\| =c\, \|f\|_{\la,-1}.
$$
We see that multiplication
with $\alpha$, regarded as an operator on $B_\lambda
\tilde{F}$, is bounded. So the operator
$f\mapsto \alpha f$ can be extended to a bounded
operator $M_\alpha: E_{\lambda, -1}\to E_{\lambda, -1}$. By approximation, we
deduce
$$
B_{\lambda,-1} (\alpha u) = \dot{\alpha} u + M_\alpha\, B_{\lambda,-1} u
$$
for all $u \in E$. In particular, if $u \in D_\lambda$, then
the right-hand side belongs to $E$ and hence $\alpha u \in D_\lambda$. This
completes the intermediate step.

\subparagraph{Step 7.} 
Set $e_\lambda(t) = e^{-\lambda t}$. For given $f\in E$
and
$\lambda \neq \mu \in \CC$ we have
$$
R_\lambda f = e_{\lambda - \mu} R_\mu(e_{\mu - \lambda}f).
$$
Due to this equation and Step 6,
 every $g=R_\lambda f \in D_\lambda$ is also an element of $D_\mu$ and vice
versa. Therefore $D_\la=:D$ is independent of $\la\in\CC$.
We define $G = -B_{\lambda = 0}$ with $D(G)=D$. Hence
$G_0\subset G\subset G_1$ so that $\la-G=B_\la$ and $R_\lambda = (\lambda -
G)^{-1}$
by  \eqref{rightinv}. With estimate \eqref{rlambda-est} 
we deduce $|\|(\lambda - G)^{-1}f|\| =(\lambda-w)^{-1} |\|f\|$ for $\lambda >w$.
Application of the Theorem of
Hille and Yosida (Theorem~\ref{th:hy}) shows that $G$ generates a semigroup
$S(\cdot)$
on $(E,|\|\cdot\||)$ and hence on $E$, as the norms are equivalent.  

On the
other hand, we define
\begin{equation}
\label{eq:howlandt}
(T(\tau)f)\,(t)=\begin{cases}
           U(t,t-\tau)f(t-\tau)&  t, t-\tau\in I,\\
           0 & t\in I,t-\tau \notin I,
         \end{cases}
\end{equation}
for $f\in E$ and $\tau\ge0$.
The continuity properties of $U(t,s)$ enable us  to check that $\tau\mapsto
T(\tau)f\in E$
is continuous and  that
the Laplace transform of $T(\cdot)f$ is equal to  $R_\lambda f$.
But the Laplace transformation is unique, so
 $T(\tau)=S(\tau)$ for all $\tau\ge 0$. Hence $T(\cdot)$ is a semigroup with
generator $G$.

The semigroup property of $(T(\sigma))_{\sigma \geq 0}$ has its counterpart on
the
level of the weak limits $U(t,s)$ of the sequence of propagators
$U_{n_{l}}(t,s)$---it corresponds to the causal factorization which we want to
prove. However, $ (T(\sigma))_{\sigma \geq 0}$ acts on an $L^{2}$ space and we
obtain the causal factorization only up to a set of measure zero.

Notice that  $(T(\sigma))_{\sigma \geq 0}$, as defined in
\eqref{eq:howlandt}, is already an evolution semigroup in the sense of Howland's
definition by construction,
but the additional condition about the continuous embedding of $D(G)$ in
$C_0(I,X)$, necessary in Theorem~\ref{char0}, is not clear. However, we are able
to derive more properties of $T(\cdot)$ and the propagators $U(t,s)$ associated
with it. Thereby we establish regularity of the solutions of the nonautonomous 
Cauchy problem \eqref{ncp}, which is not accessible in Howland's original approach.

Let $\{x_k\}$ be a dense sequence in $X$  and set $t \mapsto f_k(t) := x_k\in E$.
The weak limit $U(t,s)$ is weakly continuous in $(t,s)$, hence weakly measurable
and strongly measurable by Pettis' Theorem.
For $0\le \tau,\si\le n$ the semigroup
property of $T(\cdot)$ states $T(\tau)T(\si)f_k(t)=T(\tau+\si)f_k(t)$ in
$L^2([0,n]^2\times I, X)$. We evaluate this as an equation in $X$. Thus for the
propagator 
\begin{equation}\label{eval1}
U(t,t-\tau)U(t-\tau,t-\tau-\si)x_k= U(t,t-\tau-\si)x_k
\end{equation}
for all $k\in\NN$ and all $(t,\tau,\si)$ belonging to a set
$\Omega\subset \{(t,\tau,\si) \in I \times \RR_+ \times \RR_+ : t-\tau-\si
>a\}=:\Delta$
so that $N_k'':=\Delta\setminus \Omega$ has measure 0. This is the set, where
\eqref{eval1} fails to hold.
Equation \eqref{eval1} can be extended for fixed $(t,\tau,\si)\in\Omega$ to
all $x\in X$ by approximation. We obtain an exceptional set $\bigcup_{k \in
\NN} N_k'' = N''$ of measure $0$ as a countable union of null sets.  We
transform the set of variables in a linear way by setting $r=t-\tau$.
As the image of a null set under a Lipschitz continuous function is
again a null set \cite[Theorem~IX.5.9]{ammannescher}, the image of $N''$ under
this transformation is again a
set of measure $0$. In the following, we need another fact about null sets:  Let
$M \subset \RR^{n}$ be a set of Lebesgue measure zero. Then there exists a
set $M' \subset \RR^{n-1}$ such that for all $x \in \RR^{n-1} \setminus M'$ the sets
$\{y \in \RR : (x,y) \in M\}$
have measure zero. For the proof, see \cite[Lemma~VI.8.3]{lang}. Applied to
$N''$, we conclude that there is a
set $N'\subset D_I$ of measure $0$ such that
\begin{equation}\label{eval2}
U(t,r)U(r,r-\si)x= U(t,r-\si)x
\end{equation}
for all $x\in X$, $(t,r)\in D_I\setminus N'$, and a.e.\ $\si\in[0, r-a]$.
Varying $\si$
for fixed $(t,r)$,  we then obtain
\begin{equation}\label{eval3}
U(t,r)U(r,s)=U(t,s)
\end{equation}
for all $a \leq s \leq r$. As above, we  conclude
that there is a set $N\subset I$
of measure $0$ such that \eqref{eval3} holds for all $r\notin N$
and for a.e.\ $t\ge r$. Using the weak continuity of $t\mapsto U(t,s)$,
we extend \eqref{eval3} to
all $(t,s)\in D_I$ and $r\in[s,t]\setminus N$.

 Finally, if  $(t',s')\in D_I$ converges to $(t,s)$ with $t>s$,
 we fix $r\notin N$ between
  $\min\{s,s'\}$ and $\max\{t,t'\}$. The continuity results from
  Step 3 then show that the difference
$$U(t',s')-U(t,s)=U(t',r)(U(r,s')-U(r,s))+  (U(t',r)-U(t,r))U(r,s)$$
tends weakly to 0.
\end{proof}
Notice that in all arguments involving the admissible bounded approximations it
would be possible to use the alternative definition of an admissible bounded
approximation \emph{on $Y$}, see the remark after
Definition~\ref{admissible-def}.

For the special case of $X$ being a Hilbert space with skew-adjoint $A(t)$, we
obtain a better result. We can show that in this situation  $U(t,s)$ is
indeed independent of the choice of the approximating sequence $U_{n_l}(t,s)$.
Moreover, we derive continuity properties and unitarity of $U(t,s)$ up to a
set of measure $0$.
 
We do need the assumption of a bounded time interval no longer
and thus assume $I=\RR$ in the following. This is no loss of generality:
Given operators $H(t),\ t \in [a,b]$, fulfilling the assumptions of the next
theorem, we can extend this operator family by $H(t) = H(a)$ if $t < a$ and
$H(t) = H(b)$ if $t>b$.

 \begin{theorem}
\label{th:se}
Let $H(t)$, $t\in\RR$, be self-adjoint operators on a separable Hilbert space
$X$ and let $Y$ be a Banach space densely embedded in $X$. Assume that
\begin{enumerate}[(i)] 
\item $Y \hookrightarrow D(H(t))$ locally uniformly in $t$,
\item $t\mapsto (i+H(t))^{-1}$ is strongly
continuous,
\item $t\mapsto H(t)y$ is continuous for $y\in Y$.
\end{enumerate}
Then there are contractive 
propagators $U(t,s)$ with $U(t,s)^* = U(s,t)$ , $(t,s)\in\RR^2$, on $X$
which fulfill the causal factorization property,
\begin{equation}
\label{cfac2}
U(t,r) U(r,s) = U(t,s),
\end{equation}
for all $t,s \in \RR$ and $r \in \RR \setminus N$ for a set $N$ of
measure $0$. Moreover,
\begin{enumerate}
\item these operators are surjective for all $t \in \RR$ and $s \in
\RR\setminus
N$, isometric for all $t \in \RR \setminus N$ and $s \in\RR$, hence unitary with
$U(t,s) = U(s,t)^{-1}$ for all $t,s \in \RR \setminus N$;
\item $(t,s) \mapsto U(t,s)$ is weakly continuous on $\RR^2$;
\item $(t,s) \mapsto U(t,s)$ is strongly continuous at $(r,r)$ and on $(\RR
\setminus N) \times \RR$;
\item these operators have the differentiability properties
\begin{align}
\frac{\partial}{\partial s} \, U(t,s)y&= -U(t,s)A(s)y, \label{qapproxdiff11}\\
 \frac{\partial^+}{\partial t} \, U(t,s)y|_{t=s}&= A(t)y\label{qapproxdiff22}
\end{align}
for $t,s \in \RR$ and $y \in Y$.
\end{enumerate}
The generator of the strongly continuous semigroup $(T(\sigma))_{\sigma \geq 0}$, given by
\begin{equation}
(T(\sigma)f)(t)=
           U(t,t-\sigma)f(t-\sigma)
\end{equation}
on $E:=L^2(\RR, X)$,
is an extension of
$G_0=-d/dt+A(\cdot)$
defined on $\tilde{F}:=W^1_0(\RR,X)\cap L^2(\RR,Y)$. Moreover,
for all self-adjoint admissible bounded approximations
$H_n(t)$ of $H(t)$ with generated evolution family $U_n(t,s)$
one has $U_n(t,s)\to U(t,s)$ weakly for $(t,s)\in\RR^2$ and strongly for
$(t,s) \in (\RR \setminus N) \times \RR$ as $n\to\infty$.

The function $u=U(\cdot,s)x$
is  the unique approximative solution of the nonautonomous Cauchy problem \eqref{ncp}
with $s\in\RR$ and $x\in X$.
\end{theorem}
\begin{proof}
Set $A(t) = -iH(t)$. Again we organize the proof in several steps. 
\subparagraph{Step 1.} 
Let $A_n(t)$, $t\in\RR$,  $n\in\NN$, be  skew-adjoint 
admissible
bounded approximations of $A(t)$, e.g. as obtained in Lemma~\ref{adap}(ii).
Let $U_n(t,s)$, $(t,s)\in \RR^2$, be the unitary evolution family generated
by
$A_n(\cdot)$. Step 1--Step 3 of the proof of
Theorem~\ref{extension}
work also for our present operators
$A_n(t)$ and $U_n(t,s)$ with $(t,s)\in \RR^2$, $M=1$ and $w=0$.
The constants $c$ are uniform on compact time
intervals. Thus we obtain a subsequence $n_l$ and contractive
operators $U(t,s)$, $(t,s)\in\RR^2$, such that
$U_{n_l}(t,s) \to U(t,s)$ weakly as $l\to\infty$ for  $(t,s)\in\RR^2$.
This fact implies that $U(t,s)^*=U(s,t)$ for $t,s\in \RR$.
We further have the differentiability properties \eqref{qapproxdiff11} and
\eqref{qapproxdiff22} for $t,s\in\RR$
 and $y\in Y$. By Step 4--Step 7 of the proof of
Theorem~\ref{extension}, $U(t,s)$ satisfies the continuity properties stated  in
this theorem for all $-n<s\le t\le n$ and $n \in \NN$,  hence for all $(t,s)\in
D_\RR$. In particular, the causal factorization $\eqref{cfac}$ holds for all
$t,s
\in \RR$ and $r \in \RR \setminus N$ for a set $N$ of measure $0$. 

Now we want to examine the case $t\le s$. For this reason,  we introduce the
skew-adjoint
 operators $B(t)=-A(-t)$ for $t\in\RR$
and take the skew-adjoint admissible bounded approximation
 $B_n(t)=-A_n(-t)$ of $B(t)$. Consequently, $B_n(\cdot)$
generates the evolution family $V_n(\tau,\si)=U_n(-\tau,-\si)$, $(\tau,\si)\in
D_\RR$.
Repeating our argument, we see that the operators $U(t,s)$, $-\infty<
t\le s< \infty$,
also satisfy the causal factorization 
\eqref{cfac} up to an exceptional null set $N$ and the continuity properties
stated in Theorem~\ref{extension}. 
These facts imply that
$\RR^2\ni (t,s)\mapsto U(t,s)$ is weakly continuous and strongly measurable, and
it is strongly continuous at $(r,r), r \in \RR$.

\subparagraph{Step 2.} 

To establish the invertibility properties of $U(t,s)$,  
we define
$T(\t)f(s)=U(s,s-\t)f(s-\t)$ for $\t\ge 0$, a.e.\ $s\in \RR$ and $f\in E$.
Due to the properties obtained in Step~1, $T(\cdot)$ is a contractive,
strongly continuous semigroup.
Its generator is denoted by $G$. By reflexivity, its adjoint $G^*$ is also the
generator of a strongly continuous semigroup which coincides with
$(T(\sigma)^*)_{\sigma \geq 0}$, see \cite[\S 1.10]{pazy}. Moreover, by 
$S(\t)g=U(\cdot,\cdot-\t)g(\cdot-\t)$,
$\t\le 0$,
we define another contractive strongly continuous semigroup on $L^2(\RR,X)$
with time interval
$\RR_-$
generated by an operator $\hat{G}$. This is seen in a similar way as in 
the case of $T(\sigma)$. Note that we do not yet know that
$S(-\t)=T(\t)^{-1}$, $\tau\ge0$. 

We calculate
$$
(T(-\t)^*g)(s)= U(s-\t,s)^*g(s-\t)= U(s,s-\t)g(s-\t)=(S(\t)g)(s)
$$
for $\t\le 0$ and a.e.\ $s\in \RR$. This means that $-G^*=\hat{G}$. Hence
$G^*$ generates a contractive $C_0$--group,
see \cite[\S II.3.11]{en} or \cite[\S 1.6]{pazy}. Therefore, this
group is isometric and thus unitary. As a consequence, $G$ is skew-adjoint
and
$$
T(-\t)=T(\t)^{-1}=T(\t)^*=S(-\t), \qquad \t\ge0.
$$

The map $\RR\ni s\mapsto U(s,s-\t)x$ is weakly continuous, hence
strongly measurable, for all $\tau\in\RR$ and $x\in X$. We take a dense sequence
$\{x_k\}$
in $X$.
For $\ep>0$ set
$f=\chi_{[s-\t,s-\t+\ep]}x_k$. Then we obtain
\begin{equation}\label{isometry2}
\|x\|^2=\frac{1}{\ep}\,\|f\|^2_2 =
\frac{1}{\ep}\int_\RR\|(T(\t)f)(\si)\|^2\,d\si
= \frac{1}{\ep}\int^{s+\ep}_s \|U(\si,\si-\t)x\|^2\,d\si.
\end{equation}
Now, sending  $\ep\to0$, we obtain
$$
\|U(s,s-\t)x_k\|=\|x_k\|
$$
 for $\t\in\QQ$, $k\in \NN$ and
$s\in \RR\setminus N_1'$   with a null set
$N_1'$. Forming the countable union of these null sets, we get a null set $N_1$
such
that this equality  holds for all $\t\in\QQ$, $x\in X$ and $s\in
\RR\setminus
N_1$.
The strong continuity of $r \mapsto U(s,r)$ further implies that
$U(s,r)$ is an isometry for  $s\in\RR\setminus N_1$ and all $r\in\RR$. 
Therefore,
$t\mapsto U(t,s)$ is strongly continuous on $\RR\setminus N_1$ for all $s\in\RR$
and $U_{n_l}(t,s)$ converges strongly to $U(t,s)$ as $l\to \infty$ for
$t\in\RR\setminus N_1$ and  $s\in\RR$.

Take functions
$f_{k,n}\in C_c(\RR,X)$ equal to $x_k$ on $[-n,n]$. Then the equality
$[T(\t)T(-\t)f_{k,n}](\cdot+\t)=f_{k,n}(\cdot+\t)$ for $\t\in\RR$ yields
$$
 x_k= U(s+\t,s)U(s,s+\t)x_k
$$
for all $\t\in\QQ$, $k\in\NN$ and $s\in \RR\setminus N_2$
with a null set $N_2$. Varying $\t$ and using the density of $\{x_k\}$,
we thus obtain
$$x=U(r,s)U(s,r)x$$
for all $x\in X$, $r\in\RR$ and $s\notin N_2$. Hence, $U(r,s)$ is surjective for
$r\in\RR$ and $s\in\RR\setminus N_2$.
Consequently, $U(t,s)$ is unitary
and  $U(t,s)^{-1}=U(s,t)=U(t,s)^*$ for $t,s\notin N$, where we may assume
without loss of generality that we have the same null set as in Step~1. It is
then easy to check that
$U(t,r)U(r,s)=U(t,s)$ for $t,r,s\in\RR\setminus N$. Using the continuity
properties
of $U$, this equation holds for all $t,s\in\RR$  and $r\in \RR\setminus N$.

Let $(t',s')\to (t,s)$ in $\RR^2$ with $t\notin N$. We  fix $r\notin N$ and
write
$$
U(t',s')-U(t,s)= U(t',r)(U(r,s')-U(r,s)) + U(t',s) -U(t,s).
$$
Thus $(t,s)\mapsto U(t,s)$ is strongly continuous on $(\RR\setminus N) \times
\RR.$

\subparagraph{Step 3.} 
We now check the uniqueness of $U(t,s)$. Let
$A_l^{(1)}(t)$ and 
$A_l^{(2)}(t)$
be skew-adjoint admissible bounded approximations of $A(t)$ generating unitary
evolution
families $U^{(i)}_l(t,s)$ which
converge weakly to  $U^{(i)}(t,s)$  ($i=1,2$, $l\in\NN$, $t,s\in\RR$).
These approximations may result from different subsequences in Step~1
of the proof of Theorem~\ref{extension}
 or from different approximations of $A(t)$. The operators
$U^{(i)}(t,s)$ have the properties
established so far, in particular, they are unitary up to a set of measure $0$.
Let $G^{(i)}$ be the generators of the corresponding `evolution semigroups'
on $E(I)=L^2(I,X)$ given as in \eqref{eq:howlandt} for some $I=(a,b]$. We now
define
 $U(t,s)=\frac{1}{2} U^{(1)}(t,s)+ \frac{1}{2}U^{(2)}(t,s)$
for $t,s\in \RR$. Observe that $U(t,s)^*=U(s,t)$ and that $U(\cdot,\cdot)$ satisfies
the continuity properties stated in Theorem~\ref{extension}.

We  show that $U(\cdot,\cdot)$ is a unitary evolution family
almost everywhere,
 proceeding  as above and as
 in Theorem~\ref{extension}. This fact will lead to the
 desired equality $U(t,s)=  U^{(1)}(t,s) = U^{(2)}(t,s)$.
Equation  \eqref{eq:integral} implies that
\begin{align*}
U(t,s)y -U(t,r)y&= \frac{1}{2} \Big(U^{(1)}(t,s)y - U^{(1)}(t,r)y
   + U^{(2)}(t,s)y - U^{(2)}(t,r)y\Big) \\
     &= -\frac{1}{2} \int_r^s U^{(1)}(t,\tau)A(\tau)y\,d\tau
        -\frac{1}{2} \int_r^s U^{(2)}(t,\tau)A(\tau)y\,d\tau\\
     &= -\int_r^s U(t,\tau)A(\tau)y\,d\tau
\end{align*}
for $t\ge r,s$ and $y\in Y$. Thus \eqref{qapproxdiff11} holds for $U(t,s)$ and
$A(t)$.
Now we can deduce  \eqref{leftinv} as in Step~5 of the proof of
Theorem~\ref{extension}, defining
$R_\la$, $G_0$, $G_0'$ and $G_1$ on $E=L^2(I,X)$
as before. 
We define analogously $R_\la^{(i)}$, and we recall that $R_\la^{(i)}
=R(\la, G^{(i)})$ for $\la\in\CC$ and $G^{(i)}\subset G_1$ by the proof of
Theorem~\ref{extension}. We thus obtain
\begin{align*}
(\la R_\la f,v)&=  \frac{1}{2}\,(\la R_\la^{(1)} f,v) +  \frac{1}{2}\,(\la R_\la^{(2)} f,v)\\
                &= (f,v) + \frac{1}{2}\,(G_1 R_\la^{(1)} f,v)
                    +  \frac{1}{2}\,(G_1 R_\la^{(2)} f,v) \\
                &= (f,v) + (R_\la, G_0'v)
\end{align*}
for $f\in L^2(I,X)$ and $v\in \tilde{F}$.
Therefore also \eqref{rightinv} holds. Now we can repeat Step~6 and Step~7
in order to construct  $G=G_I$ on $E(I)$ such that $R_\la=R(\la,G)$
and $G_I$ generates a semigroup $T_I(\tau)$ given as in 
\eqref{eq:howlandt}.
Moreover, $U(t,s)$ satisfies  $U(t,r)U(r,s) = U(t,s)$ and $U(s,s) = \one$ for
$(t,s)\in D_I$, $r\in [s,t]\setminus N(I)$, and some null set $N(I)$.
 This property then holds for $I=\RR$. By duality, it also valid for
$(s,t) \in D_\RR$ and  $r\in [t,s]\setminus N$.

Reasoning as in Step~2, we derive that the `evolution
semigroup'
 $T(\tau)$,  $\tau\ge0$, associated to $U(t,s)$ on $L^2(\RR,X)$  can be embedded
into
a unitary group given by operators
$U(t,s),\ (t,s)\in\RR^2$,
which are unitary for $t,s \in \RR \setminus N$.
But then we have,
\begin{multline*}
 4 \|x\|^2= 4(U(t,s)x, U(t,s)x) \\
= 2\, \|x\|^2 + (U^{(1)}(t,s)x,U^{(2)}(t,s)x)
   + (U^{(2)}(t,s)x,U^{(1)}(t,s)x),
\end{multline*}
for $x\in X$ and $t,s\in\RR \setminus N$. This means that the numerical
range, and hence the
spectrum,
of the unitary operator
$U^{(2)}(t,s)^*U^{(1)}(t,s)=U^{(2)}(t,s)^{-1}U^{(1)}(t,s)$
is contained in the line $\Rea\la=1$,
so that $U^{(2)}(t,s)^{-1}U^{(1)}(t,s)=\one$. As a result, $U(t,s)=U^{(1)}(t,s)=
U^{(2)}(t,s)$ for $t,s \in \RR \setminus N$, and this equality extends to $t,s
\in \RR$ by the continuity properties of $U(t,s)$ and $U^{(i)}(t,s)$. Thus we
have shown that $U(t,s)$ does not depend on the
approximation,
as asserted. Consequently, the sequence $U_n(t,s)$ from Step 1 of this proof
converges strongly to $U(t,s)$  as $n\to\infty$ for all $t,s\in\RR \setminus N$
and weakly for $t,s \in \RR$.

\subparagraph{Step 4.} 
We set $u(t)=U(t,s)x$ for $t\in\RR$ and some  $s\in \RR$
and $x\in X$.
Take  the sequence $U_n(t,s)$ from Step~1  of this proof.
It is clear from the properties of $(t,s) \mapsto U_n(t,s)$ that $u$ is an
approximative solution on $\RR$
of the nonautonomous Cauchy problem \eqref{ncp}. If $ v$ is another
approximative solution corresponding
to some skew-adjoint admissible bounded approximations $A_n(t)$, then the
calculation
\begin{align}
U(t,s)v(s) - v(t) &= \lim_{n \to \infty} U_n(t,s)v_n(s) - v_n(t) \notag \\
&= \lim_{n \to
\infty} \int_s^t U_n(t,\sigma) \left( -\dot{v}_n(\sigma) -
A_n(\sigma)v_n(\sigma)\right)\;d\sigma = 0 \label{qapproxunique}
\end{align}
shows that $u=v$.
\end{proof}
Also in this theorem, the use of admissible bounded approximations on $Y$ according  to the remark after Definition~\ref{admissible-def} would be possible.

We want to point out that there might be other extensions of $G_0$
which generate evolution groups. But the corresponding evolution families give
no approximative solutions. If $G_0$ is essentially skew-adjoint on $F$,
there exists exactly one extension of $G_0$ that is an evolution generator with
the
range condition \eqref{rangecond} fulfilled. In this case the
approximative solution is actually a mild solution. 

For the proof of the independence of the construction of the admissible bounded
approximation we use the fact that only the trivial convex
combination of unitary operators is again unitary. One might have the
impression, that a similar argument works yet on the level of the semigroups.
This would allow for a generalization of the theorem to the context of
dissipative generators. However, the conjecture that only the trivial convex
combination of strongly continuous semigroups is again a strongly continuous
semigroup is wrong. There is a counterexample due to R.~Schnaubelt. Consider the
Banach space $X = \CC^2$ and the strongly continuous semigroups $(S(\tau))_{\tau
\geq 0}$ and $(T(\tau))_{\tau \geq 0}$, given by
$$
S(\tau):= \one \quad\text{and} \quad T(\tau):= 
\begin{pmatrix}
1 & \tau\\
0& 1
\end{pmatrix}
$$
for $\tau \geq 0$. Then $pT(\tau) + (1-p) S(\tau),\ p \in (0,1),$ is again a
strongly continuous semigroup in $\tau$. Hence, our argument does apply only to
the self-adjoint context.

\subsection{Application: Time evolution of the $P(\ph)_2$ model}
\label{sec:timeevop(phi)}

We fix a real, semibounded polynomial of degree $2n$,
\begin{equation*}\label{polynomial}
P(\lambda) = \sum_{j=0}^{2n} a_j \lambda^j,
\end{equation*}
and choose a real test function $g \in C^\infty_c(\RR^2)$ with $0 \leq g(t,x)
\leq
1$.
Define
the interaction Hamiltonian localized in a compact spacetime region by
\begin{equation*}\label{interaction}
V(t) = V(t;g) = \int g(t,x) :P(\ph(x)): \;dx.
\end{equation*} 

The Hamiltonian of the $P(\ph)_2$ model, given by $H(t) = H_0 + V(t;g)$, is
densely defined (according to the perturbation theory predictions mentioned in
Section~\ref{constructive}). It is essentially self-adjoint on $D(H_0) \cap
D(V(t;g))$. For details and references to the literature see 
Appendix~\ref{qftbasics}. For the  $P(\ph)_2$ model, we can prove the
existence of the time evolution by the existence theorem for approximative solutions.

\begin{theorem}\label{p(phi)}
Let $H(t)$, $t \in \RR$, be the Hamiltonian of the massive $P(\ph)_2$ model
with localized interaction, defined as above with $V(t;g)$ for a 
test function $g\in
C^\infty_c(\RR^2)$, $0\leq g \leq 1$ and a polynomial $P(\lambda)$ of order $2n,\ n\in\NN$ with $a_{2n} > 0$.  Then the time-dependent Schr\"odinger
equation \eqref{ncp}
with $A(t) = -iH(t)$ has a unique
approximative solution given by operators $U(t,s)$ with the properties
stated in Theorem~\ref{th:se}.
\end{theorem}
\begin{proof}
Denote by $Y$ the intersection $D(H_0) \cap D(N^n)$ endowed with
the sum of the graph norms of $H_0$ and $N^n$.
This space is a core for $H(t), t\in \RR$, see for example
the proof of \cite[Thm.~3.2.1]{gj}.

Consider the term of highest order in the interaction $V(t)$: 
$$V_{2n}(t) = a_{2n}\int
g(t,x) :\ph(x)^{2n}: \;dx.
$$
For
$y \in Y$ the the interaction term is dominated by a power of 
the number operator, see Theorem~\ref{th:n},
\begin{align*}
\|V_{2n}(t)y\| &\leq \|V_{2n}(t) (N+1)^{-n}\|
\|(N+1)^n y\| \notag \\
&\leq c \|w(t,\cdot)\|_{L^2(\RR^{2n})} \|(N+1)^n y\|,
\end{align*}
where $c$ is a $t$-independent constant and $w(t,k_1,\dots,k_{2n} )$ denotes the
numeric kernel of the expansion of $V_{2n}(t)$ into Wick monomials. It is given
by
\begin{equation*}
\label{eq:w}
w(t,k_1,\dots,k_{2n}) = \hat{g}(t,k_1 + \dots + k_{2n})\omega(k_1)^{-1/2} \cdot
\dots
\cdot \omega(k_{2n})^{-1/2},
\end{equation*}
where $\hat{g}$ denotes the Fourier transform of g with respect to $x$, and the
$L^2$-norm is evaluated with respect to $(k_1, \dots, k_{2n}) \in \RR^{2n}$,
see \cite[\S 6.1]{DG}
By repeated use of Young's inequality, we estimate
$$
\|w(t,\cdot)\|_{L^2(\RR^{2n})}  \leq c \|g(t,\cdot)\|^2_{L^2(\RR)}
$$
(cf.\ \cite[Lem.~6.1]{DG}). The other summands of $V(t)$ can be treated in
the same way.
These inequalities imply the
uniform boundedness of the embedding $Y \hookrightarrow D(H(t))$.
Similarly one checks the
continuity of $t \mapsto H(t)y$ for all $y \in Y$.

Because $Y$ is a core for $H(t)$, the strong continuity of $t \mapsto
-R(-i,H(t))
= (i + H(t))^{-1}$ follows. Now we can apply
Theorem~\ref{th:se} to this problem.
\end{proof}

\subsection{Application: Time evolution in curved spacetime}
\label{cst}

We  generalize the former result about the existence of the time
evolution for the $P(\ph)_2$ model on a two-dimensional Lorentzian
manifold. This extends a result of J.~Dimock \cite{dimock}, who considers
$P(\lambda) = \lambda^4$ and has to restrict himself to interactions with a
localization which is `small' compared to a given background, see
Section~\ref{sec:kis}.

Assume $(M,g)$ is a $d$-dimensional Lorentzian manifold. Consider the covariant
field equation in local coordinates,
$$
(|g|^{-1/2} \partial_\mu g^{\mu \nu} |g|^{1/2} \partial_\nu + m^2 ) \ph + 2n
\lambda \ph^{2n-1} = 0,
$$
where $|g| = |\det \{g^{\mu\nu}\}|$ and $\lambda > 0$. The coordinates are
chosen such that the hypersurfaces $x^0 = t = \text{const}$ are spacelike.
Define the canonical conjugate field $\pi$ to be the normal derivative density
$$
\pi = |g|^{1/2} g^{\nu\mu} \partial_\mu \ph.
$$
With the notation $\alpha = (g^{00})^{-1/2},\ \beta^j =
g^{j0}(g^{00})^{-1},\ \gamma_{ij} = -g_{ij}$ and $\gamma = \det
\{\gamma_{ij}\}$, the field equation
can be written as a Hamiltonian system
\begin{align}
\frac{\partial \ph}{\partial t} &= \alpha \gamma^{-1/2} \pi + \beta^i
\partial_i \ph, \\
\frac{\partial \pi}{\partial t} &= \partial_i \gamma^{ij} \alpha \gamma^{1/2}
\partial_j \ph - \alpha \gamma^{1/2} (m^2 + 2n \lambda \ph^{2n-1}) + \partial_j
\beta^j \pi,
\end{align}
where $1 \leq i,j \leq d-1$. The corresponding Hamiltonian is given by
\begin{equation}
\begin{split}
H(t,\ph,\pi) = \frac{1}{2} \int_{x_0 = t} \left( \alpha \gamma^{-1/2} \pi
^2 + \alpha \gamma^{1/2} (\gamma^{ij} \partial_i \ph \partial_j \ph + m^2 \ph)
\right. \\ 
+ \left. \alpha \gamma^{1/2} \lambda \ph^{2n} + \pi \beta^j \partial_j
\ph \right). 
\end{split}
\end{equation}

Now we specialize to $d=2$. For a two-dimensional manifold one can always choose
local coordinates in such a way that $g_{\mu\nu} = \Lambda \eta_{\mu\nu}$, where
$\eta_{\mu\nu}$ is the Minkowskian metric and $\Lambda = \Lambda(t,x)$ is a
smooth function on $\RR^2$. Hence, $|g|^{1/2} = \Lambda$ and $g^{\mu\nu} =
\Lambda^{-1} \eta^{\mu\nu}$. The field equation becomes
$$
(\Box + \Lambda m^2 ) \ph + 2n \lambda \Lambda \ph^{2n-1} = 0,
$$
the normal derivative is $\pi = \frac{\partial \ph}{\partial t}$, and the
Hamiltonian is 
$$
H(t,\ph,\pi) = \frac{1}{2} \int_{x_0 = t} \left( \pi^2 + 
(\nabla \ph)^2  + m^2 \Lambda \ph^2 + \lambda \Lambda \ph^{2n} \right).
$$
Proceeding in the same way as Dimock in \cite{dimock}, we assume that $\Lambda =
\Lambda_0 +
\Lambda_1 > 0$ with $\Lambda_0>0$ a constant and $\Lambda_1 \in
C^\infty_0(\RR^2)$. Defining $H_0$ and the Wick product with
respect to
$\Lambda_0^{1/2}$, we obtain the formal expression for the Hamiltonian which
contains a non localized term. To define the Hamiltonian as an operator on the
Fock space of the free field, we introduce an ultraviolet cut-off by restricting
the space integration in the corresponding term to a compact interval. Denote by
$\chi_l$ the characteristic function of the interval $[-l,l],\ l \in \NN$. The
Hamiltonian is given by
\begin{equation}\label{dimock} 
\begin{split}
H(t) = H_0 +  \int  \lambda( \Lambda_0\chi_l(x) + \Lambda_1(t,x) ):\ph^{2n}(x):
\,dx \\ 
+ \int \frac{m^2}{2} \Lambda_1(t,x) :\ph^2(x):\,dx. 
\end{split}
\end{equation}
\begin{theorem}
The time-dependent Schr\"odinger equation with the Hamiltonian \eqref{dimock}
has
an 
approximative solution given by an almost everywhere unitary evolution
family
$U_l(t,s)$, depending on the localization $l$.
\end{theorem}
\begin{proof}
We verify the prerequisites of Theorem~\ref{th:se} analogous to the preceding
section.
As before, the Hamiltonian is an essentially self-adjoint operator on $D(H_0)
\cap D(V(t))$, where 
$$
V(t) =    \int \left(  \lambda( \Lambda_0\chi_l(x) +
\Lambda_1(t,x) ):\ph^{2n}(x): 
+  \frac{m^2}{2} \Lambda_1(t,x) :\ph^2(x):  \right)\,dx.
$$  
Again, we set $Y=D(H_0) \cap D(N^n)
\subset D(H(t))\ \forall\ t$.
The continuity of $t \mapsto H(t)y$ for $y \in Y$ is verified as in the proof of
Theorem~\ref{p(phi)}, as well as the other requirements of
Theorem~\ref{th:se}. Thus there exists a
propagator $U_l(t,s)$ associated with an approximative solution of 
the time-dependent Schr\"odinger equation describing the time evolution
associated with $H(t)$. 
\end{proof}
The propagator depends on the ultraviolet cut-off $l$. But we can use
\cite[Theorem~2.3]{dimock} to show that the time evolution induced on local
observables is independent of $l$ if $l$ is chosen large enough. Clearly, all
the critical comments by Dimock concerning this construction apply to our
setting as well, in particular 
covariance with regard to coordinate dependence is unclear. Moreover, the
results of Torre and Varadarajan \cite{evolutioncst} show that it is not
possible to generalize a similar result to curved spacetimes of dimension
larger than two, even for free fields. Nevertheless, it might be interesting to
reexamine the free field setting with regard to the existence result of
Theorem~\ref{th:se}.

\subsection{Application: Goldstein's example}
\label{goldsteindressing}

For a direct application of Theorem~\ref{th:se} we need a sufficiently large
intersection of the domains of the time-dependent generators. As we have seen,
this is the case for $P(\varphi)_2$ models. But for models which require an
infinite renormalization, we do not expect this  to be
fulfilled. As a testing ground for a situation, where the intersection of the
domains of the generators is trivial, we examine again Goldstein's example.

As in Section~\ref{goldstein}, we define the operators $S, T, L$ and $U_1$ as
well as the test functions $\ph, \eta$ and $\psi$ and the Hamiltonian $H(t)$ of
equation \eqref{goldsteinH}.
Moreover,
we choose a smooth
function $\kappa \in C^\infty(\RR)$ such that $\kappa(t) = 0$ for $0 \leq t <
1$ and $\kappa(t) =1$ for $2 \leq t < \infty$. 

Consider the Hamiltonian 
$$
\tilde{H}(t) := (\ph(t) + \psi(t))S + (\eta(t) + \dot{\kappa}(t)) L 
$$  
with $\tilde{H}(t) = \tilde{H}(0)$ for $t \leq 0$.
It is self-adjoint for all $t \in \RR$ and $Y :=D(S) \hookrightarrow
D(\tilde{H}(t))$. Continuity of $t\mapsto \tilde{H}(t)y$ for $y \in Y$ and
strong continuity of $t \mapsto (i + H(t))^{-1}$ follow easily from the
continuity of the test functions and their support properties. Hence, by
Theorem~\ref{th:se}, there is a unique approximative solution
$\tilde{u}(t) = \tilde{U}(t,s)x$ for all $x \in X$ with a unitary propagator
(up to a set of measure zero). Associated with it there is an evolution
semigroup
$(\tilde{T}(\sigma))_{\sigma \geq 0}$ with generator $\tilde{G}$ which is an
extension of $\tilde{G}_0 = -\frac{d}{dt} -i \tilde{H}(\cdot)$ with domain
$W^{1,2}_0(\RR, X) \cap L^2(\RR, Y)$. Now set $R(t) =  e^{i \kappa(t) L}$ and
notice that $R(t) = \one$ for $t \in (0,1)$ and $R(t) = U_1$ for $t \geq 2$. It
corresponds to a unitary operator $R = R(\cdot)$ on $E = L^2(\RR, X)$.  

By Theorem~\ref{similarity}, $T(\sigma) = R \tilde{T}(\sigma)R^{-1}$ is an
evolution
semigroup with generator $G = R\tilde{G} R^{-1}$. As a result 
$(t,s) \mapsto U(t,s) =
R(t)
\tilde{U}(t,s) R(s)^{-1}$ is again a propagator with the same properties as
$\tilde{U}(t,s)$. We transform $\tilde{G}_0$
correspondingly and find formally 
$$G_0 = 
R \tilde{G}_0 R^{-1} = -\frac{d}{dt} -i (\ph(\cdot) S + \eta(\cdot) L +
\psi(\cdot) T) = -\frac{d}{dt} -iH(\cdot).
$$
In this sense $U(t,s)$ is the propagator associated with the nonautonomous Cauchy problem of
Goldstein's example.
For $x \in C^\infty(S)$ $u(\cdot)$ coincides with the solution of
Theorem~\ref{th:goldstein} by uniqueness.

\chapter{Existence of Local Scattering Operators}
\label{exlso}

In Section~\ref{existencequasiapprox} we have developed a new wellposedness
theorem
for the time-dependent Schr\"odinger equation. Furthermore, we have seen that
this
wellposedness result is suitable to show the existence of the time evolution
for $P(\ph)_2$ models with localized interaction. 

The main result of the present chapter is the proof of existence of local
scattering operators for the $P(\ph)_2$ model. However, before we address this 
topic, we indicate to which extent scattering theory is covered by the
approach of Section~\ref{sec:phi4} for $(\ph^4)_2$ with spatially localized
interaction. After this, we show the existence of the local scattering
operators for general $P(\ph)_2$ models of which the $(\ph^4)_2$ interaction is
a special case.
This result enables us to give an easy construction of the algebra of local
observables for massless bosons in two dimensions, a theory without a ground
state. We hereby demonstrate the advantage of the local construction: the
disentanglement of the infrared and the ultraviolet behavior of a theory.

In this chapter, we switch back to the notation of Chapter~\ref{ch:lso} and work in the setting of a Hilbert space $\cH$.

\section{Scattering operator for the spatially localized $(\ph^4)_2$ model}
\label{ph4background}

To establish the existence of local scattering operators, we would like to
exhibit
the existence of a solution of the time-dependent Schr\"odinger equation with an
interaction which is compactly supported in space and time.
In a first step, we investigate the example of Section~\ref{sec:phi4}.
According to Theorem~\ref{th:phi^4}, there exists a unique propagator $U(t,s)$
associated with a mild solution, but the interaction is not localized in a
compact spacetime region.
Therefore, in the  context of Theorem~\ref{th:phi^4} and using the notion of
mild solutions,
 we are not able to define the local scattering operators: in general, the
condition
\eqref{eq:gcond} in Theorem~\ref{th:phi^4} is not
satisfied for compactly supported localization functions. 

We investigate what amount of information about the scattering operator can
be gathered from the approach in Theorem~\ref{th:phi^4}.
In a first step, we can overcome the difficulty with \eqref{eq:gcond} by
convoluting $g$ in
time
with an approximated $\delta$ distribution.
\begin{example}
Given a rapidly decaying approximation of the $\delta$-distribution,
\begin{equation}
\delta_{\alpha}(t) = \frac{1}{\alpha \sqrt{\pi}} e^{- \frac{t^2}{\alpha^{2}}},
\end{equation} 
converging weakly in
$\mathcal{D}'(\mathbb{R})$
to $\delta(t)$ as $\alpha \to 0$.
Let $g \in C^{\infty}_{c}(\mathbb{R}^2)$, and set 
$g_{\alpha}(t,x) = \left(\delta_{\alpha} \ast_{t} g\right) (t,x)$. 
Then we can choose  $k > 0$ such that the positivity condition \eqref{eq:gcond} is satisfied, uniformly in $t \in [-r,r], \; r >0$. We calculate 
$$
k g_{\alpha}(t,x) - \frac{1}{2}\dot{g}_{\alpha} (t,x) = \int g(s,x)
\delta_{\alpha}(t-s) \left( 
k - \frac{1}{\alpha^{2}} (t-s)\right) \; ds
$$
and find that the choice $k:= \frac{1}{\alpha^2} (r-s')$ is sufficient, where
$s' := \inf\{s \in \RR : (s,x) \in \supp g\}$. 
\end{example}
In the following, let $g_{\alpha}$ be chosen according to the preceding example
for a $g \in C^{\infty}_{c}(\RR^{2})$.
Denote by $U_{g_\alpha}(t,s)$ the propagator solving the Schr\"odinger equation
to $H(t) = H_0 + V(t;g_\alpha)$
depending 
on $g_{\alpha}$. The Hamiltonian is an operator on the Hilbert space $\cH$, the
Fock space of the free, massive, scalar field. 

Our first result states that outside of the time support of
$g$, the evolution family
$U_{g_\alpha}(t,s)$ is close to
the evolution family of the free
Hamiltonian $U_{0}(t,s) = e^{-iH_{0}(t-s)}$. 
\begin{theorem}
Let $\rho>0$ and suppose $\supp
g \subset [-\rho,\rho]\times[-\rho,\rho]$.  Let $\psi \in
\mathcal{H}$ and $t \geq \rho$. Then 
\begin{equation} 
\label{asympfree2}
\|(U_{g_\alpha}(t+s,t) - U_{0}(t+s,t)) \psi \| \to 0 \quad
\text{for}\quad
\alpha \to 0 
 \end{equation}
and arbitrary $s>0$.
\end{theorem}
\begin{proof}
Choose a function 
$s \mapsto f(s) = u(s) \psi \in D =
C^{\infty}_{c}(\mathbb{R}) \otimes Y$ with 
$u(t) = 1$. On these functions
we can write the difference of the propagators by evaluating the image of $f$ 
under the difference of the evolution groups at the point $t+s$:
\begin{multline*}
(U_{g_\alpha}(t+s,t) - U_{0}(t+s,t)) \psi = \left(\left(
T_{g_\alpha}(s)
- T_{0}(s) \right)f \right)(t+s)\\ = i
\left(\int_{0}^{s}T_{g_\alpha}(\sigma) (G_{0} - G_{g_\alpha})
T_{0}(s-\sigma)f \, d\sigma\right) (t+s),  
\end{multline*}
where $G_0$, $G_{g_\alpha}$ are the generators of $T_0$, $T_{g_\alpha}$
respectively.
Observe that 
$T_{0}$ maps $D$ to $D$, thus the difference of the evolution generators
$G_{0} - G_{g_\alpha}$
is given by the interaction Hamiltonian. For the norm we get the estimate
\begin{equation*}
 \|(U_{g_\alpha}(t+s,t) - U_{0}(t+s,t)) \psi
\| \leq \int_{0}^{s} \| \left( T_{g_\alpha}(\sigma)
V(\cdot;g_{\alpha})T_{0}(s-\sigma) f \right)(t+s) \| \,d\sigma.
\end{equation*} 
The integrand can be calculated, 
\begin{multline} \|
\left( T_{g_\alpha}(\sigma) V(\cdot;g_{\alpha})T_{0}(s-\sigma) f
\right)(t+s) \|\\ \leq |u(t+s - 2 \sigma)| \|  V(t+s
-\sigma; g_{\alpha})) (N+1)^{-2} \| 
 \| (N+1)^{2}
\psi  \|.
\end{multline}
Thus, we may estimate
\begin{multline*}
\|(U_{g_\alpha}(t+s,t) - U_{0}(t+s,t)) \psi \| \\ \leq \sup_{t' \in
\mathbb{R}} | u(t')| \sup_{t' \in [t, t+s]} \|
V(t';g_{\alpha}))
(N+1)^{-2} \| \| (N+1)^{2} \psi \|.
\end{multline*}
 Application of the estimate \eqref{eq:nlemma}
shows,
\begin{align}
 \sup_{t' \in [t, t+s]} \| V(t';g_{\alpha})
(N+1)^{-2} \| &\leq C \sup_{t' \in [t, t+s]}
\|\hat{g}_{\alpha}(t', \sum_{l=1}^{4}k_{l})\|_{L^{2}(\mathbb{R}^{4})}
\\ &\longrightarrow 0 \quad\text{for}\; \alpha \to 0, 
\end{align}
because with $g_{\alpha} \to g$ also $\hat{g}_{\alpha} \to \hat{g}$
and $t'$ is outside of the support of $g$. 
\end{proof}

The next result shows that the scattering operators exist. 
\begin{theorem}
For fixed $\alpha$, the scattering operator 
\begin{equation}
\label{sofg}
S(g_{\alpha}) = \lim_{\substack{t\to \infty \\ s \to -\infty}}
e^{itH_{0}} U_{g_\alpha}(t, s) 
e^{isH_{0}}
\end{equation}
exists as a strong limit.
\end{theorem}
\begin{proof} 
According to \cite{howland74}, the 
existence of the scattering operator for the time-dependent Schr\"odinger
equation is equivalent to the scattering problem for the evolution
group. To show the existence of the wave operators for the
evolution generator, it is convenient to use Cook's method
\cite[Theorem~XI.4]{rsIII}.

As above, let $f \in D=C^{\infty}_{c}(\mathbb{R}) \otimes Y$ . We have $e^{\pm i
\sigma G_{0}} Y \subset Y$, 
 thus it is
sufficient to show that
\begin{equation}\label{eq:cook}
\int_{\sigma_{0}}^{\infty} \|V(\cdot;g_{\alpha} )e^{\pm  \sigma G_{0}}
f\|
\, d\sigma < \infty.
\end{equation}
To this end, observe that $t \mapsto \tilde{g}(t) = \| \hat{g}_{\alpha}(t, \sum
k_{i})\|
_{L^{2}(\mathbb{R}^{4})}$ is fast decreasing and $u$ is compactly
supported:
\begin{align}
\int \| V(t;g_{\alpha} ) (e^{\pm \sigma G_{0}}
f)(t)\|^{2} \;dt &=
\int \| V(t;g_{\alpha} ) e^{\pm i \sigma H_{0}} \psi
\|^{2} 
|u(t\pm\sigma)|^{2}\,dt \notag\\ 
&\leq c \int \tilde{g}^{2}(t) |u(t\pm \sigma)|^{2} \, dt \notag\\
&= c (g^{2} \ast |\check{u}|^{2})(\pm \sigma) 
\end{align}
is fast decreasing, and thus its square root is integrable. Hence
$\|V(\cdot;g_{\alpha} ) e^{\pm  \sigma G_{0}} f\| = (\int
\|
V(t;g_{\alpha} )(e^{\pm  \sigma G_{0}} f)(t)\|^{2} \,
dt)^{1/2}$ is integrable as a function of $\sigma$ and \eqref{eq:cook} holds.
\end{proof}
Thus the dynamics $U_{g_\alpha}(t,s)$ is asymptotically free in the sense of 
scattering theory, for
$t \to \infty, s \to -\infty$. With respect to \eqref{asympfree2}, 
one sees that the dynamics is also asymptotically free in the 
limit $\alpha \to 0$ on 
arbitrary time intervals not intersecting the time
support of $g$.

Although we have a fair amount of information on the propagators
$U_{g_\alpha}(t,s)$, it is not possible to define the local scattering
operators by a limit $g_\alpha \to g$. For smooth, compactly supported functions
the condition \eqref{eq:gcond} is not satisfied, as such a function can not
dominate its
time derivative on the boundary of its support. To perform the limit
$\lim_{\alpha \to 0}U_{g_\alpha}(t,s)$ in the strong sense, one would need
Kato-type assumptions. Weak limit points exist, but then the causal
factorization is lacking. Here we could envisage a similar strategy as in
Section~\ref{existencequasiapprox}, but then we arrive at approximative
solutions. In view of this,  it is
much simpler and more general to use admissible bounded approximations. 
Thus we start directly from the wellposedness theory leading to 
approximative solutions. We obtain the $(\ph^4)_2$ model as a special case.

\section{The existence theorem}

We give an existence theorem for local scattering operators. The underlying
spacetime is the $d$-dimensional Minkowski space, $2\leq d\leq 4$, with its
symmetry group $O(1,d-1)$. The proper, orthochronous Lorentz group is the
connection component of the identity $SO^+(1,d-1) $. We denote its universal
covering group by $G$. As in Definition~\ref{lso}, we consider scalar theories.

\begin{theorem}
\label{th:exlso}
Let $\cH$ be a separable Hilbert space which carries a unitary, irreducible
representation $\overline{\mathcal{P}} \to \cB(\cH),
(a,\alpha) \mapsto U(a,\alpha)$ of 
the universal covering group $\overline{\mathcal{P}} = \RR^d \rtimes G$  of the
Poincar\'e group on  $d$-dimensional spacetime, $2 \leq d \in \NN$. 
Let $V(t;g)$ be a 
self-adjoint operator with domain $D(V(t;g))$, depending on $ t = x_0 \in \RR$,
and $g 
\in C^\infty_c(\RR^d)$ such that
\begin{enumerate}[(i)]
\item  $V(t,g) = 0$ if $g(t,\vec{x}) = 0$ for all
$\vec{x} \in \RR^{d-1}$, 
\item $t \mapsto R(i, V(t;g))$ is
strongly continuous,
\item there is a Banach space $Y$, dense in $\cH$ and continuously embedded in
$D(V(t;g))$, locally uniformly in $t$, such that $t \mapsto V(t,g)\psi$ is
continuous for all $\psi \in Y$,
\item the time translations $U((t,\vec{0}), \id) = e^{-i H_0 t}$ fulfill
$ e^{-i H_0 t}Y
\subset Y$ and $( e^{-i H_0 t})_{t \in \RR}$ is strongly continuous in
$Y$,
\item $e^{i H_0 t}V(t;g_{\langle a, \Lambda(\alpha) \rangle}) e^{-i H_0 t}=
U(a,\alpha)^* e^{i H_0 t}V(t;g)e^{-i H_0 t}
U(a,\alpha),$ 
where $\Lambda(\alpha) \in SO^+(1,d-1)$, $\alpha \in G$, $a \in \RR^d$ and
$g_{\langle a, \Lambda(\alpha) \rangle}(x) =
g(\Lambda(\alpha)^{-1}(x-a))$.
\end{enumerate}
Then the local scattering operators $S(g)$ exist.
\end{theorem}
\begin{proof}
We organize the proof in two steps. In the first one we consider the solution of
the time-dependent Schr\"odinger equation in the interaction picture.  The
second step deals with the local scattering operators.

\subparagraph*{Step 1.}

 We transform the interaction to the \emph{Dirac} (or \emph{interaction})
\emph{picture},
$$
V^D(t;g) := e^{it H_0} V(t;g) e^{-it H_0}\quad \text{with domain}\quad
D(V^D(t;g)) := e^{it H_0}D(V(t;g)),
$$
where $H_0$ is the generator of the time translations, $e^{-it H_0} =
U((t,\vec{0}), \id)$. 
It is easy to see that $V^D(t;g)$ is self-adjoint on $D(V^D(t;g))$: Let
$\theta\in
\cH$ be arbitrary and set $\tilde{\theta} := e^{-it H_0} \theta$. Because of the
self-adjointness of $V(t;g)$ there is a $\tilde{\psi} \in D(V(t;g))$ such that
$\tilde{\theta} = (V(t;g) + i) \tilde{\psi}$. Thus $\psi := e^{itH_0}
\tilde{\psi} \in
D(V^D(t;g))$ is the pre-image of $\theta$ under $V^D(t;g) + i$. A similar
argument
shows $\Ran (V^D(t,g) - i) = \cH$.    
Because of condition (iv), $Y \subset D(V^D(t;g))$. Furthermore, condition 
(iii), together with the unitarity of $e^{\pm it H_0}$, implies that the 
embedding of $Y$ in $D(V^D(t;g))$ is also continuous, locally uniformly in $t$. The mapping
$t\mapsto R(i,V^D(t;g)) = e^{i t H_0} R(i,V(t;g)) e^{-i t H_0}$ is strongly
continuous as a product of strongly continuous functions of bounded operators. Moreover, we
find continuity of $t \mapsto V^D(t;g)\psi$ for $\psi \in Y$. We calculate
\begin{multline*}
\|(V^D(s;g) - V^D(t;g)) \psi\| \leq \| e^{is H_0} V(s;g)(e^{-is H_0} - e^{ -it
H_0})\psi\|
\\
+ \|e^{is H_0}(V(s;g)-V(t;g)) e^{-it H_0} \psi \| + \| (e^{is H_0} - e^{ it
H_0})
V(t;g)e^{-it H_0} \psi\|.
\end{multline*}
In the limit $s \to t$, the first term on the right-hand side vanishes because
$Y$ is continuously embedded in $D(V(t;g))$ locally uniformly in $t$ and $(e^{-i
t H_0})_{t \in \RR}$ is a group in $Y$, strongly continuous with respect to the
norm $\| \cdot\|_Y$. The second term converges to $0$ because of the invariance
of $Y$ under $e^{-i t H_0}$ and the continuity properties of $V(t;g)$. The third
term obviously vanishes too.

By Theorem~\ref{th:se}, there exists an approximative solution of the Cauchy
problem of the time-dependent Schr\"odinger equation with respect to $A(t) = -iV^D(t;g)$, given by a propagator
$U^D_g(t,s)$, which is unique as well as strongly continuous and unitary almost
everywhere in $t,s \in \RR$.  Furthermore,  $U^D_g(t,s)$ is the weak limit of a
sequence of propagators $\{U_{g,n}^D(t,s)\}_{n \in \NN}$, where each element
solves the time-dependent Schr\"odinger equation with respect to a bounded
admissible approximation $V^D_n(t;g)$ of $V^D(t;g)$.

\subparagraph*{Step 2.}

Now we define the local scattering operators.  Choose arbitrary, but fixed test
functions $f,g,h \in C^\infty_c(\RR^d)$ and an arbitrary Poincar\'e transformation
$(a, \Lambda(\alpha))$. Moreover, we choose $\sigma, \tau \in \RR$ such that $\sigma < \min I_g
\cup I_{g_{\langle a,\Lambda(\alpha)\rangle}} \cup I_f \cup I_h$ and $\tau > \max  I_g \cup
I_{g_{\langle a,\Lambda(\alpha)\rangle}} \cup I_f \cup I_h$ where  we set $I_g
:= \{t \in \RR :
(t,\vec{x}) \in \supp g \}$ for arbitrary test functions $g \in C^\infty_0(\RR^d)$. We define
$$
S(g) := U^D_g(\tau,\sigma),
$$
and we verify the conditions of Definition~\ref{lso}.  
Note, that $U^D_g(t,s) = \one$ if $[t,s] \cap I_g = \emptyset$. Thus, by
uniqueness of the propagator, the definition of $S(g)$ does not depend on $\tau$
and $\sigma$ as long as $I_g \subset [\sigma,\tau]$. Moreover, it is clear that
the null set $N$ is a subset of  $ I_g \cup I_{g_{\langle a,\Lambda(\alpha)\rangle}}
\cup I_f$. Hence $S(g)^{-1} = S(g)^*$ and $S(g) = \one $ if $g = 0$.

With a similar argument, involving uniqueness, we verify the causality
condition. Clearly, if two Hamiltonians $H(t)$ and $H'(t)$ coincide for $t \in I
\subset \RR$, then the admissible bounded approximation $H'_n(t)$ is  an
admissible bounded approximation of $H(t)$ at the same time. By
uniqueness of the approximative solution and the calculation
\eqref{qapproxunique} we find $U(t,s) = U'(t,s)$ for $t,s \in I$. 
Now suppose that there exists a $t_0 \in \RR$ such that $\max I_g < t_0 < \min
I_f$. Then $t_0 \not\in N$ and, using $V^D(t; g+f) = V^D(t; g)$ for $t < t_0 $
as well as $V^D(t; g+f) = V^D(t; f)$ for $t > t_0$, we see that $U^D_{g+f} (t,s)
= U^D_g(t,s)$ if $t,s \leq t_0$ and
$U^D_{g+f} (t,s) = U^D_f(t,s)$ if $t,s \geq t_0$. Hence 
$$
S(g+f) = U^D_{g+f}(\tau,t_0)U^D_{g+f}(t_0,\sigma) =
U^D_f(\tau,t_0)U^D_g(t_0,\sigma) = S(f)S(g),
$$ 
where $U^D_g(t,s)$ and $U^D_f(t,s)$ are extended by $\one$ outside the time
support of $g$ respective $f$.
Moreover we establish the causality condition
\eqref{causal} using the same argument. The test function $h \in
C^\infty_c(\RR^d)$ has arbitrary time support $I_h$. By possibly varying $t_0$
in an open neighborhood between $\max I_g$ and $\min I_f$, we achieve $t
\not\in  N$, where $N$ is the exceptional null set of the propagator
$U^D_h(t,s)$ where the causal factorization fails to hold. With this assumtion,
again the uniqueness of the approximative solution results in
$$
U_{f+h+g}^D(\tau, t_0)U_{h}^D(t_0,\sigma) = U^D_{f+h}(\tau, \sigma).
$$ 
Using this relation  and an analogous one,  we see that
\begin{align*}
S(f + h &+ g) = U^D_{f+h+g}(\tau, \sigma)\\
&=  U^D_{f+h+g}(\tau, t_0)U^D_{h}(t_0, \sigma)(U^D_{h}(t_0, \sigma))^*(U^D_{h}(\tau, t_0))^*U^D_{h}(\tau, t_0)U^D_{f+h+g}(t_0, \sigma)\\
&= U_{f+h}^D(\tau,\sigma) (U^D_h(t_0, \sigma))^*(U^D_h(\tau, t_0))^* U^D_{h+g}(\tau,\sigma)\\
&= S(f+h) S(h)^*S(h+g).
\end{align*}

Now, we consider covariance. By assumption (v), 
$$
V^D(t;g_{\langle a,\Lambda(\alpha) \rangle}) = U(a,\alpha)^* V^D(t;g)
U(a,\alpha). 
$$  
We can repeat the argument of Step~1 for $V^D(t;g_{\langle a,\Lambda(\alpha)
\rangle})$ and end up with an approximative solution of
the time-dependent Schr\"odinger equation, given by a propagator
$U^D_{g_{\langle a,\Lambda(\alpha)
\rangle}}(t,s)$, which
is unique as well as strongly continuous and unitary almost
everywhere in $t,s \in \RR$.  As above,  $U^D_{g_{\langle a,\Lambda(\alpha)
\rangle}}(t,s)$ is the weak limit of a
sequence of propagators $\{U_{g_{\langle a,\Lambda(\alpha)
\rangle},n}^D(t,s)\}_{n \in \NN}$, where each element is strongly
differentiable in $t$ and
solves the time-dependent Schr\"odinger equation with respect to an admissible
bounded
approximation $V^D_n(t;g_{\langle a,\Lambda(\alpha)
\rangle})$ of $V^D(t;g_{\langle a,\Lambda(\alpha)
\rangle})$. But, considering $U(a,\alpha)^* U^D_{g,n}(t,s)U(a,\alpha)$, we find
\begin{multline*}
i \frac{\partial}{\partial t} U(a,\alpha)^* U^D_{g,n}(t,s)U(a,\alpha)\theta =
\\ 
U(a,\alpha)^* V^D_{n}(t;g)U(a,\alpha)U(a,\alpha)^*
U^D_{g,n}(t,s)U(a,\alpha)\theta
\end{multline*}
for every $\theta \in  \cH$.
If we show that $U(a,\alpha)^* V^D_{n}(t;g)U(a,\alpha)$ is an admissible
bounded approximation of $V^D(t;g_{\langle a,\Lambda(\alpha) \rangle})$,
then uniqueness of the approximative solution implies 
$$
U^D_{g_{\langle a,\Lambda(\alpha) \rangle}}(t,s) =
U(a,\alpha)^*U^D_g(t,s)U(a,\alpha)
$$
and hence covariance of the local scattering operators,
$$
S(g_{\langle a,\Lambda(\alpha) \rangle}) = U(a,\alpha)^*S(g)U(a,\alpha).
$$
The strong continuity of $t \mapsto U(a,\alpha)^*
V^D_{n}(t;g)U(a,\alpha)$ is clear, as well as the boundedness of the
corresponding propagators.
We establish the conditions  on $U(a,\alpha)^*
V^D_{n}(t;g)U(a,\alpha)$ in the following calculations: Let $\psi \in
D(V^D(t;g_{\langle a,\Lambda(\alpha) \rangle}))$, then 
$$
\|(U(a,\alpha)^*V^D_{n}(t;g)U(a,\alpha) - V^D(t;g_{\langle a,\Lambda(\alpha)
\rangle}))\psi\| \leq \| (V^D_n(t;g) - V^D(t;g)) \tilde{\psi} \| \to 0
$$
for $n \to \infty$,
where $\tilde{\psi} = U(a,\alpha)\psi \in D(V^D(t;g))$, because $V^D_n(t;g)$ is
an
admissible bounded approximation of $V^D(t;g)$. With the same argument, 
\begin{multline*}
\|U(a,\alpha)^*V^D_{n}(t;g)U(a,\alpha) \psi \| \leq c (\|V^D(t;g) \tilde{\psi}\|
+ \|\tilde{\psi}\|)\\
\leq c(\|V^D(t;g_{\langle a,\Lambda(\alpha)
\rangle}) \psi \| + \| \psi \|).
\end{multline*}
So the family of unitary operators $\{S(g) : g \in C^\infty_c(\RR^d)\}$
satisfies the requirements of Definition~\ref{lso} and the statement follows. 
\end{proof}
For the definition of the local scattering operators $S(g)$ we use the time
evolution $U^D_g(t,s)$, generated by the Hamiltonian $V^D(t;g)$ in the
interaction picture. However, if also the Hamiltonian $H(t) = H_0 + V(t;g)$
satisfies the requirements of Theorem~\ref{th:se}, we could start with the
propagator $U_g(t,s)$ in the Schr\"odinger picture and define the local
scattering operators with its transformation
$$
\tilde{U}^D_g(t,s) := e^{itH_0}U_g(t,s)e^{-isH_0}.
$$
But this is not a good choice: Let $U_{g,n}(t,s)$ be the propagator associated with an admissible bounded approximation $H_n(t)$ of $H(t)$. Then $\tilde{U}^D_{g,n}(t,s)\theta$ is not differentiable for arbitrary $\theta \in \cH$. On a suitable subspace we find
$$
i \frac{\partial}{\partial t} \tilde{U}^D_{g,n}(t,s)\psi =  e^{it H_0}(H_n(t) - H_0)e^{-is H_0}\tilde{U}^D_{g,n}(t,s)\psi, 
$$ 
but in general $H_n(t) - H_0$ is \emph{not} an admissible bounded approximation
of $V(t;g)$. Nevertheless, it is possible to establish the equivalence of both
strategies if we require $D(H_0^p) \subset D(V(t;g))$ for a $p \in \NN$  and
specialize to $Y:=D(H_0^p)$. Then we could  choose the admissible bounded
approximation \emph{on $Y$} by $H_n(t) = H_{0,n} + V_n(t;g)$ with $ H_{0,n}$,
and $V_n(t;g)$ being admissible bounded approximations on $Y$ of
$H_0$ and $V(t;g)$ respectively, see the remark after
Definition~\ref{admissible-def}. Then
it is easy to see that the propagator in the interaction picture coincides with
the transformed time evolution operator in the Schr\"odinger picture. But we do
not go into further detail here, because we prefer the definition of the local
scattering operators starting with the propagator $U_g^D(t,s)$ generated by
$V^D(t;g)$.  In the following, we will see that it is advantageous to work
directly in the interaction picture.

\section{Local scattering operators for the $P(\ph)_2$ model}
\label{pphilso}

We establish the existence of the local scattering operators $S(g)$ for the
$P(\ph)_2$ model. As far as we know, this is the first proof of the existence 
of local scattering operators for a quantum field theory with
nonlinear field equations. The $(\phi^4)_2$ model of Section~\ref{ph4background}
is a special case.

According to Theorem~\ref{p(phi)}, there exists a unique approximative
solution of the time-dependent Schr\"odinger equation for the $P(\ph)_2$
model with interaction
localized in the support of a test function $g \in C^\infty_c(\RR^2)$, $0\leq g
\leq 1$, and $P$ a semibounded polynomial.
Associated with this solution, there is an almost everywhere unitary and strongly
continuous propagator
$U_g(t,s)$. For the definition of the local scattering operator we could
transform
the
propagator to the interaction picture, as indicated in the last section: Choose
an interval $[\sigma, \tau]$ 
large enough such that $I_g \subset [\sigma, \tau]$, where $I_g := \{t \in \RR\
: (t,\vec{x}) \in \supp g \}$.
In this case we would define
\begin{equation}
\label{eq:s(g)}
S(g) := e^{i\tau H_0} U_g(\tau,\sigma)e^{-i\sigma H_0}.
\end{equation}
However, if we aim at the local algebras, as defined in Section~\ref{algebras},
the positivity requirement on the test function $g$ is problematic. The
condition arises from the proof of essential self-adjointness of the total
Hamiltonian $H(t) = H_0 + V(t;g)$. We can avoid this kind of restriction on the
choice of the test function by working directly in the interaction picture, see
Theorem~\ref{th:exlso}. Moreover, we do not need the restriction to semibounded
polynomials $P$ and positive coupling constants any longer. These conditions,
ensuring stability, are not
necessary in a local construction.

Let $\cH$ be the Fock space of a free, massive boson field as in
Appendix~\ref{qftbasics}.  
We consider the
interaction of the $P(\ph)_2$ model,
\begin{equation*}
V(t;g) = \int g(t,x) :P(\ph(x)): \,dx,
\end{equation*} 
where $P(\lambda)$ is a real polynomial. The test function $g \in
C^\infty_c(\RR^2)$ is assumed to be real, but there are no restrictions in
other respects.
With Theorem~\ref{th:exlso} we demonstrate the existence of the local
scattering operators. 

The existence question for these
operators in the special case $(\ph^4)_2$ was formerly addressed in the work of
W. Wreszinski. In \cite{wski}, 
the Cauchy problem of the
time-dependent Schr\"odinger equation for the localized $(\ph^4)_2$ model with
\emph{factorizing test function} is
solved, using Kato's theory for the case of time-independent domains
(Theorem~\ref{timeind}). The restriction to factorizing test functions
$g(t,x) = u(t) v(x),\ u,v\in C_c^\infty(\RR)$ is necessary to achieve
time-independence of the domains of the Hamiltonians. This is a
considerable loss of
generality: For the approximation of an arbitrary compactly supported
test function, one needs linear combinations of such factorizing functions. But
for these linear combinations, the Hamiltonians do
not
have time-independent domains anymore. Recently, the question of existence of
local scattering operators
for the $(\ph^4)_2$ model is addressed \cite{Wreszinski:2003ye}, using the
Theorem of
Kisy\'nski (see Theorem~\ref{kiszynskisimple}). In this article, a scale of
Hilbert spaces $\mathcal{F}_{+2} \subset \cH \subset \mathcal{F}_{-2}$ with respect to the free Hamiltonian $H_0$ is
considered. The space $\mathcal{F}_{+2}$ equals $D(H_0)$ with the norm $\| \psi
\|_{+2} = \| (H_0 + 1) \psi \|$.  
In the course of the
argument, closedness of the sesquilinear form
$$
S(\theta,\psi) = \left((H(t) +M)^{1/2}\theta,(H(t) +M)^{1/2} \psi \right)
$$
on $\mathcal{F}_{+2}$ is assumed, where $M>0$ is a constant such that $H(t) +M
\geq 0$, see \cite[Equation~2.21]{Wreszinski:2003ye}. But if $S(\theta,\psi)$
would be closed with respect to $\mathcal{F}_{+2}$, this would imply that
$S(\psi,\psi)$ induces an equivalent norm on $\mathcal{F}_{+2}$. Or in other
words, there would exist a constant $c>0$ such that 
$$
c^{-1}(H_0+1)^2 \leq (H(t) +M) \leq c (H_0+1)^2.
$$
While the inequality on the right-hand side can be fulfilled using
Theorem~\ref{th:n}, the
left-hand inequality is \emph{not} satisfied for arbitrary $t$ in  the
\emph{localized}
$(\ph^4)_2$ model.
Hence, $S$ is not closed and Kisy\'nski's Theorem is not applicable. Notice
that this problem does not arise in Dimock's application \cite{dimock}, as in
this case the scale of Hilbert spaces is constructed with respect to the
\emph{interacting} Hamiltonian.

As we will see in the following, the notion of approximative solutions is
appropriate to
discuss the existence question for local scattering operators. This
setting results in a considerable simplification compared to previous
attempts. 
\begin{theorem}
\label{p(phi)existence}
Let $\cH$ be the Fock space of a free, massive boson field
$\ph$ in two spacetime dimensions.  
Consider the
interaction Hamiltonain of the $P(\ph)_2$ model,
\begin{equation*}
V(t;g) = \int g(t,x) :P(\ph(x)): \,dx,
\end{equation*}
where $P(\lambda) = \sum_{i=0}^n a_i \lambda^i$ is a real polynomial, not necessarily
semibounded, and $g \in C^\infty_c(\RR^2)$ is a real test function. 
Then the local scattering operator $S(g)$  exists.
\end{theorem}
\begin{proof}
The spacetime under consideration is the hyperbolical plane, that is $\RR^2$
with the Lorentzian metric $g_{\mu \nu}$. Its homogeneous isometry group is
$O(1,1)$ and the proper Lorentz group is $SO^+(1,1)$. The latter group is
simply connected and has the one-parameter form
$$
\RR \to SO^+(1,1), \quad \eta \mapsto 
\begin{pmatrix}
\cosh \eta & \sinh \eta \\
\sinh \eta & \cosh \eta
\end{pmatrix},
$$
as a group isomorphism, see \cite{hein}. Hence the universal covering group 
$G$ equals $SO^+(1,1)$ itself, and the Poincar\'e group is
$\overline{\mathcal{P}} =
\RR^2 \rtimes  SO^+(1,1)$. The Fock space $\cH$ over the one-particle space
$L^2(\RR)$ for the theory of a scalar, massive quantum field carries a unitary
representation $\overline{\mathcal{P}} \to \mathcal{B}(\cH),\ \langle a,
\Lambda\rangle \mapsto U(a,\Lambda)$ of the Poincar\'e group, see
\cite{cannonjaffe}. The interaction Hamiltonian is a sum of Wick monomials of
the free time-zero fields $\ph(x) = \ph(0,x)$, which we denote by the same
symbol. Transformed to the time $t$ by the free time evolution, the Wick
monomials transform covariantly,
\begin{multline*}
 U(a,\alpha)^* e^{i H_0 t}:\ph^n(x):e^{-i H_0 t}
U(a,\alpha) = U(a,\alpha)^* :\ph^n(t,x):U(a,\alpha) =\\
:\ph^n(\Lambda(t,x) +a):.
\end{multline*}
This implies the transformation property (v) in Theorem~\ref{th:exlso}.
 
The interaction Hamiltonian $V(t;g)$, $g = \overline{g} \in C^\infty_c(\RR^2)$,
is essentially self-adjoint on $\mathcal{D}$, the set of Fock space vectors with
finite particle number and Schwartz wavefunctions (see \cite{bqft} and
Theorem~\ref{th:qspace}). Notice that neither positivity of $g$ nor
semiboundedness of $P$ are necessary. Hence, using
Theorem~\ref{th:n}, we see that $V(t;g)$
is essentially self-adjoint on $Y:=D(N^m)$, where $m \in \NN$ is larger or
equal to $n/2$. Clearly, $V(t;g)$ vanishes outside
the time support of $g$. 

The free Hamiltonian $H_0$ and the particle number operator $N$ commute, thus
$e^{-itH_0} Y = Y$ and $(e^{-itH_0})_{t \in \RR}$ is strongly continuous in the
Banach
space norm of $Y$.

Consider the term of highest order in the interaction $V(t;g)$: 
$$V_{n}(t;g) = a_{n}\int
g(t,x) :\ph(x)^{n}: \,dx.
$$
For
$\psi \in Y$ the interaction Hamiltonian is dominated by a power of 
the number operator, see Theorem~\ref{th:n},
\begin{align*}
\|V_{n}(t;g)\psi\| &\leq \|V_{n}(t;g) (N+1)^{-m}\|
\|(N+1)^{m} \psi\| \notag \\
&\leq c \|w(t,\cdot)\|_{L^2(\RR^{n})} \|(N+1)^m \psi\|,
\end{align*}
where $c$ is a $t$-independent constant and $w(t,k_1,\dots,k_{n} )$ denotes the
numeric kernel of the expansion of $V_{n}(t;g)$ into Wick monomials. It is given
by
\begin{equation*}
w(t,k_1,\dots,k_{n}) = \hat{g}(t,k_1 + \dots + k_{n})\omega(k_1)^{-1/2} \cdot
\dots
\cdot \omega(k_{n})^{-1/2},
\end{equation*}
where $\hat{g}$ denotes the Fourier transform of g with respect to $x$, and the
$L^2$-norm is evaluated with respect to $(k_1, \dots, k_{n}) \in \RR^{n}$,
see \cite[\S 6.1]{DG}.
By repeated use of Young's inequality, we estimate
$$
\|w(t,\cdot)\|_{L^2(\RR^{n})}  \leq c \|g(t,\cdot)\|^2_{L^2(\RR)}
$$
(cf.\ \cite[Lem.~6.1]{DG}). The other summands of $V(t;g)$ can be treated in
the same way.
These inequalities imply the
uniform boundedness of the embedding $Y \hookrightarrow D(V(t;g))$.
Similarly we verify the
continuity of $t \mapsto V(t;g)\psi$ for all $y \in Y$. Strong continuity of
$t\mapsto R(i,V(t;g)) $
follows, because $Y$ is a core for $V(t;g)$. Hence, by Theorem~\ref{th:exlso},
the local scattering operators exist.
\end{proof}

\subsection{The massless boson field in two dimensions}

For the massless boson field in two dimensions, the two-point function is
divergent. The theory has no vacuum state. Thus, the massless boson field in
$d=2$ is not a field theory in the sense of Wightman's axioms in 
Section~\ref{wightman}.

However, it is possible to define the theory in the algebraic sense,
hence to construct the algebras of local observables. Problems do not arise
until
the state space is analyzed.

In Wightman's Carg\`ese lectures \cite{wightmanlecture} he addresses the
massless boson in two dimensions by using spaces with indefinite metric,
similar to Schroer's treatment of the derivative coupling model
\cite{Schroer:1963gw}. The
algebra of observables
is constructed by Streater and Wilde in \cite{Streater:1971ct}.

By our results on the existence of the local scattering operator for
$P(\ph)_2$ models, it is possible to get the algebra of local observables for
the massless boson field in two dimensions in an easier way, using the formalism
of Section~\ref{algebras}.

The classical Hamiltonian in the theory of a free massive boson field $\ph$ with
mass $m > 0$ is given by
\begin{equation}
\label{mham}
H_0 = \frac{1}{2} \int \pi(x)^2 + (\nabla \ph(x))^2 + m^2 \ph(x)^2 \,dx.
\end{equation}
 The field $\pi(x) =
\frac{d}{dt} \ph(t,x)\negthickspace\restriction_{t=0}$ is the canonically
conjugate object to
the time-zero field $\ph(x)$. We consider the formal interaction term
$$
V_\lambda(t;g) = \frac{1}{2} \lambda \int g(t,x)  \ph(x)^2 \,dx  
$$
with $g \in C^\infty_c(\RR^2)$ and $\lambda \in \RR$. If we choose
$\lambda := -m^2$ and $g(t,x) = 1$ for $(t,x)$ in  an open,
contractible region $\mathcal{O}$ of spacetime, the density of the Hamiltonian
$H_0 + 
V_\lambda(t;g)$
 coincides with the Hamiltonian density of the massless theory inside of
$\mathcal{O}$. On $\cH$, the Fock space  over $L^2(\RR)$, belonging to a free,
massive, scalar quantum
field $\ph$, we apply the unitary transformation
$R_g$, which transforms the interacting Hamiltonian $H_0 + 
V_\lambda(t;g)$ of the theory into the free
one with modified dispersion relation. That such
a transformation exists is shown by Rosen in \cite{rosenshift} for $\lambda =
-m^2 + \epsilon$, $\epsilon >0$. After addition of a vacuum energy
renormalization constant, the Hamiltonian is transformed into $H(g) = \int
a^*(k_1) \mu_{g,t}(k_1, k_2) a(k_2)\,dk_1 \, dk_2 $ with the dispersion
relation $\mu_{g,t}^2 = -\Delta + m^2 + \lambda g(t,x) = \mu^2 + \lambda
g(t,x)$. Hence, for $g \to 1$ and $\epsilon \to 0$, the dispersion relation
approaches the $(m=0)$-case $\mu^2 = -k^2$.

Since the existence result  for the local scattering
operator of the
$P(\ph)_2$ model (Theorem~\ref{p(phi)existence}) gives no restriction on the
sign of the coupling constant
respective the localization function, we can apply it to the special case
$P(\ph) = -m^2 \ph^2$. The local scattering operators generate the algebra of
observables of a massless boson field in two dimensions according to the
formalism of Section~\ref{algebras}.

Indeed we might even choose $\lambda < -m^2$. The local scattering
operators exist by Theorem~\ref{p(phi)existence} and the associated relative
scattering operators generate the local algebras of a bosonic field theory with
negative squared mass. As in the massless case, for these theories there are no
vacuum states. But they are well defined theories in the algebraic sense.
By coupling of another field or addition of a suitable self\mbox{}interaction,
these
\emph{tachyonic fields} with imaginary mass may acquire a stable vacuum state.
This process is often called \emph{tachyon condensation}.

\subsection{Going further}

Up to this point, we have established the existence of the local scattering
operators for a quantum field theory with nonlinear field equation, but without
infinite renormalization (apart from Wick ordering). The next step would be to
investigate models which require this technique.

The simplest model which shows a UV divergence is the exactly solvable
\emph{van Hove}
model \cite{derezinski}. The question of the existence of propagators
for time-dependent van Hove models was addressed in
\cite{Schlegelmilch:2001mk}. The strategy is similar to our approach to
Goldstein's example in Section~\ref{goldsteindressing} and uses dressing
transformations. 

Consider a $3$-dimensional spacetime and the theory of a free, massive, scalar
field $\ph(\vec{x})$ on the Fock space $\cH$. The Hamiltonian of the
time-localized van Hove model is formally given as
$$
H(t) = H_0 + g(t) \ph(\vec{0}),
$$
with $g \in C^\infty_c(\RR)$ and the time-zero field $\ph$. To get a 
meaningful expression, we choose test functions $\rho_l \in
C^\infty_c(\RR^2),\ l \in
\NN$, such that $\hat{\rho}_l \to 1$ for $l \to \infty$. The function
$\vec{k} \mapsto 
\hat{\rho}_l(\vec{k})$ serves as a UV cut-off. We define $H_l(t) = H_0 +
V_l(t)$,
where 
$$
V_l(t):= g(t)\int \rho_l(\vec{x}) \ph(\vec{x})\,d^2x.
$$
In the limit $l \to \infty$, it turns out that $H_l(t)$ is not defined as an
operator on $\cH$. It is necessary to renormalize the Hamiltonian by addition
of a $c$-number
$E_l(t)$,
$$
H_{\text{ren}, l}(t) := H_l(t) + E_l(t),
$$
where $E_l(t) \to \infty$ but $H_{\text{ren}, l}(t)$ is a well defined
operator for $l \to \infty$. Perturbation theory suggests 
$E_l(t) := g(t)^2 \int (2 \mu(\vec{k}))^{-2} |\rho_l(\vec{k}) |\,d^2k$.

The renormalized Hamiltonian is unitarily equivalent to the free Hamiltonian.
There exists a unitary dressing transformation $R_l(t)$, such that
$$
R^{-1}_l(t)H_{\text{ren}, l}(t) R_l(t)=  H_0.
$$ 
The dressing transformation is exact.
It is defined by $R_l(t) = e^{- \Gamma V_l(t)}$, involving the
inverse adjoint action of $H_0$, that is $ \Gamma V_l(t)$ is the operator which
satisfies $[H_0 , \Gamma V_l(t)] = V_l(t)$. The $\Gamma$ operation produces an
additional factor $\mu(\vec{k})^{-1}$ in the numerical kernel of $V_l(t)$, hence
it leads to a faster decrease of the kernel for large $|\vec{k}|$. Thus, $R(t)
:= \lim_{l \to \infty} R_l(t)$ is well defined as a unitary operator and
we define $H_{\text{ren}}(t) := R(t) H_0 R^{-1}(t)$ with domain
$D(H_{\text{ren}}(t)) = R(t) D(H_0)$. The intersection of the domains of $H_0$
and $ H_{\text{ren}}(t)$ is trivial.

To discuss the time evolution, we formally define the candidate for the
generator of an evolution group by $i G_0 := H_{\text{ren}}(t) -i\frac{d}{dt}$
on $E = L^2(\RR,\cH)$ and use the dressing transformation
as an operator on $E$ to calculate
$$
iG_0 = R\left( H_0 -i \Gamma \dot{V} -i\frac{d}{dt} -\tilde{E} \right)R^{-1},
$$
where $\tilde{E}$ is a bounded, c-number-valued function.
In \cite{Schlegelmilch:2001mk} we discussed essential self-adjointness of
$i\tilde{G}_0 = H_0 -i \Gamma \dot{V} -i\frac{d}{dt} -\tilde{E}$. This is
possible due to the simple structure of the model. But our results of
Section~\ref{existencequasiapprox} show that it is sufficient to deal with
approximative solutions which belong to a self-adjoint extension of
$i\tilde{G}_0$, which exists under more general assumptions. The existence
theorem for approximative solutions (Theorem~\ref{th:se}) and hence
Theorem~\ref{th:exlso} make it possible to envisage a similar strategy for
the proof of existence of local scattering operators for more complicated
models. Locality of the interaction ensures $R(t) = \one$ for sufficiently
large $t$, thus it is sufficient to investigate existence of local scattering
operators for the regularized interaction $H_0 -iR(t)^{-1}(\frac{d}{dt}
R(t))$.
We obtain a regularization procedure similar to the one developed by
Mickelsson and Langmann in \cite{langmann}. They work on the level of
the one-particle space and then discuss the implementability of the scattering
operators in Fock space. This is possible for models with linear field
equations, but the quantization of the scattering operators leaves a phase
factor undefined. Our Theorem~\ref{th:exlso} allows for a direct discussion of 
the existence question of the local scattering operators in Fock space, thereby
avoiding the mentioned ambiguity and extending the scope of application of the
regularization procedure to models with non-linear field equations. 

Possible candidates for a further investigation of models with linear field
equations are fermions in external fields with a dressing transformation as in
\cite{fredenhagen}
 or massive bosons with a localized $\ph^2$ interaction in
 four dimensions \cite{ginibrevelo, rosenshift}.
Let $\cH$ be the Fock space over $L^2(\RR^3)$ and let $\ph(\vec{x})$ be a
scalar, massive time-zero field. For a real test function $g \in
C^\infty_c(\RR^4)$ we consider formally $H(t) := H_0 + V(t)$, where
$$
V(t) := \int g(t,\vec{x}) :\ph^2(\vec{x}):\,d^3x.
$$ 
The interaction term is the sum of terms with zero, one and two creation
operators. We introduce an ultraviolet cut-off $\sigma>0$ in the numerical
kernel by the
characteristic function $\chi_\sigma$ of the set $[-\sigma,\sigma]^3 \subset
\RR^3$ and define
\begin{align*}
V_{0,\sigma}(t) &= V_{2,\sigma}(t)^* = \int \tfrac{\hat{g}(t,\vec{k} +
\vec{k'})\chi_\sigma(\vec{k})\chi_\sigma(\vec{k'})}{\sqrt{\mu(\vec{k})
\mu(\vec{k'})}}
a(\vec{k}) a(\vec{k'}) \,d^3k\,d^3k',\\
V_{1,\sigma}(t) &= 2 \int \tfrac{\hat{g}(t,\vec{k} -
\vec{k'})\chi_\sigma(\vec{k})\chi_\sigma(\vec{k'})}{\sqrt{\mu(\vec{k})
\mu(\vec{k'})}}
a^*(\vec{k}) a(\vec{k'}) \,d^3k\,d^3k',
\end{align*}
where $\hat{g}(t,\cdot)$ is the Fourier transform of $g$ with respect to
$\vec{x}$. 
To define the Hamiltonian, a counter term is necessary. By
$\underset{n}{\wick[u]{1}{<*A>*B}}$ we denote the term with
$n$ contractions in the product of two Wick monomials $A$ and $B$. Perturbation
theory predicts that addition of 
$$
E_\sigma(t) := \underset{2}{\wick[u]{1}{<*{V_{0,\sigma}(t)}>*{\Gamma
V_{2,\sigma}(t)}}}
$$
to $H(t)$ leads to a renormalized Hamiltonian,
$$
H_{\text{ren},\sigma}(t) := H_0 + 
V_{0,\sigma}(t)+ V_{1,\sigma}(t) + V_{2,\sigma}(t) + E_\sigma(t) = H_0 +
V_\sigma(t) + E_\sigma(t),
$$
which is well-defined as an operator in the limit $\sigma \to \infty$. Notice
that in this limit the counter term $E_\sigma(t)$ diverges. As above, $\Gamma
= \ad_{H_0}^{-1}$.  

We define the dressing transformation in the following way: By Nelson's theorem
of analytical vectors, we can establish essential skew-adjointness of $
W_\sigma(t) := \Gamma (V_{0,\sigma}(t) + V_{2,\sigma}(t))$ and set
$R_\sigma(t):= e^{-W_\sigma}$. This dressing transformation is unitary. We investigate the candidate for an evolution generator, which we obtain by the transformation of $iG_{0,\sigma} :=
H_{\text{ren},\sigma}(\cdot) - i\frac{d}{dt}$ on $E$, and calculate
\begin{multline*}
R^{-1}\left(  H_{\text{ren},\sigma} - i\frac{d}{dt}\right) R =
H_0 -i\frac{d}{dt}\\+ \sum_{n=1}^\infty \frac{1}{n!}
\left(\ad_{W_\sigma} \right)^{n-1}\left[ \ad_{W_\sigma}(V_\sigma) - V_{0,\sigma}
- V_{2,\sigma} -i\ad_{W_\sigma}\left(\frac{d}{dt}\right) \right],
\end{multline*}
again using analytical vectors.
The counter term as well as the interaction term containing two creators  but no
$\Gamma$-factor cancel exactly. Expanding the terms up to second order explicitly, we find
\begin{align}
R^{-1}&\left(   H_{\text{ren},\sigma} - i\frac{d}{dt} \right) R = H_0 -i
\frac{d}{dt} + V_{1,\sigma}\notag\\
&+ \underset{1}{\wick[u]{1}{<*{\Gamma V_{0,\si}}>*{V_{2,\si}}}} - 
\underset{1}{\wick[u]{1}{<*{V_{0,\si}}>*{\Gamma V_{2,\si}}}}
+ \frac{1}{2}\underset{1}{\wick[u]{1}{<*{V_{0,\si}}>*{\Gamma V_{2,\si}}}}
- \frac{1}{2}\underset{1}{\wick[u]{1}{<*{\Gamma V_{0,\si}}>*{V_{2,\si}}}}\notag\\
&+ \underset{1}{\wick[u]{1}{<*{V_{0,\si}}>*{V_{1,\si}}}}
-\underset{1}{\wick[u]{1}{<*{V_{1,\si}}>*{\Gamma V_{2,\si}}}}
+ i \Gamma \dot{V}_{0,\si} +i \Gamma \dot{V}_{2,\si}\notag\\
&+\frac{1}{2} \left[ (\Gamma V_{0,\sigma} + \Gamma V_{2,\sigma}) , 
\left(\underset{1}{\wick[u]{1}{<*{\Gamma V_{0,\si}}>*{V_{2,\si}}}}
- \underset{1}{\wick[u]{1}{<*{V_{0,\si}}>*{\Gamma V_{2,\si}}}}
- \underset{1}{\wick[u]{1}{<*{V_{1,\si}}>*{\Gamma V_{2,\si}}}}\right) \right]\notag \\
&+ \frac{1}{2} [\Gamma V_{0,\si}, \Gamma\dot{V}_{2,\si}] 
+ \frac{1}{2} [\Gamma V_{2,\si}, \Gamma\dot{V}_{0,\si}]\notag\\
&+\sum_{n=1}^\infty \frac{1}{n!}
\left(\ad_{W_\sigma} \right)^{n-1}\left[ \ad_{W_\sigma}(V_\sigma) - V_{0,\sigma}
- V_{2,\sigma} -i\ad_{W_\sigma}\left(\frac{d}{dt}\right) \right]. 
\label{sum} 
\end{align}
There are no dangerous terms left and one could envisage the discussion of the
existence of local scattering operators for the regularized interaction of the
right-hand side. 
To this end, we first notice that the right-hand side is symmetric. All terms
have at most two external creators or annihilators which are not contracted.
The higher orders can be discussed in terms of connected Friedrichs' graphs
\cite{friedrichsbook}. For superrenormalizable models, the results of Hepp
\cite[Theorem~4.5]{hepp} and Glimm \cite[Theorem~2.2.1]{glimm68a}
 indicate
that, as the number of vertices in a graph increases, the kernel decreases more
rapidly. Therefore, because there are no dangerous terms left in lower orders,
the situation becomes even better in higher orders and we could envisage an
analysis of summability of \eqref{sum} by similar arguments as in
\cite{schrader}. Then applicability of Theorem~\ref{th:exlso} for the existence
of local scattering operators could be investigated and we conjecture an
affirmative outcome.  Furthermore, we expect that this strategy can be extended
to cover other superrenormalizable models like $Y_2$ or even $(\ph^4)_3$.
However, it might be possible to choose the dressing transformation in a more
convenient way. As it is important to retain unitarity of the dressing
transformation, constructions similar to Federbush's in \cite{federbush,
federbushgidas} or a Pad\'e approximation of the formal wave operator may be
interesting starting points. 
Unfortunately,  we are not able to present results in this direction in the present work.

\chapter*{Conclusions and Outlook}
\addcontentsline{toc}{chapter}{Conclusions and Outlook}

In this thesis, we have established the existence of local scattering operators
for
$P(\ph)_2$ models. To this end, we investigated the Cauchy problem for the
time-dependent Schr\"odinger equation in a general context. We used the theory
of evolution semigroups to attack the wellposedness problem. The strategy is as
follows: We investigated the extensions of a certain operator and found
sufficient conditions that an extension exists which is the generator of an
evolution semigroup. For Hilbert spaces and self-adjoint Hamiltonians we saw
that this extension corresponds to a unique approximative solution of the Cauchy
problem of the time-dependent Schr\"odinger equation. This is a new
wellposedness
result requiring considerably less assumptions than Kato's classical
wellposedness
theory. We sacrifice some regularity of the solution. 

In future investigations, it may be possible to
eliminate the exceptional set $N$ of measure $0$, where strong continuity of $t
\mapsto U(t,s)$ fails to hold. One may have the impression that this set is a
technical artefact. If it can be ruled out, the notion of approximative
solutions may be modified to involve strong instead of weak convergence of the
approximating solutions to an admissible bounded approximation of the
generators.

However, at the moment we do not see if it is possible to develop an even more
general wellposedness theory without requiring fundamentally new concepts. In
special situations it is possible to use similarity transformations to extend
the scope of application of our wellposedness result, as we demonstrated with
Goldstein's example. 

In the course of our discussion, we investigated the time evolution of a
$\ph^4$ theory on a two-dimensional, curved spacetime and extended a result of
Dimock. We do not expect that it is possible to extend similar results to
higher dimensions because of the results of Torre and Varadarajan
\cite{evolutioncst}: It is problematic to formulate the concept of time
evolution from Cauchy surface to Cauchy surface even for free theories.
But there are various other fields of physics where our solvability result for
the
time-dependent Schr\"odinger equation may be useful: In applications to
quantum optics \cite{qoptics},  a formal method due to Lewis and Riesenfeld
\cite{lr} is used. 
In the context of general relativity, the influence of an expanding universe on
a quantum system is investigated in  \cite{genrel}.  Yajima uses techniques
close to Howland's ideas to investigate the Stark effekt \cite{yajima}, compare
also \cite{yajima2}.
The approach of Asch et.al. \cite{ahs} for the examination of the adiabatic
properties of a Landau Hamiltonian correspond to the mild solutions in the
present work, an extension to approximative solutions might give a considerable
simplification.

Our main application aims at quantum field theory.
We establish the existence of local scattering operators for $P(\ph)_2$
models. To our knowledge, this is the first proof of existence of these 
in a quantum field theoretical model with nonlinear field equation. The local
scattering operators enable us to access the algebras of local observables by
the general construction of R.~Brunetti and K.~Fredenhagen \cite{FB}, even for
models
having no
ground state. This demonstrates the disentanglement of the ultraviolet and the
infrared problem, the main advantage of the approach.
Examples are models with a non-semibounded polynomial $P(\lambda)$ or the case
of
a massless boson in two dimensions. For future investigations, we find it
interesting to examine the
isomorphy of the algebra of local observables in this case as constructed using
local scattering operators or by the indefinite-metric approach of Streater and
Wilde explicitly. Furthermore, the direct and easily accessible construction of
the local algebras via local scattering operators may have some benefit for the
investigation of other properties of interacting fields as for example
nuclearity in the sense of Buchholz and Wichmann \cite{buchholzw}.

Evidently, future investigations should focus on local scattering operators
for models with infinite renormalizations. We conjecture that our approach
together with the technique of dressing transformations is sufficiently general
to include superrenormalizable models which are accessible via a Hamiltonian
construction, for example $Y_2$. However, we also see the limitations of the
approach: On the side of the existence result, a further generalization seems to
be out of reach. From a quantum field theoretical point of
view, it is the manifestly Hamiltonian formulation which makes it difficult to
use advanced methods from constructive quantum field theory. We think that an
attempt to analyze the existence of local scattering operators for
renormalizable models, as the Gross-Neveu model in two dimensions, may show
whether
the approach to the local net via local scattering operators is powerful enough
to revitalize the interest in constructive quantum field theory.

\cleardoublepage

\thispagestyle{plain}
\subsection*{Acknowledgments}

I would like to express my gratitude to the supervisor of this thesis, Professor
Klaus Fredenhagen. He not only initiated this work, but also his constant support and encouragement as well as his patient explanations and guidance were indispensable to bring it to an end. 

Furthermore, I am highly indepted to Dr.~Roland Schnaubelt. I
would like to thank him for freely sharing his ideas with me
during our collaboration and in particular for his patience and friendly support.

I would like to thank Professor W.~F.~Wreszinski and Dr.~Romeo Brunetti for 
helpful discussions. I am indepted to  Dr. Martin Porrmann and Ralf
\mbox{Diener}, who
carefully proofread the manuscript. Their
numerous remarks were of indispensable benefit.

Thanks to all the members of the local quantum physics group at the II. Institut
f\"ur Theoretische Physik, Universit\"at Hamburg.

My special thanks to my companion in life,  Nicole Levai, and to our
beloved son Iven for the day-to-day sacrifices they made in order to help me
and for everything else.

The financial support of the Evangelisches Studienwerk e.V.~Villigst is
gratefully acknowledged. 

\clearpage
\thispagestyle{plain}

\begin{appendix}

\renewcommand{\theequation}{A.\arabic{equation}}
  \setcounter{equation}{0}

\chapter{Scalar quantum field theory in two dimensions}
\label{qftbasics}

We give a short summary of some basic facts about polynomial
self-interacting, scalar massive quantum field theory in two dimensional
spacetime, mainly
to fix our notation. The proofs of all results which are not mentioned
explicitly can be found in the standard reference \cite{bqft}. We follow the
exposition given there.

Let $\cH_{0} = \mathbb{C}$ and $\cH_{n} = \{ \psi \in
L^{2}(\mathbb{R}^{n}) : \psi(k_{1}, \dots ,k_{n}) \;\text{is
symmetric}\}$. The Fock space is defined as
$\cH= \bigoplus_{n=0}^{\infty} \cH_{n}$.
The subspace $\cH_{n}$ is the ``$n$-particle subspace'' of
$\mathcal{H}$, the vector $\Omega = 1 \subset \mathcal{H}_{0}$ is called the
``vacuum''.
For $\psi = \{\psi_{0}, \psi_{1}, \dots ,\psi_{n}, \dots\} \in
\mathcal{H}$, the number operator $N$ is defined by
$$
\psi \mapsto N\psi = \{0,
\psi_{1}, 2\psi_{2}, \dots ,n \psi_{n}, \dots \}
$$
on the domain $D(N)
= \{ \psi \in \mathcal{H} : \sum_{n=0}^{\infty} \|n \psi_{n}
\|_{2}^{2} < \infty \}$.

Let $\mu(k) = (k^{2} +
m^{2})^{1/2}$ for $ k \in \mathbb{R}$ and $m >0$. The free Hamiltonian
$H_{0}$ maps $\psi$ on  $H_{0}\psi$ given by
\begin{equation}
\label{eq:h0}
\psi \mapsto H_{0} \psi = \{0,
\mu(k_1) \psi_{1}, (\mu(k_1) + \mu(k_2))\psi_{2}, \dots ,\textstyle
\sum_{i=1}^{n} \mu(k_{i}) \psi_{n},
\dots \}
\end{equation}
for $\psi$ in $D(H_{0}) = \{ \psi \in \mathcal{H} : 
\sum_{n=1}^{\infty} \| \sum_{i=0}^{n} \mu(k_i) \psi_{n} \|_{2}^{2} < \infty \}$.

Denote by $S_{n}$ the projection of $L^{2}(\mathbb{R}^{n})$ on its
symmetric part $\mathcal{H}_{n}$. The finite linear combinations of
$\Omega$ and of vectors of the form $S_{n} f_{1}(k_{1}) \dots
f_{j}(k_{j})$ for $f_{i} \in \mathcal{S}(\mathbb{R})$ span the dense
subset $\mathcal{D}$ of $\mathcal{H}$. It is an invariant domain for
the annihilation operator $a(k), k \in \mathbb{R}$, defined by
$$
(a(k)\psi)_{n}(k_{1}, \dots, k_{n}) = (n+1)^{1/2}\psi_{n+1}(k, k_{1},
\dots, k_{n}).
$$
The operator $a(k)$ is not closable, its formal adjoint $a^{\ast}(k)$ is
defined as a quadratic form on $\mathcal{D}\times\mathcal{D}$. The
commutation  relations of $a$ and $a^{\ast}$ are
$[a(k), a^{\ast}(k')] = \delta(k-k').$
For $\psi_{1}, \psi_{2} \in \mathcal{D}$, $(\psi_{1},
a^{\ast}(k_{1})\dots a^{\ast}(k_{m}) a(k'_{1}) \dots a'(k_{n}) \psi_{2})$ is
a function in $\mathcal{S}(\mathbb{R}^{m+n})$. Therefore, every
distribution $w \in \mathcal{S}'(\mathbb{R}^{m+n})$ defines a
bilinear form $W$ on $\mathcal{D}\times \mathcal{D}$ by
$$
W= \int a^{\ast}(k_{1})\dots a^{\ast}(k_{m}) w(k_{1}, \dots, k_{m};
k'_{1}, \dots, k'_{n}) a(k'_{1})\dots a(k'_{n}) \, d^mk\,d^nk'.
$$
It is called a Wick monomial of order $(m,n)$.

An important property of Wick monomials is that they can be dominated by powers
of $N$ if
their kernels are sufficiently regular.
\begin{theorem} \label{th:n}
Let $W$ be a Wick monomial with kernel $w$ which is a bounded
operator from $S_{n}L^{2}(\mathbb{R}^{n}) \to S_{m}L^{2}(\mathbb{R}^{m})$ with
norm
$\|w\|$. Then  $(N + 1)^{-m/2} W (N+ 1)^{-n/2}$ is a
bounded operator
and
\begin{equation}\label{eq:nlemma}
\|(N+1)^{-m/2} W (N+1)^{-n/2}\| \leq  \|w\|.
\end{equation}
Moreover, let $a+b \geq m+n$. Then
$$
\|(N + 1)^{-a/2} W (N + 1)^{-b/2}\| \leq  (1+ |m-n|)^{|a-m|/2}j^{-(a+b)/2} \| w
\|.
$$
\end{theorem}
\begin{cor}
\label{cor:operator}
With the same assumptions on the kernel of $W$, the bilinear form
defines an operator in the domain $D(N^{(m+n)/2})$.
\end{cor}

Denote by $\mathcal{V}(\mathcal{S}')$ the set of bilinear forms on $\mathcal{D}
\times
\mathcal{D}$ which are given as
$$
V = \sum_{j=0}^{p} \binom{p}{j} \int v(k_{1}, \dots, k_{p}) a^{\ast}(k_{1})
\dots
a^{\ast}(k_{j})a(-k_{j+1}) \dots a(-k_{p}) \; d^p\!k
$$
with $v\in \mathcal{S}'$ symmetric with real Fourier transform. These are
special Wick
monomials. For $\mathcal{X} \subset \mathcal{S}'$ let $\mathcal{V}(\mathcal{X})$
be the
subset of $\mathcal{V}(\mathcal{S}')$ with kernels restricted to $\mathcal{X}$.

There is a different realization of $\mathcal{H}$ as square-integrable functions
on a finite
measure space $(Q,dq)$, the so-called Schr\"odinger or $Q$-space. From
$L^{2}(Q,dq)$,
Fock  space is recovered as a Hermite expansion. The advantage of the $Q$-space
realization is
 that every $V \in \mathcal{V}(L^2)$ is represented by a multiplication
operator:
\begin{theorem}
\label{th:qspace}
Let $V \in \mathcal{V}(L^2)$. Then $V$ is essentially selfadjoint on $\mathcal
{D}$. Moreover, $V \in L^r(Q, dq)$ for all $r < \infty$. If
$v = \ker V$
is the integral kernel of $V$, then $\|V\|_r \leq \|v\|_2$.
\end{theorem}

The free Hamiltonian $H_{0}$ from \eqref{eq:h0} can be written as a Wick
monomial
$H_{0} =\int \mu(k) a^{\ast}(k) a(k) \; dk$
with a kernel from $\mathcal{S}'(\mathbb{R}^{2})$. Nevertheless, it is an
operator in
contrast to the free field, which is defined by
$\varphi(x) = (4\pi)^{-1/2} \int e^{-ikx} \mu(k)^{-1/2}(a^{\ast}(k) + a(-k)) \;
dk$ for $x
\in \mathbb{R}$ as a bilinear form.

The free Hamiltonian and the free field are the building blocks for
the  $P(\varphi)_{2}$ model. Let $P(\lambda) := \sum_{i=0}^{n} a_i \lambda^i$
be a real polynomial. The Hamiltonian is formally given as $H(t) := H_0 +
V(t;g)$ with $V(t;g) := \int g(t,x) :P(\ph(x)): \,dx$. The function $g:\RR^2
\to \RR$ is chosen such that the Wick monomials in  $V(t;g)$
are elements in $\mathcal{V}(L^2)$. In particular, the choice $g \in
C^\infty_c(\RR^2)$ is possible. In the following, we concentrate on the special
case $P(\lambda) = \lambda^p,\ p\in \NN$, for simplicity. Then the interaction
Hamiltonian is given by
\begin{equation}
\begin{split}
\label{eq:interaction}
V(t;g) &= \int g(t,x):\varphi^{p}(x): \,dx\\ &= (4\pi)^{-p/2}
\sum_{j=0}^{p} \binom{p}{j} \int
\hat{g}(t, {\textstyle \sum_{i=1}^{p}}k_{i}) {\textstyle \prod_{i=1}^{p}}
\mu(k_{i})^{-1/2}
\times \\
&\phantom{(4\pi)^{-p/2} \sum_{j=0}^{p} \binom{p}{j} \int}
\times a^{\ast}(k_{1})\dots
a^{\ast}(k_{j}) a(-k_{j+1}) \dots a(-k_{p}) \, d^{p}k,
\end{split}
\end{equation}
where $\hat{g}(t,k) = \int e^{-ikx} g(t,x) \; dx$ is the Fourier transform of
$g$ with
respect to $x$. For $x \mapsto g(t,x) \in L^2(\RR)$ the Hamiltonian $V(t;g)$ is
an element from
$\mathcal{V}(L^{2}(\mathbb{R}^{p}))$. This
is
seen by writing the kernel as a convolution and using Young's inequality
\cite{RSII}.
Hence, by
Corollary
\ref{cor:operator},
$V(t;g)$  is a densely defined
operator on $\mathcal{H}$ and Theorem \ref{th:qspace} implies essential
self-adjointness on $\mathcal{D}$.

To deal with the operator sum $H(t) = H_0 + V(t;g)$, further assumptions are
necessary. In the following, let $p$ be even and $g \geq 0$.
Introducing the free field with an ultraviolet cut-off $\kappa$ by
$$\varphi_{\kappa} = (4\pi)^{-1/2} \int_{|k| \leq \kappa} e^{-ikx}
\mu(k)^{-1/2}(a^{\ast}(k)
+ a(-k)) \, dk,
$$
we define the cut-off interaction Hamiltonian:
$V_\kappa(t;g) = \int :\varphi_{\kappa}^{p}(x): g(t,x) \,dx.$
It has an expansion in Wick monomials similar to \eqref{eq:interaction} with the
numerical kernels
replaced by
$$
v_{\kappa}(k_1, \dots, k_p) = \textstyle \prod_{i=1}^{p}
\chi_{\kappa}(k_i)\mu(k_{i})^{-1/2}
\hat{g}(t, \sum_{i=1}^{p} k_i),
$$
where $\chi_{\kappa}$ is the characteristic function of the interval $[-\kappa,
\kappa]$.
For fixed $t \in \RR$ and $\kappa >0$,
 $V_\kappa(t;g)$ is essentially self-adjoint on
$\mathcal{D}$. This follows directly from the above result on
Wick monomials with $L^2$-kernels.

Whereas $V(t;g)$ is an unbounded operator, $V_\kappa(t;g)$ is
semibounded
from below.
\begin{lemma}
\label{lemma:semib}
Given a function $g \geq 0$ on $\mathbb{R}^{2}$ such that  $g(t,\cdot)
\in L^{1}(\mathbb{R}) \cap L^{2}(\RR)$ and $\sup_{t \in \mathbb{R}}
\|g(t,\cdot)\|_{1} < \infty$,
the operator $V_\kappa(t;g)$
is semibounded from below uniformly in $t$. The lower bound is
$\mathcal{O}((\log
\kappa)^{p/2})$.
\end{lemma}
\begin{proof}
Let $c_{\kappa} := (4\pi)^{-1} \int_{|k| \leq \kappa} \mu^{-1}(k) \;dk =
\mathcal{O}(\log \kappa)$.
Wick's theorem states
$$
:\varphi^{p}_{\kappa}(x): = \sum_{j=0}^{[p/2]} (-1)^{j}
\frac{p!}{(p-2j)!j!2^{j}}
c_{\kappa}^{j} \varphi^{p-2j}(x)
$$ as an operator identity on $\mathcal{D}$, see \cite{bqft}. Since $p$ is even,
the right-hand side is a semibounded polynomial in $\varphi$. Therefore
$:\varphi^{p}_{\kappa}(x):
\geq -M c_{\kappa}^{p/2}$ and
$H_{I}^{g}(t;\kappa) \geq -M c_{\kappa}^{p/2} \sup_{t \in \mathbb{R}} \|
g(t,\cdot)\|_{1}$
\end{proof}

The semiboundedness of $V_\kappa(t;g)$ enables one to prove the next theorem.
\begin{theorem}
$H_\kappa(t):= \overline{H_{0} + V_\kappa(t;g)}$ is essentially self-adjoint on
$D(H_{0}) \cap D(V_\kappa(t;g))$. Moreover,
$\Ran e^{-r H(t;\kappa)} \subset D(H_{0}) \cap D(\tilde{V})$ for every $r>0$ and $\tilde{V} \in
\mathcal{V}(L^{2}(\mathbb{R}^{p}))$.
\end{theorem}

By this theorem $V_\kappa(t;g)$ fullfills the prerequisites for the
application of the following theorem in $Q$-space.
\begin{theorem}
\label{th:kernel}
Let $V \in L^r(Q,dq)$ for some $r \geq 1$ such that $H_0 +V$ is essentially
self-adjoint
on $D(H_0) \cap D(V)$. Suppose that $-M \leq V$ for some $M\geq 0$. Then the
integral kernels fulfill
\begin{equation}
\label{eq:semib}
0 \leq \ker e^{-r (H_0 + V) } \leq \ker e^{rM} e^{-r H_0}.
\end{equation}
\end{theorem}
The proof uses the Trotter product formula.
The Duhamel formula,
\begin{equation}
\label{eq:duhamel}
e^{-r H_\kappa(t)} = e^{-r H_{\kappa'}(t)} - \int_{0}^{r} e^{-s H_{\kappa'}(t)}
(H_\kappa(t)
- H_{\kappa'}(t)) e^{-(r-s)H_\kappa(t)} \,ds,
\end{equation}
can be used to generate an expansion of $e^{-r H_\kappa(t)}$ in $\kappa$. This
expansion
gives essential self-adjointness and a lower bound for $H(t)$.
\begin{theorem}
\label{th:sa}
$H(t) = \overline{H_0 + V(t;g)}$ is essentially self-adjoint on $D(H_0) \cap
D(V(t;g))$ and semibounded from below. Moreover,
$\Ran e^{-r H(t)} \subset D(H_0) \cap D(\tilde{V})$ for $r>0$ and $\tilde{V} \in
\mathcal{V}(L^2(\mathbb{R}^p))$.
\end{theorem}
\begin{remark}
\label{cores}
Other cores for $H(t)$ are $D(H_{0})\cap D(N^{p/2})$ and $C^\infty(H_0) =
\bigcap_{n=0}^\infty D(H_0^n)$, see \cite{gj}.
\end{remark}

\begin{theorem}
Let $0 \leq g \leq 1$  and let $(t,x) \mapsto h(t,x)$ be a real function with $\|h\|_\infty  \leq 1$. Set $D = \diam \supp h <
\infty$.
\label{th:pertofh(g)}
Then there exists a constant $M$, independent of $g$ and $h$, such that
$$
|\inf \sigma(H_0 + V(t;g+h))| \leq MD.
$$
\end{theorem}
For the proof, see \cite{pertofham}.

The full
Hamiltonian is  defined by $H(t) = H_0 +V(t;g) +E$ for every $t \in \RR$, where
$E>0$ is a finite renormalization constant. As we see in the following, it can
be chosen in such a way that $H(t)$ is positive on compact time intervals.
\begin{lemma}\label{semib}
Let $I = [a,b] \subset \RR$ be an arbitrary compact interval. Then there is a
constant $E>0$ such that $H(t) >0 $ for all $t \in I$.
\end{lemma}
\begin{proof}
By \cite[Theorem 5.8]{DG}, there are
$c >0$ and $p \in \NN$ independent of the
interaction Hamiltonian such that $H(t)$ is semibounded. Set $\tilde{H}(t):=
H_0 + V(t;g)$. Explicitly one finds
$$
\tilde{H}(t) \geq -c -\ln \| e^{-V(t;g)}\|_{L^p(Q,d\mu)},
$$
where the measure space $Q$ refers to the Q-space representation of
the Fock space, see above or \cite[Section 5]{DG}. For $\lambda>0$ sufficiently
large, the
resolvent $R(\lambda, \tilde{H}(t))$ of $\tilde{H}(t)$ is defined and, as $Y$
is a core, we conclude norm continuity of $t\mapsto R(\lambda,
\tilde{H}(t))$. In turn we have norm continuity of the semigroup
$e^{i\lambda' \tilde{H}(t)}$ in the parameter $t$ (which is not the
semigroup parameter).
The formula
$$
\inf \sigma( \tilde{H}(t)) = \lambda'^{-1} \ln \|e^{i\lambda'
\tilde{H}(t)}\|
$$
shows that the infimum of the spectrum of $\tilde{H}(t)$ is continuous in $t$.
As we have $t$ in a compact interval,
we can define $E = 1 + | \min_t \inf \sigma(\tilde{H}(t))|$ and  get $H(t) >
0$.
\end{proof}
\newpage
\thispagestyle{empty}
\cleardoublepage

\chapter{Maximal accretivity of an operator sum}

\renewcommand{\theequation}{B.\arabic{equation}}
  \setcounter{equation}{0}

For the proof of the Theorem of Sohr we need a lemma which goes back to
\cite{sohr77}.
 This lemma shows in particular the closedness of the sum
$B+C$ on $D(B)\cap D(C)$ provided
that the maximally accretive operators $B$ and $C$ satisfy the `angle condition'
\eqref{angle}.
\begin{lemma} \label{sohr}
Let $B$ and $C$ be maximally accretive operators on a Hilbert space
$X$ such that $C^{-1} \in \cB(X)$. Let $\cD$ be a core
of $B^{\ast}$. If there exists $0 \leq a < 1$ such that
\begin{equation}\label{angle}
\Rea (B^{\ast}x, C^{-1} x) + a \|x\|^{2} \geq 0
\end{equation}
for every $x \in\cD$, then $B+C$ is maximally accretive
on $D(B+C) = D(B) \cap D(C)$.
\end{lemma}
With respect to our application we remark that in general the angle condition is
violated for $B$ being the time derivative and $C$ being the
generator of a \emph{hyperbolic} evolution equation.
\begin{proof}
On Hilbert
spaces maximal accretivity of an operator $A$ is equivalent to
accretivity and $\Ran A+ \beta = \mathcal{H}$ for $\beta > 0$.
Clearly, the sum $B+C$ is accretive, so it remains to prove $\Ran
B+C+\beta = \mathcal{H}$.
 
Given $y \in \mathcal{H}$, we are looking for  $x \in D(B) \cap D(C)$
with $(B+C+\beta)x = y$. The Yosida approximants $B_{n}= B(1
+ \frac{1}{n}B)^{-1}$ and $C_{n}= C(1 + \frac{1}{n}C)^{-1}$
are bounded and maximally accretive operators for every $n \in
\mathbb{N}$, which converge strongly on $D(B)$ respectively $D(C)$.
 
Assume that 
\begin{equation} 
\label{eq:assumption} \|B_{n}x + C_{n}x\|
\geq (1-a)\|C_{n}x\| \end{equation} 
for all $x \in \mathcal{H}$ and $n
\in \mathbb{N}$. By the boundedness and accretivity of $B_{n} +
C_{n}$, there exists for every $n$ an $x_{n}$ such that $(B_{n} + C_{n}
+ \beta) x_{n} = y$. The sequences $\{B_{n}x_{n}\}_{n \in
\mathbb{N}}$, $\{C_{n}x_{n}\}_{n \in \mathbb{N}}$ and $\{x_{n}\}_{n
\in \mathbb{N}}$ are bounded. This follows from 
\begin{align*}
\|y\|^{2} &= \|B_{n}x_{n} + C_{n}x_{n}\|^{2} + \beta^{2} \|x_{n}\|^{2}
+ 2 \beta \Rea (B_{n}x_{n} + C_{n}x_{n}, x_{n}) \notag\\ &\geq (1-a)
\|C_{n}x_{n}\|.  
\end{align*}
 Without loss of generality we can assume
the sequence itself to be weakly convergent.  Let $x$ be the weak
limit of   $\{x_{n}\}_{n \in \mathbb{N}}$. One shows $x \in D(B^{\ast
\ast}) = D(B)$. For $z \in D(B^{\ast})$ one finds 
\begin{equation*} |
\lim_{n \to \infty} (x_{n}, (B_{n}^{\ast}- B^{\ast})z | \leq \sup_{n
\in \mathbb{N}}\|x_{n}\| \lim_{n \to \infty} \|(B_{n}^{\ast}
-B^{\ast})z\| = 0 
\end{equation*} 
because of the strong convergence of
$B_{n}^{\ast}z$ to $B^{\ast}z$. Therefore, one has 
\begin{equation*}
\lim_{n \to \infty}(B_{n}x_{n},z) = \lim_{n \to \infty}(x_{n},
B^{\ast}z) = (x, B^{\ast}z) \end{equation*}
and it follows $|(x, B^{\ast}z)| \leq \sup_{n \in \mathbb{N}} \|B_{n}x_{n} \|
\|z\|$. This
is the continuity of the linear functional $z \mapsto (x , B^{\ast}
z)$, so $ x \in D(B^{\ast\ast}) = D(B)$ and the equation $\lim_{n \to
\infty} (B_{n}x_{n}, z) = (Bx,z)$ extends to all $z \in \mathcal{H}$.
With the same reasoning one shows $x \in D(C)$ and $ \lim_{n \to
\infty} (C_{n}x_{n}, z) = (Cx,z) $ for all $z \in \mathcal{H}$. One
concludes from $y = (B_{n} + C_{n} +\beta) x_{n}$ for every $n$ that $y=
(B + C +\beta)x$ and $x \in D(B)\cap D(C) =D(B+C)$.
 
It remains to prove the estimate \eqref{eq:assumption}:  $\|B_{n}x +
C_{n}x\| \geq (1-a)\|C_{n}x\|$.
 
Let $y_{n} := C_{n} x$ and $z_{n} := (1 + \frac{1}{n}
B^{\ast})^{-1} y_{n}$. By the accretivity of $B_{n},\ B^{\ast}$ and $C^{-1}$,
we find 
\begin{align*} 
\Rea(B_{n}x, C_{n}x) &=
\Rea(B_{n}C^{-1}_{n}y_{n}, y_{n}) = \Rea( B_{n}( C^{-1} +
\tfrac{1}{n}) y_{n}, y_{n}) \\ &\geq \Rea( B_{n} C^{-1}y_{n}, y_{n}) =
\Rea(C^{-1}(1 + \tfrac{1}{n} B^{\ast}) z_{n}, B^{\ast}
z_{n}) \\ &\geq \Rea( C^{-1}z_{n}, B^{\ast}z_{n}). 
\end{align*}
Besides
we have $\|y_{n}\|^{2} \geq \|z_{n}\|^{2}$ and conclude 
\begin{align*}
\|(B_{n} + C_{n})x\| \|C_{n}x\| &\geq \Rea((B_{n} + C_{n})x,C_{n}x) =
\Rea (B_{n}x, C_{n}x) + \|C_{n}x\|^{2}\\ &= \Rea(B_{n}x, C_{n}x) + a
\|y_{n}\|^{2} + (1-a) \|C_{n} x\|^{2}\\ &\geq \underbrace{\Rea(
C^{-1}z_{n}, B^{\ast}z_{n}) + a \|z_{n}\|^{2}}_{\geq 0 \; \text{by
assumption}} + (1-a) \|C_{n}x\|^{2}\\ &\geq (1-a)\|C_{n}x\|^{2}.
\qedhere 
\end{align*} 
\end{proof}

\end{appendix}

\addcontentsline{toc}{chapter}{Bibliography}
\bibliography{thesis}
\bibliographystyle{hplain}

\end{document}